\documentclass[times,twocolumn,final]{elsarticle}

\usepackage{xcolor}
\usepackage{medima}
\usepackage{framed,multirow}
\usepackage{graphicx}
\usepackage{booktabs}
\usepackage{tabularx}
\newcolumntype{Y}{>{\centering\arraybackslash}X}
\usepackage{amsmath,amsfonts,bm}
\usepackage{amssymb}
\usepackage{latexsym}
\usepackage{tabularx, booktabs}
\usepackage{url}
\usepackage[pdfauthor=icovid consortium]{hyperref}
\usepackage{float}
\begin{document}

\begin{frontmatter}

\title{Comparative study of deep learning methods for the automatic segmentation of lung, lesion and lesion type in CT scans of COVID-19 patients}

\author[1,2,7]{Sofie Tilborghs\corref{cor1}}
\author[1,3,8]{Ine Dirks\corref{cor1}}
\author[1,4]{Lucas Fidon\corref{cor1}}
\author[1,2,7]{Siri Willems\corref{cor1}}
\author[1,2,7]{Tom Eelbode\corref{cor1}}
\author[1,2,7]{Jeroen Bertels\corref{cor1}}
\author[1,5]{Bart Ilsen}
\author[1,6]{Arne Brys}
\author[1,13]{Adriana Dubbeldam}
\author[1,5]{Nico Buls}
\author[1,3,8]{Panagiotis Gonidakis}
\author[1,3,8]{Sebasti\'an Amador S\'anchez}
\author[1,9,10]{Annemiek Snoeckx}
\author[1,9,11,12]{Paul M. Parizel}
\author[1,5]{Johan de Mey}
\author[1,2,7]{Dirk Vandermeulen}
\author[1,4]{Tom Vercauteren}
\author[1,2,6,7]{David Robben}
\author[1,6]{Dirk Smeets}
\author[1,2,7]{Frederik Maes}
\author[1,3,8]{Jef Vandemeulebroucke}
\author[1,2,7]{Paul Suetens}

\cortext[cor1]{These authors contributed equally. Order was chosen randomly.}

\address[1]{icovid consortium}
\address[2]{KU Leuven, Department of Electrical Engineering, ESAT/PSI, Leuven, Belgium}
\address[7]{UZ Leuven, Medical Imaging Research Center, Leuven, Belgium}
\address[3]{Vrije Universiteit Brussel (VUB), Department of Electronics and Informatics (ETRO), Brussels, Belgium}
\address[8]{imec, Leuven, Belgium}
\address[4]{School of Biomedical Engineering \& Imaging Sciences, King's College London, London, United Kingdom}
\address[5]{Vrije Universiteit Brussel (VUB), Universitair Ziekenhuis Brussel (UZ Brussel), Brussels, Belgium}
\address[6]{icometrix, Leuven, Belgium}
\address[13]{KU Leuven - UZ Leuven, Department of Imaging and Pathology, Leuven, Belgium}
\address[9]{University of Antwerp (UA), Antwerp, Belgium}
\address[10]{Antwerp University Hospital (UZA), Antwerp, Belgium}
\address[11]{Royal Perth Hospital, Perth, Australia}
\address[12]{University of Western Australia, Perth, Australia}

\begin{abstract}
Recent research on COVID-19 suggests that CT imaging provides useful information to assess disease progression and assist diagnosis, in addition to help understanding the disease.
There is an increasing number of studies that propose to use deep learning to provide fast and accurate quantification of COVID-19 using chest CT scans.
The main tasks of interest are the automatic segmentation of lung and lung lesions in chest CT scans of confirmed or suspected COVID-19 patients.
In this study, we compare twelve deep learning algorithms using a multi-center dataset, including both open-source and in-house developed algorithms.
Results show that ensembling different methods can boost the overall test set performance for lung segmentation, binary lesion segmentation and multiclass lesion segmentation, resulting in mean Dice scores of 0.982, 0.724 and 0.469, respectively. The resulting binary lesions were segmented with a mean absolute volume error of 91.3~ml. In general, the task of distinguishing different lesion types was more difficult, with a mean absolute volume difference of 152~ml and mean Dice scores of 0.369 and 0.523 for consolidation and ground glass opacity, respectively.
All methods perform binary lesion segmentation with an average volume error that is better than visual assessment by human raters, suggesting these methods are mature enough for a large-scale evaluation for use in clinical practice.
Code for all the methods developed in this paper will be made publicly available. 
\end{abstract}


\end{frontmatter}


\section*{List of abbreviations}
\noindent CT = computed tomography;
COVID-19 = coronavirus disease 2019;
RT-PCR = reverse-transcription polymerase chain reaction;
GGO = ground glass opacity;
CON = mix of consolidation and linear opacity;
CPP = crazy paving pattern;
COM = combined pattern (mix of GGO, CON and CPP);
OAT = other abnormal tissue;
SEN = sensitivity;
DSC = Dice score coefficient;
HD95 = Hausdorff distance at 95\% percentile;
ASD = average surface difference;
AVD = average volume difference;
CNN = convolutional neural network;
HU =  Hounsfield units.

\section{Introduction}

Early-stage research studies on coronavirus disease 2019 (COVID-19) have consistently reported the presence of ground glass opacity, crazy paving pattern and consolidation in chest CT scans of COVID-19 patients~\citep{pan2020time,revel2020covid}.
Despite not being specific to COVID-19, the presence, location and extent of those lesion types correlate with different stages of the disease and possibly outcome.
The European Society of Radiology (ESR) and the European Society of Thoracic Imaging (ESTI) have reported that ground glass opacity with predominant peripheral, basal and subpleural location is associated with early stage of the disease, while crazy paving pattern
is associated with a more advanced disease, and consolidation is associated with poorer prognosis in older patients~\citep{revel2020covid,Salehi2020,Ye2020}.

The role of chest CT for the diagnosis and management of COVID-19 patients is still controversial and debated~\citep{wang2020role}.
In a statement in March 2020, the British Society of Thoracic Imaging warned against the use of CT for COVID-19 as a stand-alone diagnostic tool, as CT performed in an early stage of the disease might present no abnormality~\citep{li2020ct,rodrigues2020update}.

In April 2020, an international panel of 15 radiologists and 10 pulmonologists from the Fleischner Society recommended to reserve the use of chest CT for COVID-19 based on symptoms (mild or severe), pre-test probability and resource constraints in the clinical environment~\citep{rubin2020role}.
\cite{rubin2020role} indicate the use of chest CT for confirmed COVID-19 patients with moderate-to-severe features or worsening respiratory status, for suspected COVID-19 patients with co-morbidities and risk factors for disease progression, and for patients with functional impairment after recovery from COVID-19.
Chest CT findings and quantification are considered useful for the assessment of COVID-19 progression and for future long-term follow-up studies of survivors~\citep{revel2020covid,rubin2020role,simpson2020radiological}.

Regarding the diagnosis of COVID-19, reverse-transcription polymerase chain reaction (RT-PCR) remains the gold standard~\citep{revel2020covid}.
However, the time required to obtain the results of RT-PCR tests may be problematic in a resources-constrained clinical environment in which rapid triage of patients is needed.
In this situation, chest CT can be useful for diagnosis and rapid triage~\citep{revel2020covid,rubin2020role}.

Unfortunately, large-scale and accurate quantification of lung lesions that can be visible in the chest CT scans of COVID-19 patients is practically impossible to obtain manually. Indeed, it can take hours for experienced radiologists to accurately delineate the lesions in a single chest CT volume.
As a result, the analysis of chest CT findings has been mainly limited so far to qualitative or semi-quantitative evaluation~\citep{colombi2020well,li2020ct,pan2020time,prokop2020co}.
The development of automated methods for the segmentation of the lungs and lesions visible in CT of COVID-19 patients appears of great importance to unlock the full potential of CT in helping the management of COVID-19 patients and enable more clinical research.

\subsection{Related work}
The application of deep learning methods to chest CT scans in the context of COVID-19 is an emerging field of research that is evolving rapidly.
Most previous research on machine learning-based methods exploiting CT scans of COVID-19 patients have focused on automated diagnosis, i.e. classifying a subject as COVID-19 or not COVID-19
~\citep{shi2020review,dong2020role}.
This is due to the difficulty to gather large datasets with manually segmented CT scans that are required for the application of deep learning methods for segmentation.
Recently, a few works have directed attention towards using Convolutional Neural Networks (CNNs) for automated segmentation of lungs, lung lobes, and the segmentation of lung lesions that can be caused by COVID-19.

\paragraph*{COVID-19 lungs and lung lobes segmentation}
The first segmentation problem to which deep learning was applied using CT of COVID-19 patients was the automatic segmentation of lungs and lung lobes.
This is an essential task, as it is required to localize the lesions and compute the infection ratio of the lungs and lung lobes.

\cite{Hofmanninger2020} proposed to train a 2D U-Net~\citep{Ronneberger2015} on a slice by slice basis for lung segmentation using a combination of several datasets covering various diseases. 
Despite not being designed for COVID-19 cases originally, they recently reported that their method can be applied on COVID-19 chest CT data in their git repository\footnote{\url{https://github.com/JoHof/lungmask}}.
However, they have not evaluated quantitatively the lung segmentations predicted by their method for COVID-19.
\cite{chaganti2020quantification} proposed a two-step approach for the segmentation of lungs and lung lobes of COVID-19 chest CT scans. First, a deep reinforcement learning method~\citep{ghesu2017multi} is used to detect the lung region.
Second, a deep image-to-image network~\citep{yang2017automatic} segments lungs and lung lobes in the detected lung region.
\cite{Xie2020} proposed a CNN for lung lobe segmentation in COVID-19 patients. They introduced a non-local neural network module in their CNN to capture structured relationships. The CNN was initially trained on a large dataset of non-COVID patients and subsequently tuned to COVID-19 pathology by using transfer learning.

\paragraph*{Binary COVID-19 lesion segmentation}
Binary COVID-19 lesion segmentation using CT consists in automatically predicting a mask for all the types of lesion indiscriminately.
This allows to quantify the extent of lesion in general in the lung of a confirmed or suspected COVID-19 patient.

\cite{chaganti2020quantification} proposed to automatically segment ground glass opacities and areas of consolidation together, using a DenseUNet~\citep{Ronneberger2015}.
\cite{wu2020jcs} proposed an encoder-decoder CNN architecture with an attentive feature fusion strategy and deep supervision.
\cite{chassagnon2020ai} proposed CovidENet: an ensemble of 2D and 3D CNNs based on AtlasNet~\citep{vakalopoulou2018atlasnet} for total lesion segmentation and achieved human-level segmentation performance in terms of Dice Score and Hausdorff distance. Combining the predicted total lesion segmentation with other clinical features, they achieved promising results in predicting short-term outcome for confirmed COVID-19 patients.
However, methods for binary lesion segmentation cannot differentiate between different types of lesion.
This is in contrast with the radiology literature that has reported that the type of lesion is related to the progression and the severity of COVID-19~\citep{revel2020covid}.

\paragraph*{Multiclass COVID-19 lesion segmentation}
Methods for the automatic multiclass COVID-19 lesion segmentation in CT aims at predicting a mask for different types of lesion simultaneously.
The methods differ in the set of lesion types that they can distinguish.
At the time of writing, ground glass opacity, consolidation and crazy paving pattern appear as the clinically most important types of lesion to identify and quantify~\citep{revel2020covid,simpson2020radiological}.

\cite{fan2020inf,zhang2020clinically} proposed two-step approaches for the segmentation of different lesion classes.
A first CNN segments the total lesion in~\citep{fan2020inf}, and the total lesion and healthy lung tissue in~\citep{zhang2020clinically}, while a second CNN exploits the predicted segmentation of the first network to segment ground glass opacity (GGO) and consolidation (CON) in~\citep{fan2020inf} and GGO, CON, pulmonary fibrosis, interstitial thickening, pleural effusion and healthy lung tissue in~\citep{zhang2020clinically}.
However, in the work of \cite{zhang2020clinically}, the segmentation task is just an intermediate step for predicting diagnosis, and they performed only limited quantitative segmentation evaluation.

\subsection{Goal of this study and contributions}
The scientific contributions in this work have evolved during the outbreak of COVID-19 in Europe.
In order to support radiologists with quantitative information on lesion tissue inside the lungs, the authors started developing novel algorithms and comparing existing algorithms for lung segmentation, binary lesion segmentation and multiclass lesion segmentation, thereby working in parallel to allow a natural selection and development of algorithms for the three tasks.

Currently, there is no quantitative comparison of the different deep learning methods that have been proposed for the three aforementioned segmentation tasks in chest CT scans of confirmed or suspected COVID-19 cases.
In this study, we aim to compare twelve deep learning methods, both open-source and in-house developed methods, on those segmentation problems using a common multi-center dataset for evaluation.
%
%
In particular, we pay special attention to multiclass lesion segmentation, which so far has received limited attention but has shown clinical potential to stage the progress of the disease~\citep{revel2020covid}.

\section{Material}
\label{sec:data}
The icovid consortium collected non-contrast enhanced chest CT scans that were acquired for triage and staging of potential COVID-19 patients at different centers in Europe and Latin-America. The dataset includes both COVID-19 positive and COVID-19 negative patients, confirmed via RT-PCR testing, and patients with unknown status. 

We randomly selected cases for the training and validation set from 7 centers and we used data from two other centers for the test set. Exclusion criteria are insufficient image quality and comorbidities that have a large impact on the image (in particular lung tumors and pneumothorax).

The dataset used for training and validation contains 74 subjects (33 female, 24 male and 17 unknown). Among them, 42 have a positive RT-PCR test, 24 have a negative PCR test and the remaining have an unknown status.
The test set contains 7 subjects (5 female and 2 male) with suspected but unconfirmed COVID-19 status and is enlarged with 10 extra COVID-19 confirmed cases from a public dataset\footnote{\url{https://zenodo.org/record/3757476\#.XtzlaC2B1QJ}} \citep{COVID-19-CT-Seg-Dataset}.

\subsection{Annotation workflow}
Based on the literature review, we decided to create ground truth segmentations for the lungs, lung lobes and following lesion types: ground glass opacity, consolidation, crazy paving pattern (CPP), linear opacity (LO), reverse halo sign (RHS), combined pattern (COM, used if distinguishing between the mentioned types is impossible), and other abnormal tissue (OAT).
A consensus document was created with guidelines on how to delineate different lesions to ensure consistency and examples of the various labels were provided.
Two different annotation workflows are used. The first is used for the training and validation set. The second workflow was established later in time and is used for the independent test set. 
Using for the test set a different annotation workflow and CT scans of different centers, minimises the bias towards methods trained on our own training/validation set.

\subsubsection{Training and validation set}
\label{subsec:training_data}
The MD.ai platform\footnote{\url{https://matrix.md.ai}} (MD.ai, Inc., US) is used to create manual segmentations for different lesion types on 2D slices of 3mm slice thickness and a varying in-plane resolution. Several radiologists were assigned different cases. Each case was thereafter carefully reviewed by a highly experienced radiologist (BI). In addition, for 51 of these cases lung segmentations were provided by an experienced researcher using 3D Slicer and starting from the manual lesion segmentations.

\subsubsection{Test set}
\label{subsec:test_data}
The Mimics software (Materialise, BE) is used to create semi-automatic 3D segmentations of lungs, lobes and lesion types on the original CT scans by experienced support engineers. The segmentations are reviewed by a highly experienced radiologist (either BI or AD) using the Mimics Webviewer. Based on the comments, corrections are performed by the support engineers and if needed another round of review and corrections is performed. 

The ten cases of the online available dataset only contain ground truth segmentations for the lungs and the binary lesions. For our study, the lesions were assigned to the combined pattern class.

\section{Benchmarking methodology}
\label{sec:icovid_challenge}

In this section, we describe the evaluation method used for each of the three tasks on the aggregated left-out sets (Section~\ref{subsec:training_data}) and the independent test set (Section~\ref{subsec:test_data}).

For the development of in-house methods, all authors had access to the same data (Section~\ref{sec:data}) and they agreed on the same 5-fold split for cross-validation.
%
%
Test set predictions for all three tasks were calculated by averaging the probabilistic predictions obtained by the five models trained during cross-validation.

\subsection{Lung segmentation assessment}
The manual lung masks are compared with their automated counterparts in terms of Dice score (DSC), Hausdorff distance at percentile 95\% (HD95), average surface distance (ASD) and absolute volume difference (AVD). The first three metrics are often used in segmentation challenges~\citep{Menze2015, Winzeck2018} and the AVD is of specific clinical interest, since guidelines are typically based on the lesion volumes~\citep{colombi2020well,li2020ct}. 

The segmentation of the lung tissue at the border of the lung in regions containing consolidation is challenging and ambiguous due to their location and their intensity that is similar to the one of the surrounding extra-pulmonary tissues.
To evaluate the ability of deep learning methods for lung segmentation to tackle this problem, we compared their sensitivity (SEN$^{class}$), i.e. the percentage of lesion in the manual lung segmentation that is correctly covered by the predicted lung segmentation for all the lesion classes defined in Section~\ref{subsec:multiclass_lesion_segmentation}.


\subsection{Binary lesion segmentation assessment}
The manual segmentations contain both pneumonia-specific and general lesion patterns. Therefore, the task of binary lesion segmentation was defined as to automate the segmentation of all abnormalities inside the lungs (BIN),
i.e. 
the union of all lesion types as defined in Section~\ref{sec:data}. Similarly to the lung segmentation task, 
we compare the manual binary lesion masks to their automated counterparts by calculating the DSC, HD95, ASD and AVD. The DSC for cases for which there is no binary lesion present in the manual delineation is not included in the comparison. Furthermore, when there is no manual or automated binary lesion segmentation present, we respectively fill the entire manual or predicted image volume for calculating the HD95 and ASD measures.

\subsection{Multiclass lesion segmentation assessment}
\label{subsec:multiclass_lesion_segmentation}

Due to the variety of lesion types that have been observed in the lungs of COVID-19 patients~\citep{revel2020covid}, several definitions of the multiclass lesion segmentation problem are possible.
Ground glass opacity, consolidation and crazy paving pattern appear as the most important tissue types for the management of COVID-19 patients~\citep{simpson2020radiological}.
As a result, linear opacity has been grouped with consolidation to form the CON class, and reversed halo sign, which is a specific combination of ground glass opacity and consolidation~\citep{Chiarenza2019}, has been grouped with the combined tissue to form the COM class. 
In addition, both linear opacities and reversed halo signs were rare in our dataset, making the evaluation of their segmentation impossible.

We can summarize how the different lesion types were grouped and abbreviated in the following diagram:

\[
    \overbrace{\underbrace{\substack{\text{consol-}\\\text{idation}}+\substack{\text{linear}\\\text{opacity}}}_{
    \let\scriptstyle\textstyle\substack{\text{CON}}}+\text{CPP}+\text{GGO}+\underbrace{\substack{\text{combined}\\\text{tissue}}+\substack{\text{reversed}\\\text{halo sign}}}_{
    \let\scriptstyle\textstyle\substack{\text{COM}}}+\substack{\text{other abn-}\\\text{ormal tissue}}}^{
    \let\scriptstyle\textstyle\substack{\text{BIN}}}
\]

When a specific lesion type is not present in the manual delineation for a given case, the DSC for this case and this lesion type is not included in the comparison.
When there is no delineation present for a certain lesion type, either for the manual or the automatic segmentation, we fill the entire image volume for calculating the HD95 and ASD measures for that specific lesion type. 
Furthermore, all algorithms are trained to differentiate between the different lesion types present in the COM class. As a result, the COM voxels are ignored when calculating the DSC, HD95, ASD and AVD metrics.
%

\subsection{Statistical testing methodology}
Final comparisons for superiority were done in a pair-wise setting between methods and tests for statistical significance were performed using non-parametric bootstrapping, putting only minimal assumptions on the distribution of the test statistic.
A difference was considered significant if p $<$ 0.05.

This approach has been used in many other biomedical benchmarks, including in BRATS~\citep{Menze2015} and ISLES challenges~\citep{Winzeck2018}. The implementation details can be found in \citep{bakas2018identifying}, with the ranks substituted for metric scores.

\section{Automated segmentation methods}

In this section, we describe the deep learning methods for segmentation that are compared in this study.

\begin{table*}[tb]
	\centering
	\caption{Summary of the benchmarked methods and the segmentation tasks they solve.}
	\label{tab:overview}
	\begin{tabular}{l | c  c  c } 
		\toprule
		Method & Lung segmentation & Binary lesion segmentation & Multiclass lesion segmentation\\
		\cmidrule(l){1-4}
		JoHof~\citep{Hofmanninger2020} & \checkmark & & \\ 
		DMLu                                & \checkmark & &\\ 
		DMLo                                & \checkmark & &\\ 
		CTA \citep{Chen2020}                & \checkmark & \checkmark & \\
		CovidENet~\citep{chassagnon2020ai}  & & \checkmark & \\ 
		2DS                                 & & \checkmark & \\
		InfNet~\citep{fan2020inf}           &  & \checkmark & \checkmark \\ 
		DMmc                                & & & \checkmark \\ 
		2DRnx                                 & & & \checkmark\\ 
		WASS                                & \checkmark & \checkmark & \checkmark\\ 
		UNWM                                & \checkmark & \checkmark & \checkmark \\ 
		MAJ                                 & \checkmark & \checkmark & \checkmark \\ 
		\bottomrule
	\end{tabular}
\end{table*}

Table \ref{tab:overview} gives an overview of all benchmarked methods and the tasks they solve, i.e. lung segmentation, binary lesion segmentation, multiclass lesion segmentation or a combination of these tasks.

\subsection{JoHof: Lung segmentation for severe pathologies ~\texorpdfstring{\citep{Hofmanninger2020}}{Hofmanninger et al., 2020}}

\cite{Hofmanninger2020} claim that data diversity, i.e covering multiple diseases in the training dataset, has more impact than the choice of the CNN architecture on lung segmentation performance.
As a result, they used a large training dataset of 231 CTs coming from three different public datasets and their own clinical data that covers a large range of diseases and includes cases with severe pathologies.
The CNN architecture was the 2D U-Net architecture as originally proposed by~\cite{Ronneberger2015}.
We used the pre-trained model (R231) made publicly available\footnote{\url{https://github.com/JoHof/lungmask}} by the authors out of the box for evaluation.
\subsection{DMLu: 3D Lung segmentation using DeepMedic}
\label{siri_method_lung}
A 3D neural network based on DeepMedic architecture \citep{kamnitsas2017} is re-implemented from scratch to perform automatic lung segmentations for COVID-19 suspected patients. 
\paragraph*{Data preprocessing}
The chest CTs were resampled with an integer factor and linear interpolation to be as close as possible to 3 mm slice thickness but with varying in-plane resolution. The Hounsfield Units (HU) intensities were clipped with -1100 HU and 100 HU before rescaling to [0, 1].

\paragraph*{Neural network architecture}
The DeepMedic architecture, first published by \cite{kamnitsas2017} is used and consists of four parallel pathways, which operate on different resolutions: (1, 1, 1), (3, 3, 1), (5, 5, 3) and (9, 9, 3). Each pathway contains ten convolutional (CONV) layers: five with kernel size (3, 3, 1) and five with kernel size (3, 3, 3). This principle allows the network to process fine details while taking a broader context into account with the last layer processing almost the whole region of interest, i.e. the lung region, and therefore preserving spatial context. Each convolutional layer is followed by batch normalization (BN) \citep{Ioffe2015} and ReLU as non-linear activation function. The four pathways are subsequently concatenated followed by a common pathway of two extra CONV-BN-ReLU layers, one layer with kernel sizes (3, 3, 3) and one layer with kernel size ($1 \times 1 \times 1$). The final prediction layer consists of a $1 \times 1 \times 1$ convolution followed by a sigmoid activation function.
 
\paragraph*{Training}
The network is trained using a patch-based approach in which segments of (43, 43, 13) voxels are predicted simultaneously. Data augmentation is performed by adding Gaussian noise and performing random affine transformations on the chest CT images and corresponding ground truth. The 3D neural network is trained using the Adam optimizer~\citep{kingma2014adam} and a batch-size of 8. The objective function to train the network is the binary cross-entropy. A learning rate schedule is used to automatically reduce the learning rate by monitoring the validation loss during training. Early stopping is subsequently used to stop training and avoid overfitting.
 
\paragraph*{Postprocessing}
Connected component analysis is used as a final post processing step to keep the two largest components, i.e. the two lungs. The final outputs of the network were upsampled using the inverse pooling scheme for the out of plane direction using nearest neighbour interpolation.

\subsection{DMLo: Lobe segmentation using DeepMedic ensemble}
\label{sofie_method}
An ensemble of three 3D CNNs, all trained from scratch, is used to perform lung lobe segmentation for COVID-19 suspected patients. 

\paragraph*{Data}
Since no ground truth lung lobe segmentation was available for the training set specified in Section \ref{sec:data}, the training set was ignored and an alternative training set was created using the publicly available LUNA16 dataset \citep{Setio2017} and the recent lung lobe segmentation algorithm of \cite{Xie2020}. 40 subjects for which a ground truth lung lobe segmentation is available \citep{Tang2019} were randomly selected from the LUNA16 dataset. The online lobe segmentation tool\footnote{\url{https://grand-challenge.org/algorithms/pulmonary-lobe-segmentation/}} of \cite{Xie2020} was additionally used to augment the dataset with scans of COVID-19 patients. A total of 40 scans, collected by the icovid consortium but not in the training or test set, with variable level of pathology from two centers was processed and the results were visually checked to not contain large, clear errors. 27 scans for which the lobe segmentation was considered visually acceptable were included as a semi ground truth for training. No manual correction was performed. 

\paragraph*{Data preprocessing}
The chest CTs were resampled into three different resolutions: $1 \times 1 \times 3 \,\textup{mm}^3$, $1 \times 3 \times 1 \,\textup{mm}^3$, and $3 \times 1 \times 1 \,\textup{mm}^3$, maintaining high resolution in two dimensions and subsampling to a lower resolution in the third dimension. The resampled CTs are each used in a different CNN. This way, the details of the fissures are maintained in two out of three volumes for all dimensions. The HU intensities were clipped between $-1250$ HU and 250 HU before rescaling to $[-1, 1]$.

\paragraph*{Neural network architecture}
 Each CNN has a 3D DeepMedic architecture \citep{kamnitsas2017}, identical to DMLu (Section~\ref{siri_method_lung}) with two minor changes: (1) the final prediction layer has changed to a $1\times1\times1\times6$ convolutional layer followed by a softmax activation function, predicting five lobes and background. (2)~The resolutions of the four DeepMedic pathways as well as the convolutional kernel sizes are adapted according to the input resolution, i.e. the dimension with lowest resolution ($3\,mm$) has the properties of the third dimension in DMLu.  

\paragraph*{Training}
The networks are trained using a patch-based approach in which segments of (43, 43, 13), (43, 13, 43) or (13, 43, 43) voxels are predicted simultaneously. We use a class-weighted sampling approach in which the center voxels of the training patches are equally sampled from each class. Data augmentation is performed by addition of Gaussian noise and by applying random affine transformations on the chest CT images and corresponding ground truths. The CNNs are trained for 400 epochs using the Adam optimizer~\citep{kingma2014adam} with a learning rate of 0.001 and a batch size of 8, minimizing the sum of the categorical cross-entropy loss on the lobe segmentation and the binary cross-entropy of the total lungs.

\paragraph*{Postprocessing}
The predicted probability maps are resampled to the original resolution and the predictions of the three CNNs are subsequently averaged. Each voxel is assigned to the class with maximal probability.  

\subsection{CTA: CT Angel software for lung segmentation and binary lesion segmentation~\texorpdfstring{\citep{Chen2020}}{Chen et al., 2020}}
\label{ctangel_method}
The CT Angel method by~\cite{Chen2020} was one of the earlier methods for detecting viral pneumonia, driven specifically by the importance of CT in the diagnosis of COVID-19. The training set consisted of axial slices from 106 patients, selected and annotated by three expert radiologists in consensus. At the core, the models share the U-Net++ architecture~\citep{unetpp}, which was trained for 2D lung segmentation and binary lesion segmentation. their code is open source and includes pre-trained models for lung segmentation and binary lesion segmentation available\footnote{\url{https://github.com/endo-angel/ct-angel}}. Using their preprocessing, we used this system out of the box for generating predictions.
 
\subsection{CovidENet: AtlasNet ensembling for binary lesion segmentation~\texorpdfstring{\citep{chassagnon2020ai}}{Chassagnon et al., 2020}}
\label{covidenet_method}

CovidENet is an ensemble of 2D and 3D CNNs based on the AtlasNet framework~\citep{vakalopoulou2018atlasnet}.
AtlasNet combines CNN ensembling and spatial normalization by registration to a template.
Each CNN is associated with a template CT scan, such that all CT scans are registered to this template before being given as input to the network.
In addition, different networks use different CT scan templates and different 2D and 3D CNN architectures were used.

The authors of the CovidENet~\citep{chassagnon2020ai} kindly accepted to run and share with us the segmentation predictions for our CT scans using the model that they trained on their data.
%




\subsection{2DS: 2D U-Net for binary lesion segmentation}
\label{ine_method}
A 2D U-Net is trained from scratch on patches extracted from the axial plane to perform binary lesion segmentation.

\paragraph*{Data}
Besides the data described in Section \ref{subsec:training_data}, additional CT scans, taken from the publicly available MosMedData~\citep{mosmed} set, were included for training. This dataset contains 1110 anonymised CT scans, both with and without COVID-19 related findings. It includes 50 CTs that are completed with the ground truth lesion segmentations, mainly indicating ground glass opacities and consolidations. These are used to supplement the training data for this method. As we are working with 2D patches extracted from the axial plane, the spacing of 8~mm in the z-direction does not pose a problem here. 

\paragraph*{Data preprocessing}
The chest CTs were resampled to $1 \times 1 \times 3 \,\textup{mm}^3$ and the HU intensities were clipped between -1000 HU and 400 HU before rescaling the intensities to $[-1, 1]$. The segmentation output was upsampled using nearest neighbour interpolation.
The CT images were masked using the automated lung segmentation model DMLu, described in Section~\ref{siri_method_lung}. Patches of (128, 128) were extracted from this masked CT within the bounding box around the lungs.

\paragraph*{Neural network architecture}
We used a 2D U-Net~\citep{Ronneberger2015} with 5 levels and $16$ features after the first convolution layer. Each convolutional block consisted of 2 convolutions followed by a ReLU activation. During training, a 5\% dropout was implemented after each maxpooling to improve generalisability of the model. The upsampling in the expansion path was performed through transposed convolutions. This architecture has a total of 2.2 M parameters. 

\paragraph*{Training}
We used a patch based training strategy by sampling patches of (128, 128) from the bounding box around the lungs. This corresponds to approximately 16 patches per axial slice. Only patches comprising a lesion were used for training as the number of negative voxels in these patches was sufficient to represent this class. We trained with a batch size of 32, the Adam optimizer~\citep{kingma2014adam} and a dice loss function. During training, the learning rate was reduced by a factor of 10 with a patience of 3 and a minimum value of $10^{-5}$. Early stopping was implemented with a patience of 10 and a maximum number of epochs of 500 to reduce the chance of overfitting. 

\subsection{DMmc: 3D multiclass lesion segmentation using DeepMedic}
\label{siri_method_lesion}
The DeepMedic network for lung segmentation in Section~\ref{siri_method_lung} is slightly adapted to perform multiclass lesion segmentation on chest CT images and assuming already available binary lesion segmentations. The output of the network is then multiplied with this binary lesion mask both at training and inference time, effectively making this a two-step approach. For  the  first  step, the  result  of  majority  voting for binary lesion segmentation in 
Section~\ref{sec:lung_segmentation_MAJ}
is used.

\paragraph*{Data preprocessing}
 Data is preprocessed in the same way as for DMLu (Section~\ref{siri_method_lung}).
 
\paragraph*{Neural network architecture}
The network architecture is identical to the DMLu model except for the output layer. The output layer is a softmax layer with three output nodes for each of the classes: CON, CPP and GGO.

\paragraph*{Training}
The network is trained using a patch-based approach in which segments of (43, 43, 13) voxels are predicted simultaneously. Data augmentation is performed by adding Gaussian noise and performing random affine transformations on the chest CT images and corresponding ground truth. The 3D neural network is trained using the Adam optimizer \citep{kingma2014adam} and a batch-size of 8. The objective function used is the weighted categorical cross-entropy, for which the weights are defined depending on the volumes of each lesion present in the training set. The gradients for pixels not belonging to GGO, CPP or CON are masked to 0. A learning rate schedule is used to automatically reduce the learning rate by monitoring the validation loss during training. Early stopping is subsequently used to stop training and avoid overfitting on the validation set.
\subsection{2DRnx: 2D lesion classification using U-Net}
\label{sec:lesion_classification_2DU}
A 2D U-Net is applied for multiclass segmentation of the lesion types on each axial slice separately. The output of the network is then multiplied with a binary lesion mask both at training and inference time, effectively making this a two-step approach. For the first step, the result of majority voting for binary lesion segmentation described in 
Section~\ref{sec:lung_segmentation_MAJ}
is used.

\paragraph*{Data preprocessing}
The chest CTs were resampled axially using an average pooling scheme to be as close as possible to 3 mm$^3$ voxel size.  The HU intensities were clipped between $-1300$ HU and -100 HU before rescaling to [0, 1]. The outputs of the network were upsampled axially using the inverse pooling scheme, but using nearest interpolation instead and padding with zeros at the bottom.

\paragraph*{Neural network architecture}
We use a 2D U-Net~\citep{Ronneberger2015} with 4 levels, 16 features after the first convolution layer, ReLU activations, batch normalization, max pooling, linear upsampling and dropout at the deepest level. The input size is (512, 512, 1) which corresponds to the full resolution, grayscale image and same padding is used for all convolutional layers. The output layer is a softmax layer with three output nodes for each of the classes: CON, CPP and GGO.

\paragraph*{Training}
The network is trained with full images in a batch size of 32. Data augmentation consists of rotation, shift, shear, zoom and horizontal flipping. The network is trained with the Ranger optimizer, a synergistic optimizer combining Rectified Adam~\citep{liu2019variance} and LookAhead~\citep{zhang2019lookahead}. The objective function that is optimized is the mean soft-dice loss where the gradients for non-lesion pixels and COM are masked to 0. A fixed learning rate of $10^{-4}$ is used for 200 epochs.

\subsection{Inf-Net: binary and multiclass lesion segmentation for COVID-19~\texorpdfstring{\citep{fan2020inf}}{Fan et al., 2020}}
\cite{fan2020inf} uses a two-step approach for multiclass COVID-19 lesion segmentation: a first CNN performs binary lesion segmentation and a second CNN uses this result for classifying lesion voxels into GGO or CON. 
A 2D CNN encoder-decoder architecture using reverse attention modules is used for the first step that aims at binary lesion segmentation.
The second step consists of an ensemble of two CNNs, a U-Net~\citep{Ronneberger2015} and a FCN8s~\citep{long2015fully}, that take the CT and the binary segmentation predicted in the first step as input.
They trained their CNNs on 100 manually segmented CT slices from 19 patients and 1600 unlabeled CT slices using a semi-supervised learning method to exploit unlabeled data. 

We used the pre-trained model made publicly available\footnote{\url{ https://github.com/DengPingFan/Inf-Net}} by the authors for evaluation. We converted the CT volumes of our dataset to 2D slices using a lung window of [-1250, 250] HU. All slices of one volume were cropped using the largest lung bounding box of that volume. Additionally, we masked the predictions with the lung masks of UNWM (Section \ref{jeroen_method}). Since InfNet does not separately predict CPP, which is a special type of GGO~\citep{Rossi2003}, we evalutated the InfNet GGO class as being a combination of GGO and CPP.

\subsection{WASS: multiclass lesion segmentation using the Generalized Wasserstein Dice loss and Distributionally Robust Optimization}
\label{lucas_method}
This method is part of the method developed for this study and trained using the training dataset described in Section~\ref{subsec:training_data}.
It is designed to address jointly the multiclass lesion segmentation problem and the lung segmentation problems.
The main difference with respect to the other methods is the use of the Generalized Wasserstein Dice loss~\citep{fidon2017generalised}.

\paragraph*{Data preprocessing}
The chest CT were resampled to $1 \times 1 \times 3 \,\textup{mm}^3$ and the HU intensities were clipped between -1000 HU and 1000 HU before rescaling the intensities to $[-1, 1]$.
In addition, lungs masks were used to crop the CT around the lungs with a margin of $5$ voxels. The lung masks used were obtained using the automatic lung segmentation method DMLu.
After cropping around the lungs, we split the lungs into right and left parts with respect to the center and with an overlap of 5 mm along the right-left axis.

\paragraph*{Neural network architecture}
We used a 3D U-Net~\citep{cciccek20163d} with $4$ levels, $32$ features after the first convolution layer, leaky ReLU, instance normalization~\citep{ulyanov2016instance}, max pooling, and linear up-sampling. To take into account the lower resolution along the z-axis, we used $3 \times 3 \times 1$ convolution in the first and last blocks and only the x-axis and y-axis were downsampled in the the first level. This architecture has a total of 13.3~M parameters.
A patch-based approach with a patch size of $(144,\, 208,\, 80)$ was used.

\paragraph*{Training}

The anisotropic 3D U-Net was trained to segment: GGO, CON, CPP, and an additional \textit{healthy lung} class that contains all voxels in the lung mask that are not labelled as any lesion or abnormal tissue type.
%
%
To tackle the presence of super-class in the manual ground truth, we used the Generalized Wasserstein Dice Loss~\citep{fidon2017generalised} which is a loss function that is able to exploit classes with a hierarchical structure.
As a result, the network will be penalized if it labels a \textit{combined pattern} voxel as either \textit{healthy lung} or \textit{background}, but not if it labels it as GGO, CON or CPP.
%
%
We refer the reader to the appendix for more details about the hyperparameters used for the Generalized Wasserstein Dice loss.

To deal with the high variability of pathologies, we used the \textit{hardness weighted sampling} described in~\citep{fidon2020sgd} for training deep neural networks with distributionally robust optimization. 

We used gradient checkpointing~\citep{chen2016training} during training to reduce the memory consumption of the 3D U-Net, allowing for a batch size equal to $3$.
We used right-left flipping for data augmentation. The 3D U-Net was trained for $400$ epochs using the Adam optimizer~\citep{kingma2014adam} with default parameters, and a fixed learning rate of $0.0003$.
For the segmentation results reported on the 5-fold cross-validation we used the model corresponding to the last epoch, while for the evaluation on the test set we used early-stopping on the validation fold and the ensemble of five models trained on the different training folds.

\subsection{UNWM: a U-Net with Waterfall Masking for lung, binary lesion and multiclass lesion segmentation}
\label{jeroen_method}
The idea of the UNWM method is to learn to segment all lung and lesion tissue, and detect CON, CPP and GGO lesion types. \textit{Waterfall masking}, as described below, is applied to the network outputs to take into account inter-class relationships to provide consistency and favor semantically meaningful predictions, and to let each output focus on a learned subset of voxels during training.

\paragraph*{Data preprocessing}
The chest CTs were resampled axially using an average pooling scheme to be as close as possible to 3 mm$^3$ voxel size. The HU intensities were clipped between $-1100$ HU and 100 HU before rescaling to [-1, 1]. The outputs of the network were upsampled axially using the inverse pooling scheme, but using nearest interpolation instead and padding with zeros at the bottom.

\paragraph*{Neural network architecture}
The first part of the network uses a 3D version of U-Net~\citep{Ronneberger2015} with the following modifications: the first convolution had eight kernels, every second convolution in the left arm had twice the number of kernels of the first convolution, linear upsampling, batch normalization~\citep{Ioffe2015} and leaky ReLU activations~\citep{Maas2013}. To take into account the lower axial resolution, we used $3 \times 3 \times 1$ instead of $3 \times 3 \times 3$ convolutions in every first and second convolution in the left and right arm, respectively. As the unique characteristic of the UNWM method and following the U-Net structure, the second part of the network has six parallel pathways, each having one $1 \times 1 \times 1$ convolution with 16 kernels followed by a final sigmoid layer, and with the number of kernels defined by the task. The first two pathways are devoted to lung segmentation. The second two pathways are devoted to binary lesion segmentation and are masked with the corresponding pathways from the lung segmentation. The third two pathways are devoted to multiclass lesion segmentation (CON, CPP and GGO) and are masked with the corresponding pathways from the binary lesion segmentation. This masking scheme is what we call: \textit{waterfall masking}. Due to the sigmoid activation, this network is flexible enough to allow multiple lesion types to co-exist and potentially group them into the COM tissue class at test time. This architecture has a total of 4.7~M parameters. The output patch size was $84 \times 84 \times 34$, which allows to fit approximately three patches at inference in a GPU with 10~GB of memory.

\paragraph*{Training}
We used a patch based training strategy by sampling 36 patches from inside the manual lung mask every epoch to optimize the randomly initialized weights. We added some Gaussian noise, small translations and in-plane rotations, and right-left flipping for data augmentation. For each task we have one pathway optimizing the cross-entropy objective and one pathway optimizing the soft Dice objective, each shown to have its own advantages~\citep{Bertels2019a,Bertels2019b}. For computing the multiclass losses, only voxels manually labeled as CON, CPP or GGO were taken into account. We train for 1200 epochs, used a batch size of three and the Adam optimizer~\citep{kingma2014adam} with default parameters at an initial learning rate of $10^{-3}$. After 800 and 1000 epochs we reduced the learning rate by a factor of 10. At test time, for now only the soft Dice optimized prediction maps were used, notwithstanding more exotic combinations were possible.
\def\rvx{{\mathbf{x}}}
\def\rvy{{\mathbf{y}}}
\def\vmu{{\bm{\mu}}}
\def\vtheta{{\bm{\theta}}}

\subsection{MAJ: Majority voting}
\label{sec:lung_segmentation_MAJ}
In this paragraph, we present the ensembling method that was used.
The maximum-a-posteriori segmentation of the ensembling model is computed directly by applying a majority voting to every voxel of the segmentations predicted by the individual models after argmax.
In case different classes obtained the same score after majority voting, the class predicted by the ensemble is chosen randomly among them.

This majority voting ensembling strategy is applied to all three tasks: lung segmentation, lesion segmentation and multiclass lesion segmentation. All the methods described were included in the majority voting.



\section{Results}
\label{sec:results}
We evaluate the performance of all methods for each of the icovid benchmark tasks on two different datasets: the 5-fold cross-validation set and an additional independent test set. This section is therefore divided in six parts, one for each of the icovid benchmark tasks (i.e. lung segmentation, binary lesion segmentation and multiclass lesion segmentation) both for the validation set and test set. For each task, the mean DSC, HD95, ASD and AVD are reported in a table. The highest scoring method for each metric is highlighted in bold and a significantly superior performance compared to all other methods is followed by an asterisk. For more details on the icovid benchmark setup, metrics and statistical testing, we refer the reader to Section~\ref{sec:icovid_challenge}.

These tables are complemented by a series of boxplots in the appendix to provide more information about the distribution of the data and to allow an easier visual comparison between the methods' performances. For each of the benchmark tasks, we also provide example predictions from all the methods for a qualitative appreciation in Figures~\ref{fig:qualitative_master_lungs_testset},~\ref{fig:qualitative_master_bin_testset} and~\ref{fig:master_qualitative_mc}. These figures show for each task a good, a median and a bad prediction. These CT scans are selected with respectively the third quartile, median and first quartile value of the AVD of the predictions by the majority voting method. The slice shown is that with the third quartile, median and first quartile DSC across slices (with foreground) in the chosen CT scan.

\subsection{Lung segmentation - cross-validation}
\label{sec:results_lung_segmentation_cv}

Table~\ref{tab:master_lung} gives the performance metrics for all lung segmentation methods: DMLo, JoHof, UNWM, DMLu, CTA, WASS and MAJ. No individual method outperforms the others on all metrics, but majority voting does outperform all individual methods with the highest value for all metrics and significantly higher DSC, HD95 and ASD.

\begin{table}[H]
    \caption{Mean of metrics for lung segmentation task for cross-validation set. The best value for each metric is highlighted in bold and indicated with an * if significantly superior than all the other methods. Methods indicated with \# did not use the dataset in Section~\ref{subsec:training_data} for training.}
    \begin{tabularx}{\linewidth}{l|YYYYYYY}
    \toprule
    \emph{method} $\rightarrow$ & DMLo\textsuperscript{\#}    &      & UNWM   &   & CTA\textsuperscript{\#}   &   & MAJ   \\
    metric                      &       &  JoHof\textsuperscript{\#} &       &  DMLu  &       & WASS      &       \\
    \midrule
    DSC &  0.968 &  0.971 &  0.971 &  0.969 &  0.966 &  0.969 &  \textbf{0.978\textsuperscript{*}} \\
    HD95 &  4.00 &  3.430 &  7.28 &  3.79 & 16.0 &  8.69 &  \textbf{2.43\textsuperscript{*}} \\
    ASD &  0.860 &  0.644 &  0.677 &  0.758 &  0.636 &  0.626 &  \textbf{0.559\textsuperscript{*}} \\
    AVD & 73.1 & 49.3 & 61.7 & 74.1 & 56.6 & 56.6 & \textbf{33.6} \\
    \midrule
    SEN\textsuperscript{CON} & 0.701 &  0.813 & 0.811 & 0.739 & 0.686 & \textbf{0.845} &  0.806 \\
    SEN\textsuperscript{CPP} & 0.991 &  0.993 & 0.992 & 0.993 & 0.991 & 0.994 &  \textbf{0.995} \\
    SEN\textsuperscript{GGO} & 0.968 &  \textbf{0.978} & 0.966 & 0.965 & 0.964 & 0.974 &  0.977 \\
    \bottomrule
\end{tabularx}
    \label{tab:master_lung}
\end{table}

An additional experiment is performed in order to show to what extent the lung mask covers the different types of lesions. The sensitivity for a lesion type quantifies the percentage of lesion in the ground truth that is actually covered by the lung segmentation. The sensitivity of the different methods for CON, CPP and GGO is reported in Table~\ref{tab:master_lung} and accompanying boxplots can be found in appendix. 


For all methods, the sensitivity for CON is clearly lower compared to their sensitivities for CPP and GGO. WASS, which has the highest sensitivity for CON, uses a hardness weighted sampling strategy which effectively enforces the difficult consolidated lungs to be sampled more often.


\subsection{Lung segmentation - test set}
\label{sec:results_lung_segmentation_testset}
\begin{figure*}[htb]
    \centering
    \resizebox*{\linewidth}{!}{
    
    \def\arraystretch{0}
    \setlength{\tabcolsep}{0pt}
    \begin{tabularx}{\linewidth}{l @{\hspace{2pt}} YYYYYYYY}

        & CT & DMLo    &  JoHof    & UNWM   & DMLu  & CTA   & WASS  & MAJ   \\

        \rotatebox{90}{\hspace{20pt} IQR25} &
        \includegraphics[width=\linewidth]{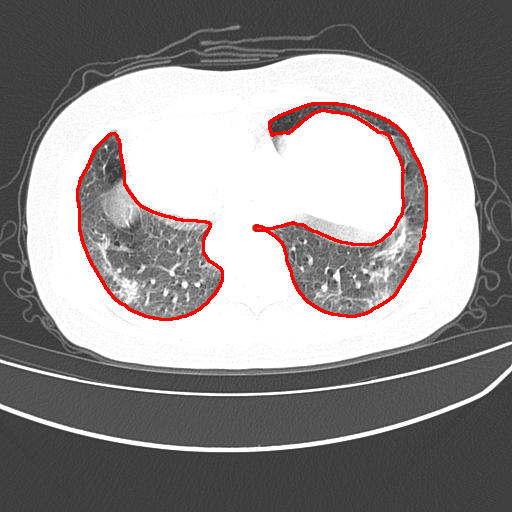} &  
        \includegraphics[width=\linewidth]{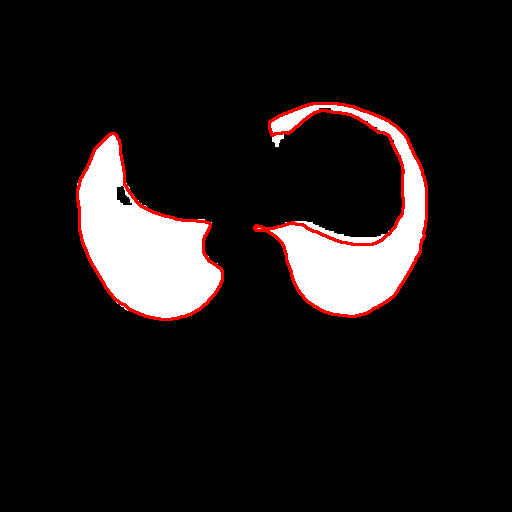} &  
        \includegraphics[width=\linewidth]{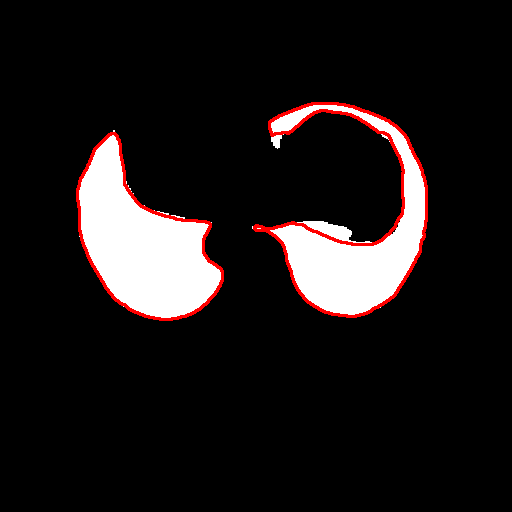} &  
        \includegraphics[width=\linewidth]{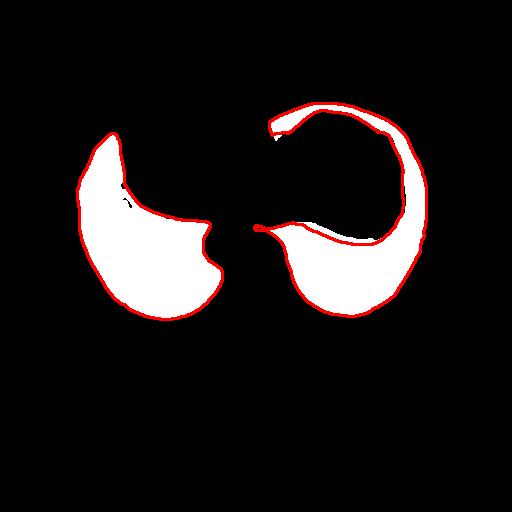} &  
        \includegraphics[width=\linewidth]{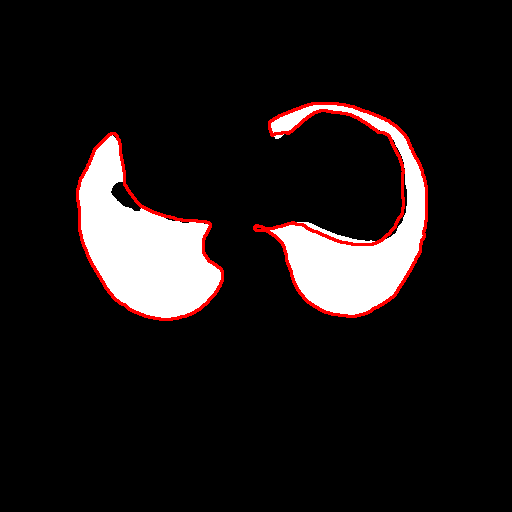} &  
        \includegraphics[width=\linewidth]{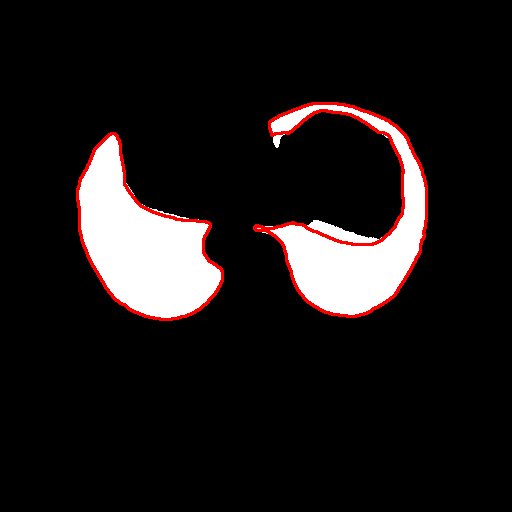} &  
        \includegraphics[width=\linewidth]{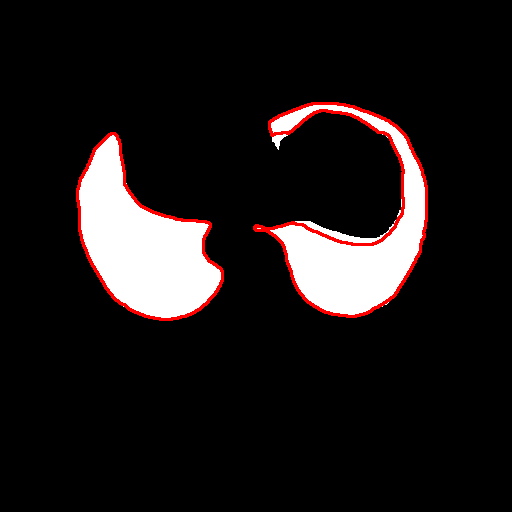} &  
        \includegraphics[width=\linewidth]{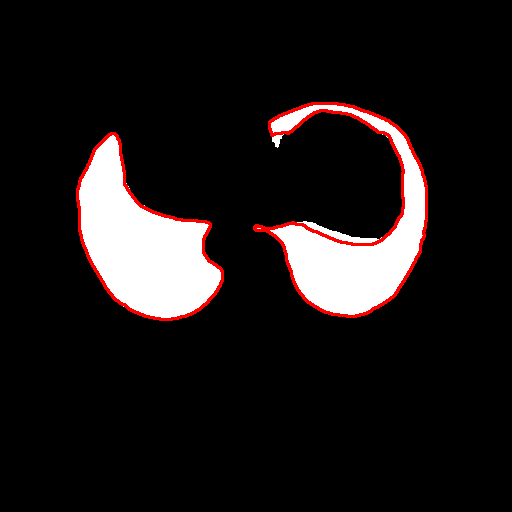} \\ 
        
        \rotatebox{90}{\hspace{20pt} MEDIAN} &
        \includegraphics[width=\linewidth]{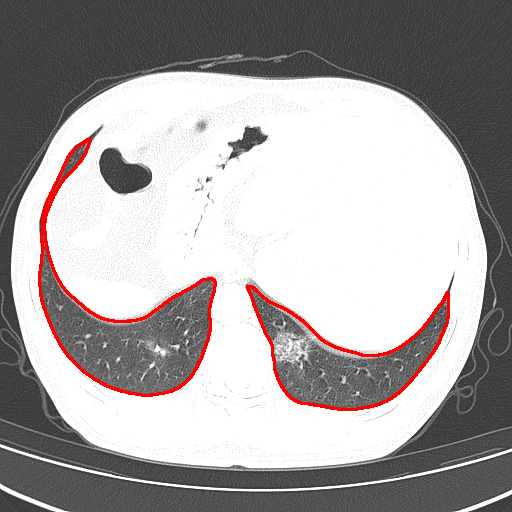} &  
        \includegraphics[width=\linewidth]{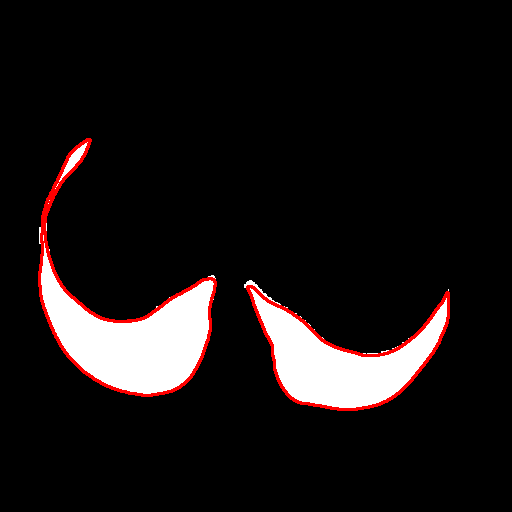} &  
        \includegraphics[width=\linewidth]{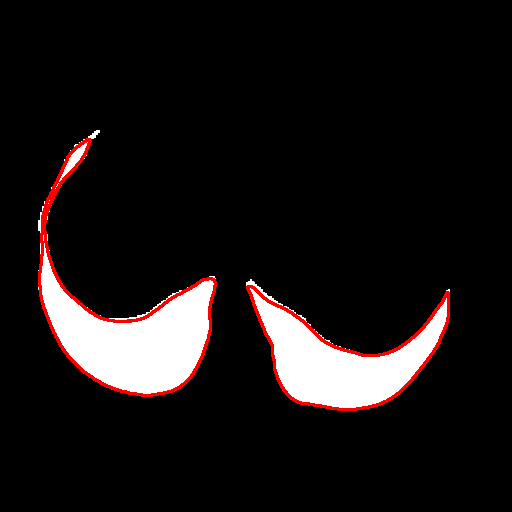} &  
        \includegraphics[width=\linewidth]{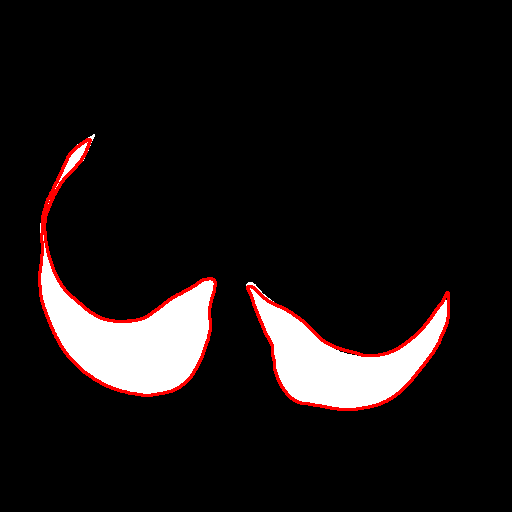} &  
        \includegraphics[width=\linewidth]{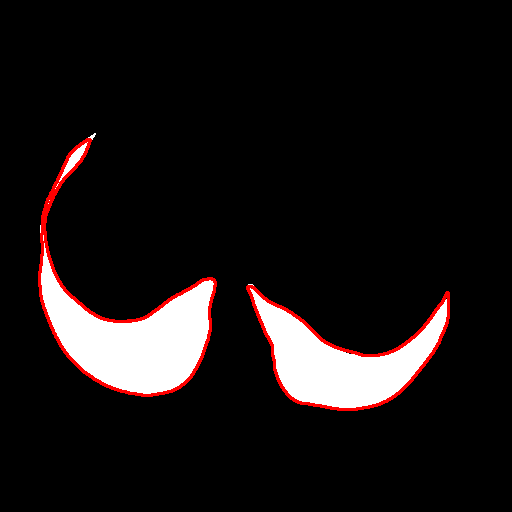} &  
        \includegraphics[width=\linewidth]{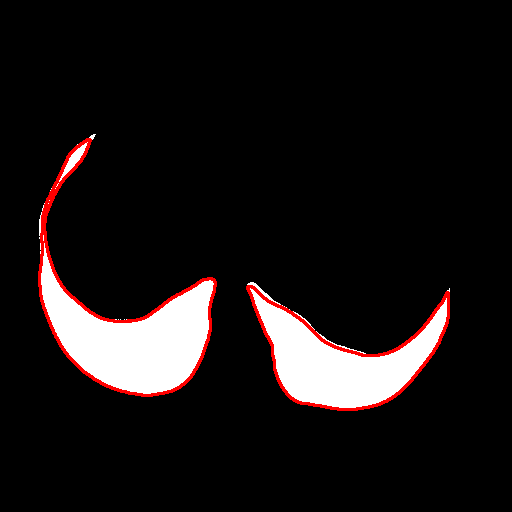} &  
        \includegraphics[width=\linewidth]{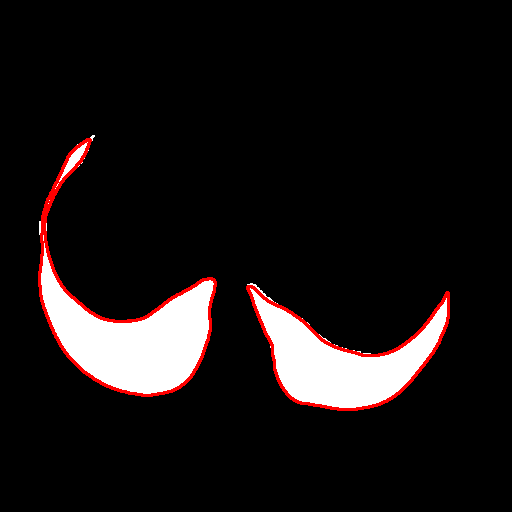} &  
        \includegraphics[width=\linewidth]{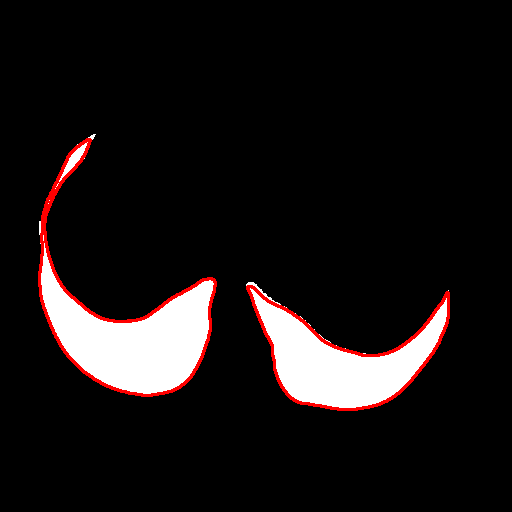} \\ 
        
        \rotatebox{90}{\hspace{20pt} IQR75} &
        \includegraphics[width=\linewidth]{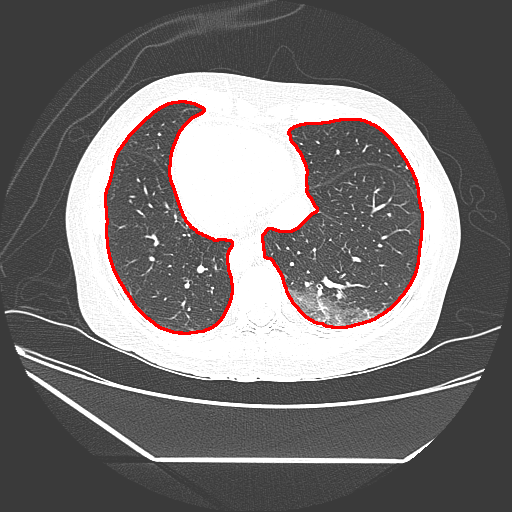} &  
        \includegraphics[width=\linewidth]{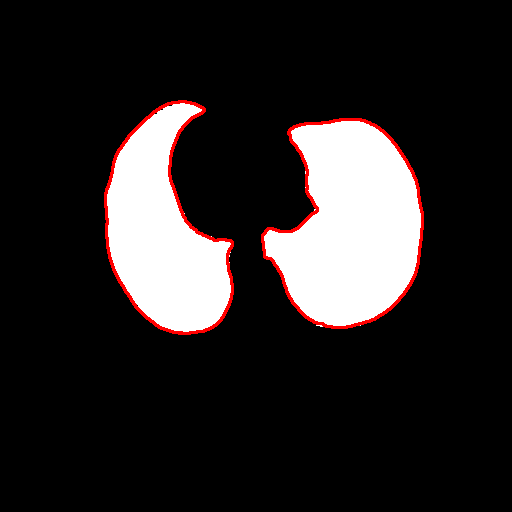} &  
        \includegraphics[width=\linewidth]{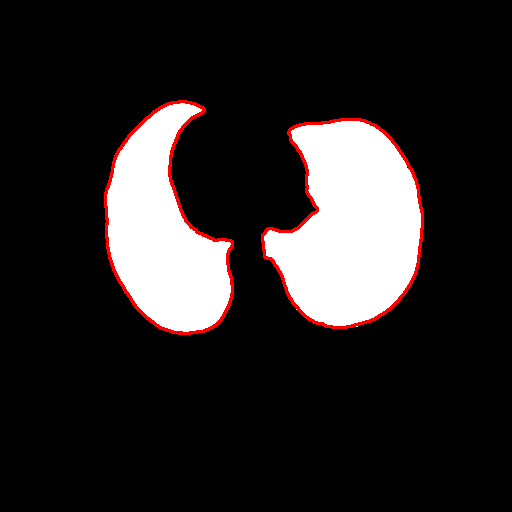} &  
        \includegraphics[width=\linewidth]{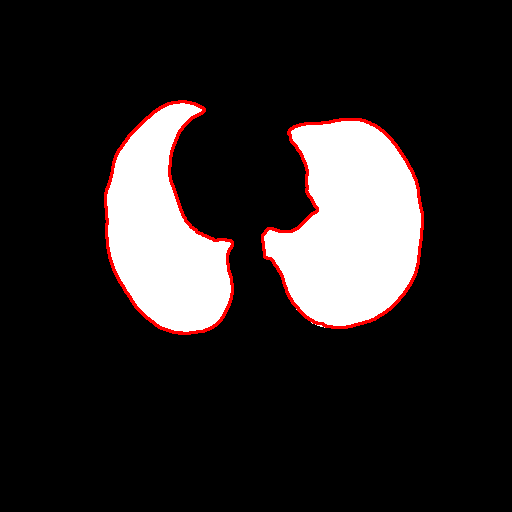} &  
        \includegraphics[width=\linewidth]{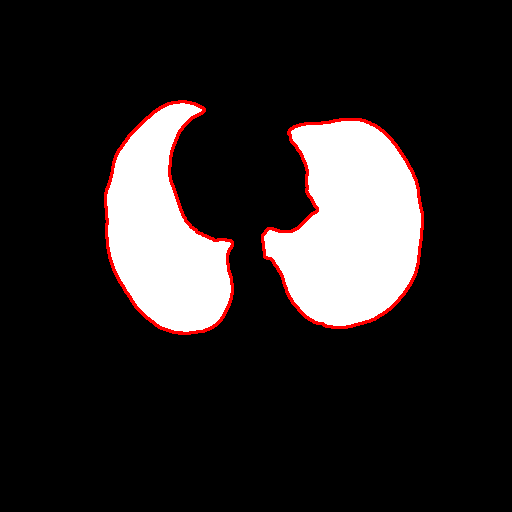} &  
        \includegraphics[width=\linewidth]{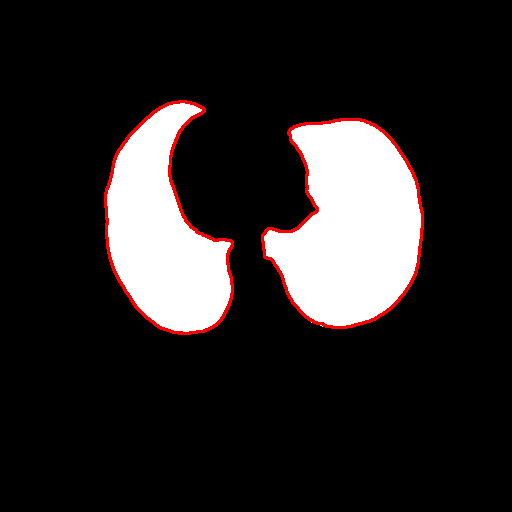} &  
        \includegraphics[width=\linewidth]{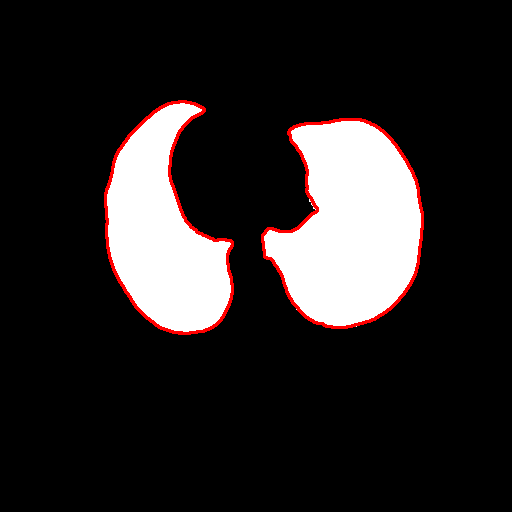} &  
        \includegraphics[width=\linewidth]{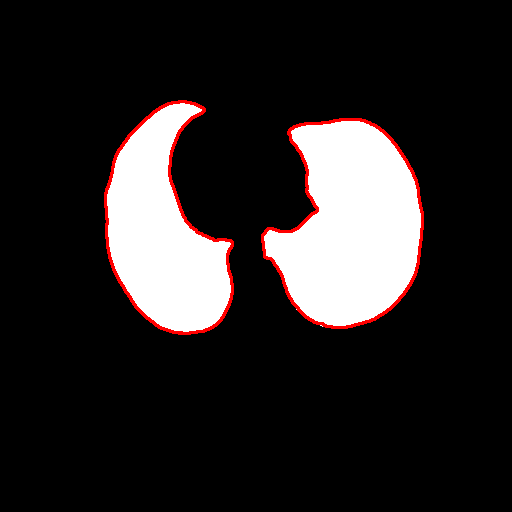} \\

    \end{tabularx}
    }
        \caption{A qualitative evaluation of the resulting predictions for lung segmentation on the test set from each method (columns, from left to right: original CT, and predictions from DMLo, JoHof, UNWM, DMLu, CTA, WASS and MAJ). The predicted delineations are obtained after thresholding at 0.5 and each image is overlayed with the manual ground truth in red. The three rows are slices from different CT scans for which the AVD is respectively the median, first and third quartile value.}
    \label{fig:qualitative_master_lungs_testset}
\end{figure*}

Table \ref{tab:master_lung_testset} gives the performance metrics for the independent test set for all lung segmentation methods: DMLo, JoHof, UNWM, DMLu, CTA, WASS and MAJ. The ensembling method performs best for DSC and AVD, and is significantly better for DSC. JoHof performs best for HD95 and ASD, and is significantly better for HD95. UNWM and JoHof have the highest sensitivities for CON and GGO, respectively. Sensitivity for CPP is not applicable since there are no CPP lesions in the testset. Figure \ref{fig:qualitative_master_lungs_testset} shows representative predictions for the considered methods.

\begin{table}[H]
    \caption{Mean of metrics for lung segmentation task for the test set. The best value for each metric is highlighted in bold and indicated with an * if significantly superior than all the other methods. Sensitivity for CPP is not applicable.}
    \begin{tabularx}{\linewidth}{l|YYYYYYY}
    \toprule
    \emph{method} $\rightarrow$ & DMLo    &      & UNWM   &   & CTA   &   & MAJ   \\
    metric                      &       &  JoHof &       &  DMLu  &       & WASS      &       \\
    \midrule
    DSC &  0.976 &   0.984 &   0.972 &   0.974 &   0.977 &  0.979 &  \textbf{0.987\textsuperscript{*}} \\
    HD95 &  1.98 &   \textbf{1.39\textsuperscript{*}} &   5.76 &   2.05 & 168 &  1.93 & 1.71 \\
    ASD &  0.643 &   \textbf{0.395} &   0.629 &   0.684 &  83.5 &  0.574 &  0.403 \\
    AVD & 97.0 & 37.8 & 150 & 164 &  95.5 & 94.2 & \textbf{30.3} \\
    \midrule
    SEN\textsuperscript{CON} & 0.921 &  0.938 & \textbf{0.947} & 0.904 & 0.896 & 0.931 & 0.939 \\
    SEN\textsuperscript{CPP} & - &  - & - & - & - & - & - \\
    SEN\textsuperscript{GGO} & 0.988 &  \textbf{0.999} & 0.990 & 0.988 & 0.987 & 0.989 & 0.998 \\
    \bottomrule
\end{tabularx}
    \label{tab:master_lung_testset}
\end{table}

\subsection{Lesion segmentation - cross-validation}
\label{sec:results_lesion_segmentation_cv}
Table \ref{tab:master_bin} gives the performance metrics for lesion segmentation methods: WASS, 2DS, UNWM, CTA, CENet, InfNt and MAJ. No individual method outperforms the others on all metrics, but majority voting does have a significantly higher DSC than all other methods and the highest score for HD95. CENet and InfNt have the best score for ASD and AVD respectively. The reader is referred to the appendix for the boxplots visualising the data distributions of the metrics over the cases.

\begin{table}[H]
    \caption{Mean of metrics for lesion segmentation task for validation set. The best value for each metric is highlighted in bold and indicated with an * if significantly superior than all the other methods. Methods indicated with \# did not use the dataset in Section~\ref{subsec:training_data} for training.}
    \begin{tabularx}{\linewidth}{l|YYYYYYY}
    \toprule
    \emph{method} $\rightarrow$ &    WASS &      &    UNWM &      &   CENet\textsuperscript{\#} &    &     MAJ \\
    metric &         &    2DS     &         &    CTA\textsuperscript{\#}     &         &    InfNt\textsuperscript{\#}     &         \\
    \midrule
    DSC &   0.536 &   0.497 &   0.520 &   0.467 &   0.481 &   0.497 &   \textbf{0.554\textsuperscript{*}} \\
    HD95 &  48.0 &  67.3 &  46.7 &  93.9 &  56.0 &  54.5 &  \textbf{42.6} \\
    ASD &   5.62 &   3.74 &   5.56 &   7.00 &   \textbf{3.54} &   4.50 &   4.71 \\
    AVD & 129 & 141 & 157 & 138 & 249 & \textbf{112} & 126 \\
    \bottomrule
\end{tabularx}

    \label{tab:master_bin}
\end{table}

\subsection{Lesion segmentation - test set}
\label{sec:results_lesion_segmentation_testset}
\begin{figure*}[htb]
    \centering
    \resizebox*{\linewidth}{!}{
    
    \def\arraystretch{0}
    \setlength{\tabcolsep}{0pt}
    \begin{tabularx}{\linewidth}{l @{\hspace{2pt}} YYYYYYYY}

        & CT &   WASS &     2DS &     UNWM &   CTA & CENET & InfNet  &   MAJ \\
        
        \rotatebox{90}{\hspace{20pt} IQR25} &
        \includegraphics[width=\linewidth]{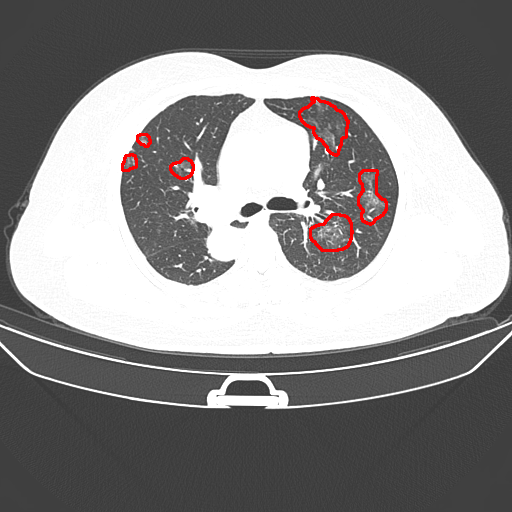} &  
        \includegraphics[width=\linewidth]{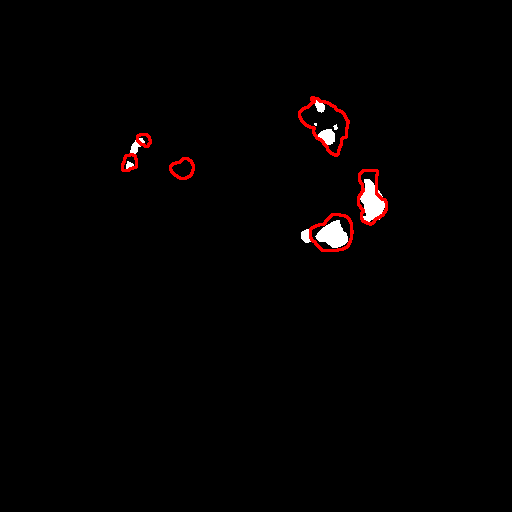} &  
        \includegraphics[width=\linewidth]{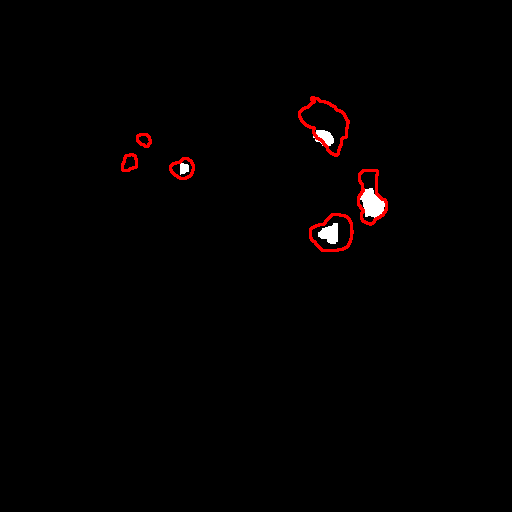} &  
        \includegraphics[width=\linewidth]{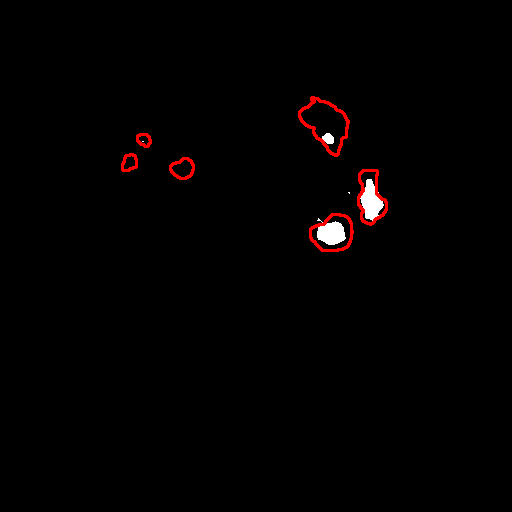} &  
        \includegraphics[width=\linewidth]{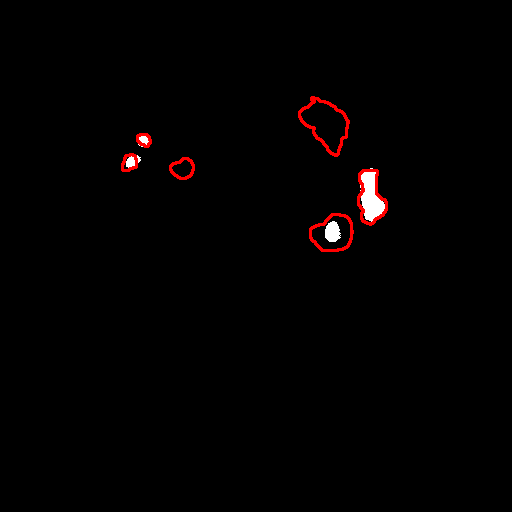} & 
        \includegraphics[width=\linewidth]{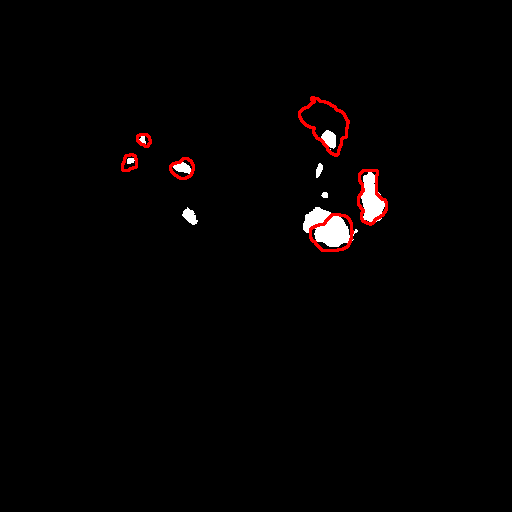} &  
        \includegraphics[width=\linewidth]{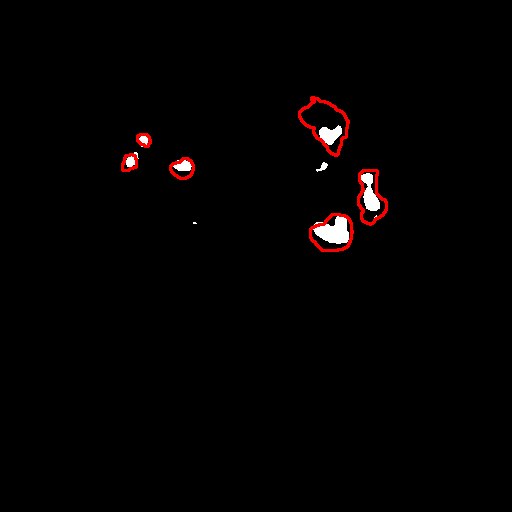} &  
        \includegraphics[width=\linewidth]{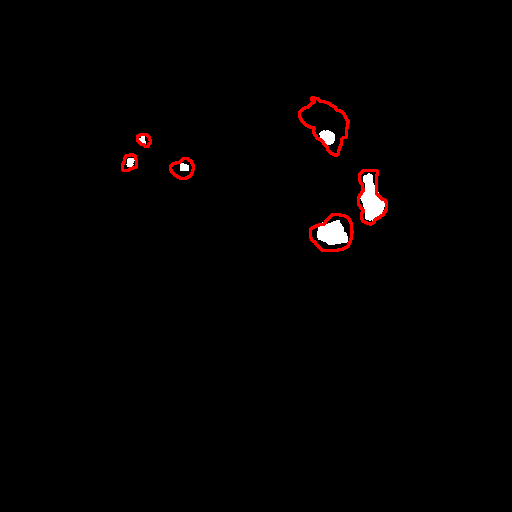} \\ 
        
        \rotatebox{90}{\hspace{20pt} MEDIAN} &
        \includegraphics[width=\linewidth]{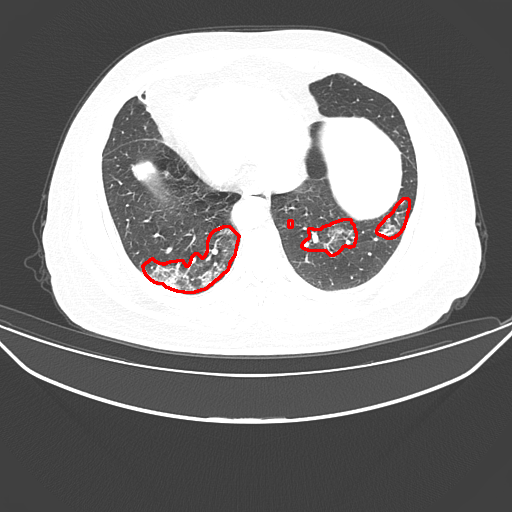} &  
        \includegraphics[width=\linewidth]{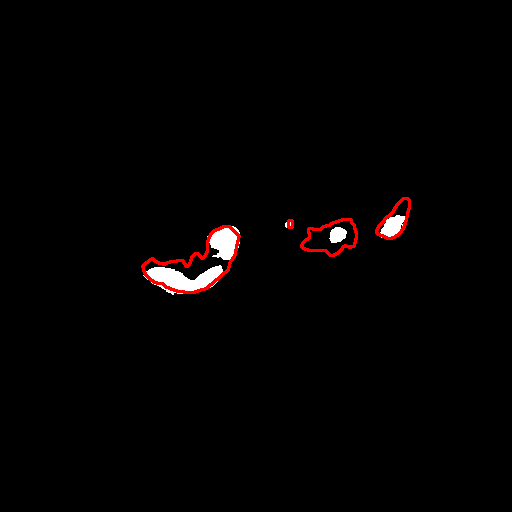} &  
        \includegraphics[width=\linewidth]{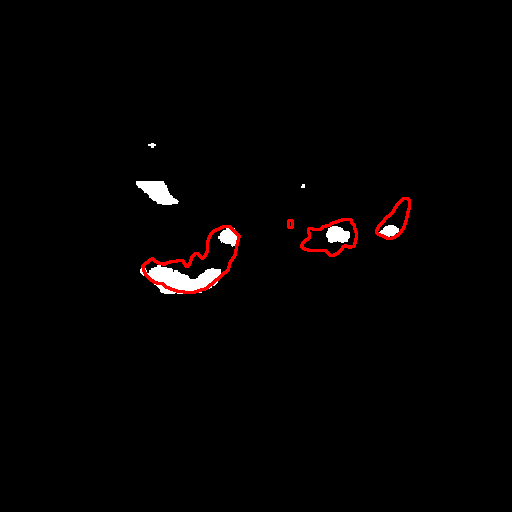} &  
        \includegraphics[width=\linewidth]{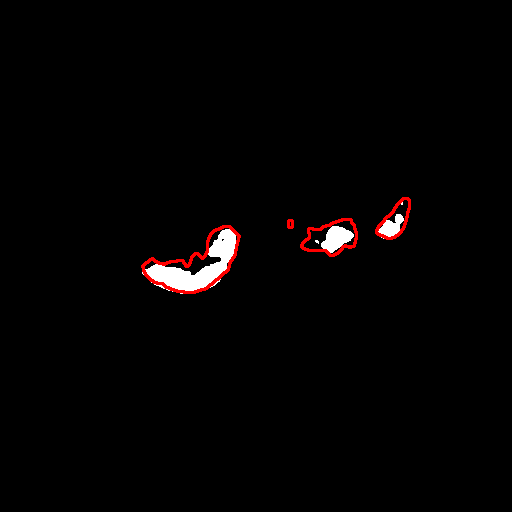} &  
        \includegraphics[width=\linewidth]{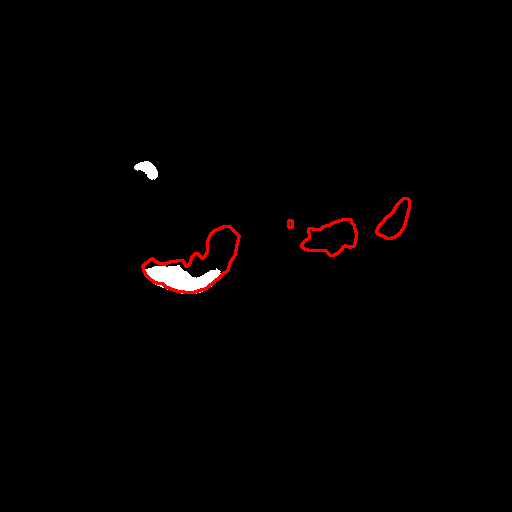} & 
        \includegraphics[width=\linewidth]{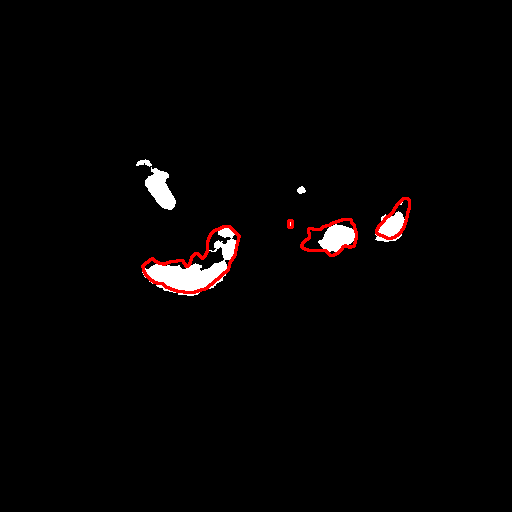} &  
        \includegraphics[width=\linewidth]{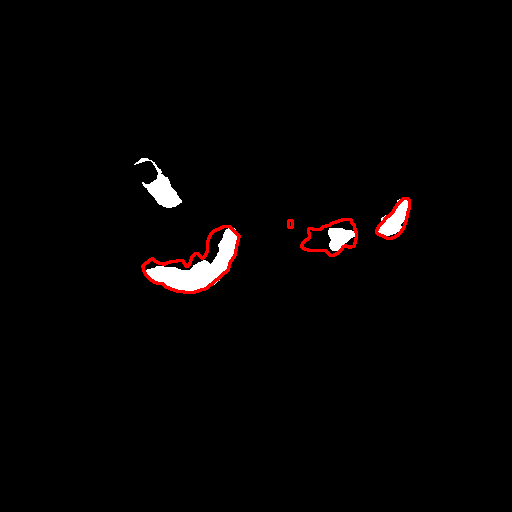} &  
        \includegraphics[width=\linewidth]{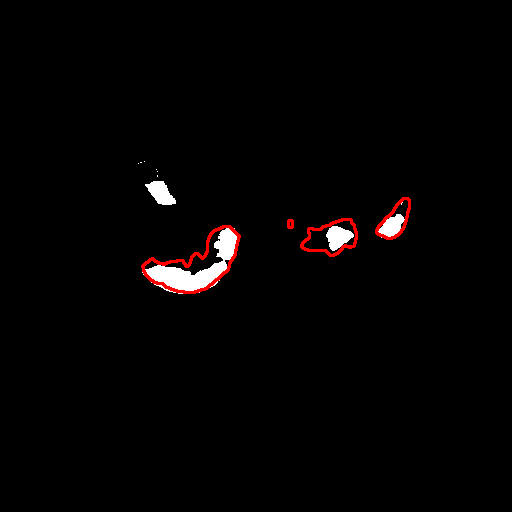} \\ 
        
        \rotatebox{90}{\hspace{20pt} IQR75} &
        \includegraphics[width=\linewidth]{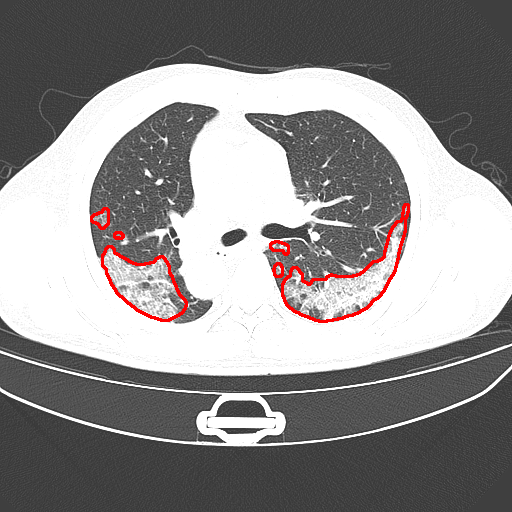} &  
        \includegraphics[width=\linewidth]{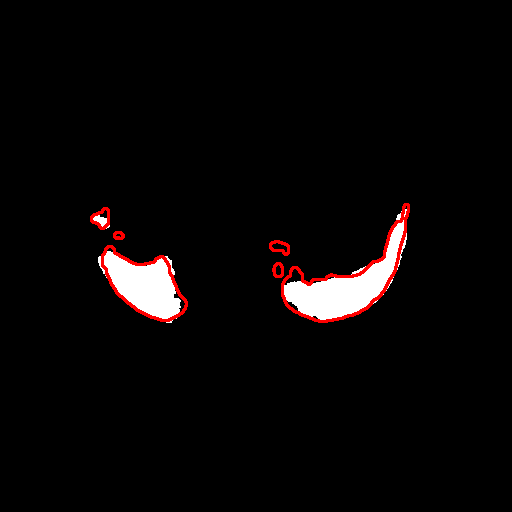} &  
        \includegraphics[width=\linewidth]{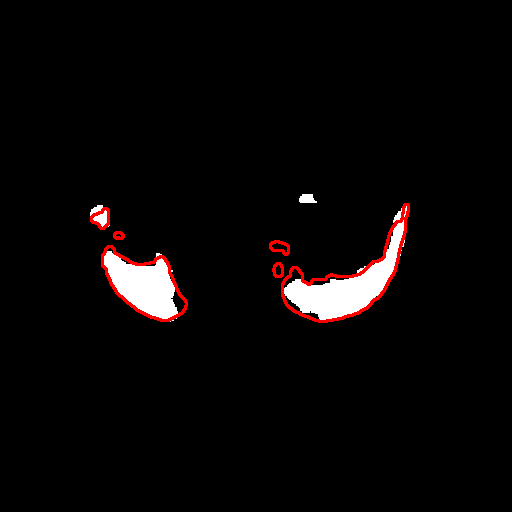} &  
        \includegraphics[width=\linewidth]{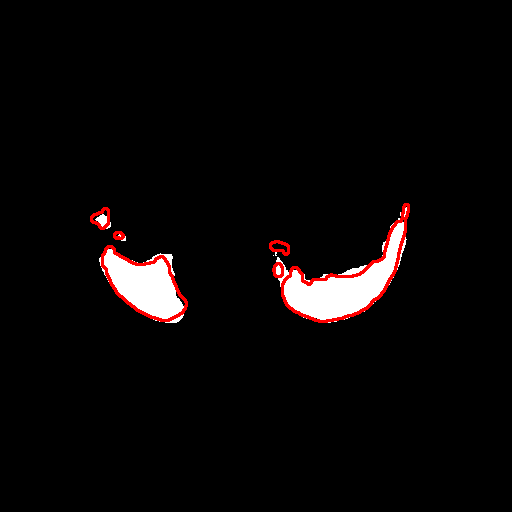} &  
        \includegraphics[width=\linewidth]{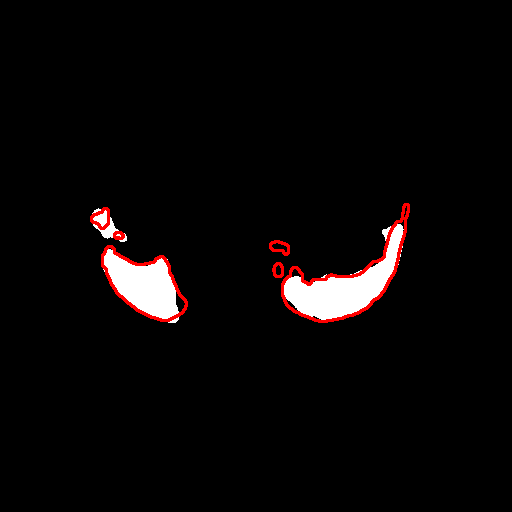} &
        \includegraphics[width=\linewidth]{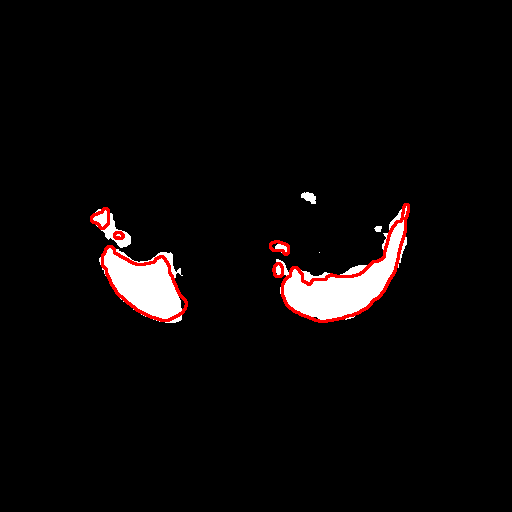} &  
        \includegraphics[width=\linewidth]{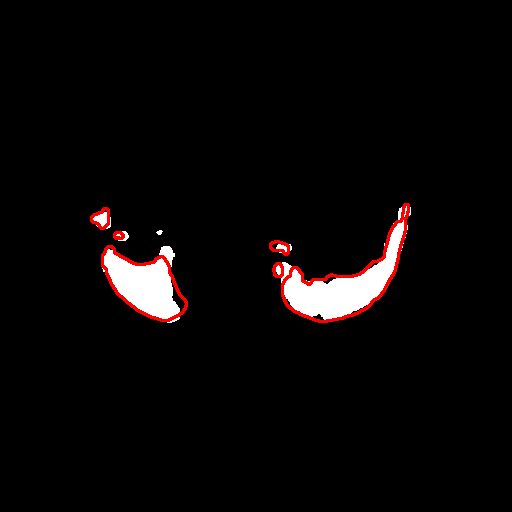} &  
        \includegraphics[width=\linewidth]{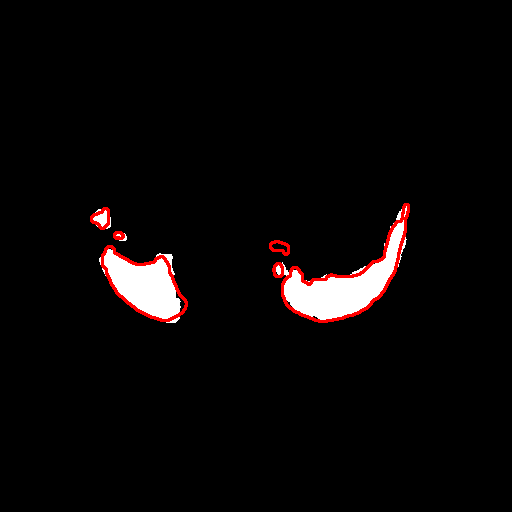} \\

    \end{tabularx}
    }
        \caption{A qualitative evaluation of the resulting predictions for lesion segmentation on the test set from each method (columns, from left to right: original CT, and predictions from WASS, 2DS, UNWM, CTA, CENET, InfNet and MAJ). The predicted delineations are obtained after thresholding at 0.5 and each image is overlayed with the manual ground truth in red. The three rows are slices from different CT scans for which the AVD is respectively the median, first and third quartile value.}
    \label{fig:qualitative_master_bin_testset}
\end{figure*}

Table \ref{tab:master_bin_testset} gives the performance metrics for the independent test set for all lesion segmentation methods: WASS, 2DS, UNWM, CTA, CENet, InfNt and MAJ. No individual method outperforms the others on all metrics. 
Fig. \ref{fig:qualitative_master_bin_testset} shows representative predictions for the considered methods. We observe that lesions often have very irregular shapes and diffuse boundaries, which partly causes the much higher HD95 and ASD values compared to the lung segmentation. Another factor that causes these metrics to be much higher for this task is that not every image contains lesions and so the distance to the edges of the CT scan is calculated even for very small predicted segmentation masks.

\begin{table}[H]
    \caption{Mean of metrics for lesion segmentation task for test set. The best value for each metric is highlighted in bold and indicated with an * if significantly superior than all the other methods.}
    \begin{tabularx}{\linewidth}{l|YYYYYYY}
    \toprule
    \emph{method} $\rightarrow$ &    WASS &      &    UNWM &      &   CENet &    &     MAJ \\
    metric &         &    2DS     &         &    CTA     &         &    InfNt     &         \\
    \midrule
    DSC &   0.663 &   0.639 &   0.649 &   0.653 &   0.646 &   0.625 &  \textbf{0.724\textsuperscript{*}} \\
    HD95 &  \textbf{77.6} & 100 & 102 & 122 &  99.9 &  99.2 & 83.6 \\
    ASD &  25.3 &  \textbf{24.6} &  46.4 &  26.3 &  28.1 &  28.3 & 37.0 \\
    AVD & 176 & 132 &  94.2 & 199 & 120 & 105 & \textbf{91.3} \\
    \bottomrule
\end{tabularx}

    \label{tab:master_bin_testset}
\end{table}

\subsection{Multiclass lesion segmentation - cross-validation}
\label{sec:results_multiclass_lesion_segmentation_cv}

Table \ref{tab:master_mc} gives the performance metrics for the five tested methods on multiclass lesion segmentation and the majority voting. No results for InfNt are shown for the classes CPP, GGO and MEAN since this method is not capable of predicting these. No individual method outperforms the others on all metrics, though majority voting achieves the highest DSC for CPP, GGO (significantly better), mean (significantly better) and GGO + CPP. For the other metrics, it depends on the lesion type which model performs best. 
The data distributions of the metrics over the cases is shown in the boxplots in appendix.
Fig. \ref{fig:master_qualitative_mc} shows representative results for the considered methods.

\begin{table}[H]
    \caption{Mean of metrics for multiclass lesion segmentation task for validation set. The best value for each metric is highlighted in bold and indicated with an * if significantly superior than all the other methods. Methods indicated with \textsuperscript{\#} did not use the dataset in Section~\ref{subsec:training_data} for training.}
    \begin{tabularx}{\linewidth}{ll|YYYYYY}
    \toprule
     & \emph{method} $\rightarrow$   &    WASS &      &  DMmc &   &     InfNt\textsuperscript{\#} & \\
    class & metric &         &  UNWM       &         &   2DRnx      &        & MAJ \\
    \midrule
    CON & DSC  &   \textbf{0.277}   &   0.215   &   0.248   &   0.259   &   0.228   &   0.272 \\
         & HD95 &   102     & 107       & 102       & 105       & 106       &   \textbf{97.1} \\
         & ASD  &  26.5     &  19.4     &  29.1     &  \textbf{18.3\textsuperscript{*}}     &  23.4     &   25.8 \\
         & AVD  &  51.5     & 200       &  \textbf{50.2}     &  75.1     &  59.0     &   73.5 \\
    CPP  & DSC  &   0.270   &   0.281   &   0.160   &   0.236   &     -     &   \textbf{0.303} \\
         & HD95 & 179       & 165       & 180       & \textbf{165}       &     -     &   172 \\
         & ASD  & 104       &  \textbf{81.0}     &  97.1     &  81.2     &     -     &   89.0 \\
         & AVD  &  \textbf{68.0\textsuperscript{*}}     & 320       & 107       &  91.3     &     -     &   121 \\
    GGO  & DSC  &   0.298   &   0.306   &   0.270   &   0.282   &     -     &   \textbf{0.339\textsuperscript{*}} \\
         & HD95 &  86.5     &  90.6     &  \textbf{83.6}     &  87.0     &     -     &   84.0 \\
         & ASD  &  23.6     &  21.9     &  27.8     &  \textbf{21.2}     &     -     &   24.0 \\
         & AVD  &  92.4     & 288       & 130       &  \textbf{81.1}     &     -     &   121 \\
    MEAN & DSC  &   0.294   &   0.275   &   0.255   &   0.274   &     -     &   \textbf{0.323\textsuperscript{*}} \\
         & HD95 & 122       & 121       & 121       & 119       &     -     &   \textbf{117} \\
         & ASD  &  51.4     &  40.7     &  51.3     &  \textbf{40.2}     &     -     &   45.7 \\
         & AVD  &  \textbf{70.5\textsuperscript{*}}     & 269       &  95.6     &  82.6     &     -     &   110 \\
    \midrule
    GGO  & DSC  &   0.403   &   0.380   &   0.428   &   0.404   &   0.364   &     \textbf{0.424} \\
    +    & HD95 &  77.1     &  81.5     &  \textbf{68.6}     &  74.3     &   82.4    &     69.9 \\
    CPP  & ASD  &  16.6     &  16.0     &  16.8     &  16.2     &   \textbf{13.3\textsuperscript{*}}    &     19.9 \\
         & AVD  &  78.8     & 212       &  90.2     &  96.4     &   108     &     \textbf{75.6} \\

    \bottomrule
\end{tabularx}

    \label{tab:master_mc}
\end{table}

\subsection{Multiclass lesion segmentation - test set}
\label{sec:results_multiclass_lesion_segmentation_testset}

Table \ref{tab:master_mc_testset} gives the performance metrics for the independent test set for all multiclass lesion segmentation methods: WASS, UNWM, DMmc, 2DRnx, InfNt and MAJ. No individual method outperforms the others on all metrics and majority voting only obtains the highest DSC for segmenting GGO and for the MEAN of all classes.

\begin{table}[H]
    \caption{Mean of metrics for multiclass lesion segmentation task for test set. The best value for each metric is highlighted in bold and indicated with an * if significantly superior than all the other methods.}
    \begin{tabularx}{\linewidth}{ll|YYYYYY}
    \toprule
     & \emph{method} $\rightarrow$   &    WASS &      &  DMmc &   &     InfNt & \\
    class & metric &         &  UNWM       &         &   2DRnx      &        & MAJ \\
    \midrule
    CON     & DSC &   0.162 &   0.368 &   \textbf{0.420} &   0.335 &   0.325 &   0.369 \\
            & HD95 & 231 & 267 & 190 & \textbf{176} & 201 & 208 \\
            & ASD & 158 & 188 & 133 & \textbf{120} & 141 & 146 \\
            & AVD &   9.79 &  93.4 &   \textbf{4.87} &   6.99 & 111 &  10.8 \\
    CPP     & DSC &     - &     - &     - &     - &     - &     - \\
            & HD95 & \textbf{125} & 239 & 243 & 207 &     - & 170 \\
            & ASD &  \textbf{93.3} & 168 & 177 & 144 &     - & 124 \\
            & AVD & 176 & 451 &  \textbf{37.0} & 131 &     - & 256 \\
    GGO     & DSC &   0.330 &   0.514 &   0.504 &   0.406 &     - &   \textbf{0.523} \\
            & HD95 & 133 & 118 & 120 & \textbf{117} &     - & 120 \\
            & ASD &  58.6 &  69.1 &  58.5 &  \textbf{55.8} &     - &  58.8 \\
            & AVD & 528 & \textbf{122} & 239 & 308 &     - & 190 \\
    MEAN    & DSC &   0.253 &   0.465 &   0.460 &   0.383 &     - &   \textbf{0.469} \\
            & HD95 & \textbf{163} & 208 & 184 & 167 &     - & 166 \\
            & ASD & \textbf{103} & 142 & 123 & 107 &     - & 109 \\
            & AVD & 238 & 222 &  \textbf{93.6} & 149 &     - & 152 \\
    \midrule
    GGO     & DSC &   0.440 &   0.514 &   \textbf{0.517} &   0.461 &   0.352 &   0.482 \\
    +       & HD95 & 132 & 118 & 119 & 119 & 148 & \textbf{113} \\
    CPP     & ASD &  57.9 &  69.0 &  58.2 &  \textbf{56.4} &  66.9 &  78.1 \\
            & AVD & 352 & \textbf{121} & 202 & 178 & 285 & 204 \\
    \bottomrule
\end{tabularx}

    \label{tab:master_mc_testset}
\end{table}

\begin{figure*}[htb]
    \centering
    \resizebox*{\linewidth}{!}{
    
    \def\arraystretch{0}
    \setlength{\tabcolsep}{0pt}
    \begin{tabularx}{\linewidth}{l @{\hspace{2pt}} YYYYYYY}

        &  CT &  WASS &     UNWM &  DMmc &  2DU &  InfNt &  MAJ \\

        \rotatebox{90}{\hspace{20pt} IQR25} &
        \includegraphics[width=\linewidth]{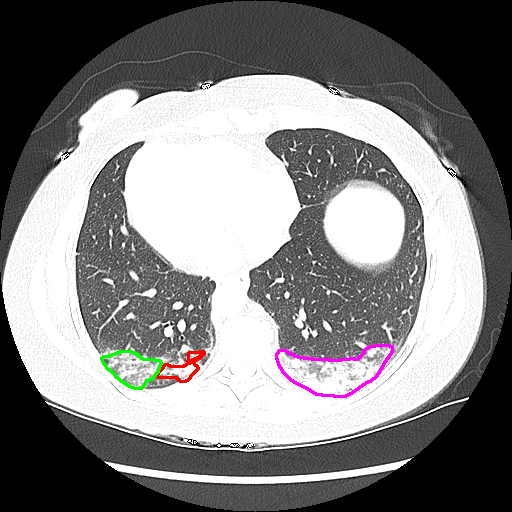} &  
        \includegraphics[width=\linewidth]{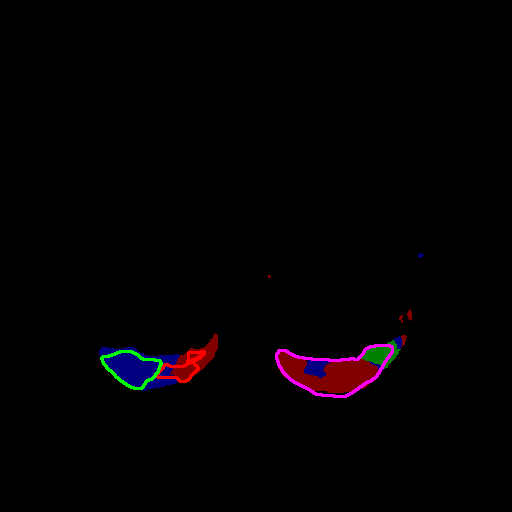} &  
        \includegraphics[width=\linewidth]{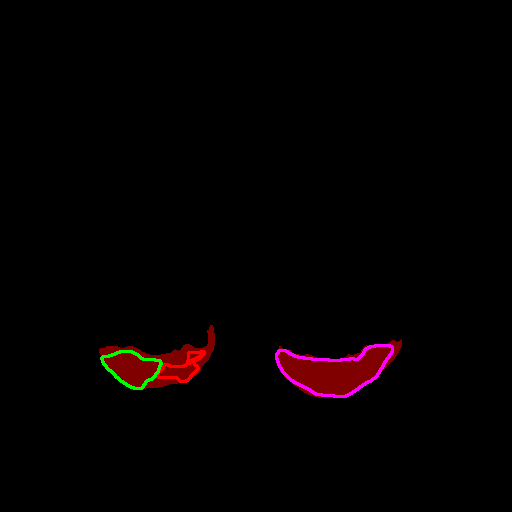} &  
        \includegraphics[width=\linewidth]{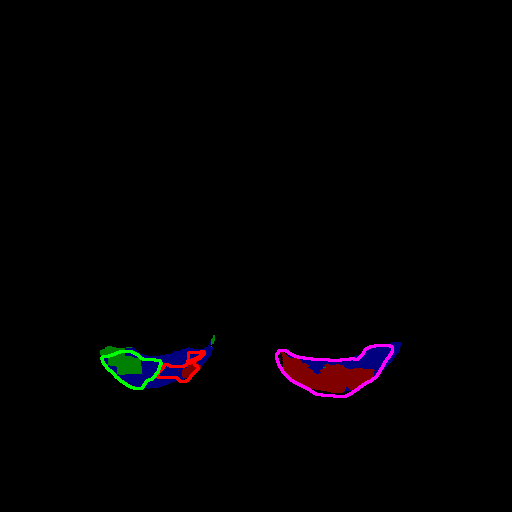} &  
        \includegraphics[width=\linewidth]{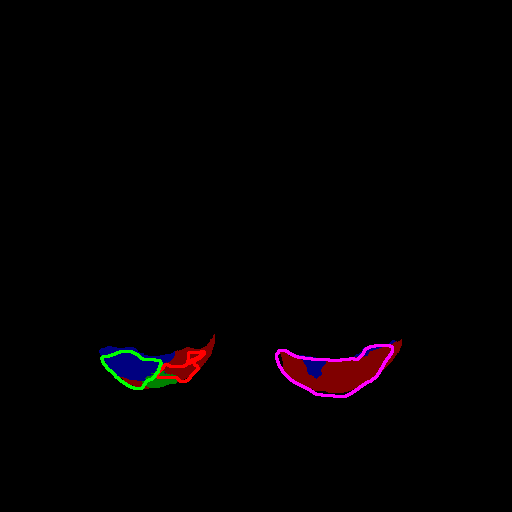} &  
        \includegraphics[width=\linewidth]{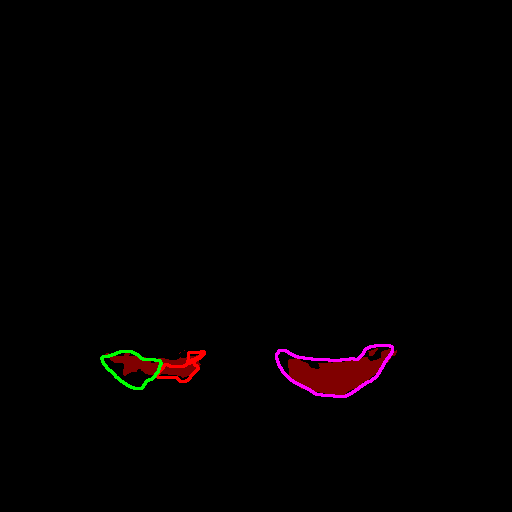} &  
        \includegraphics[width=\linewidth]{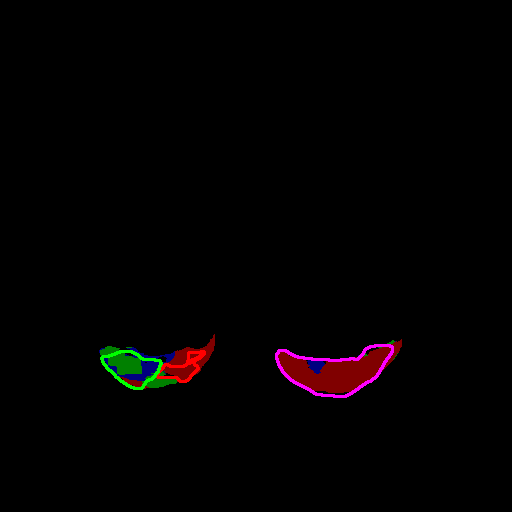}   \\ 
        
        \rotatebox{90}{\hspace{20pt} MEDIAN} &
        \includegraphics[width=\linewidth]{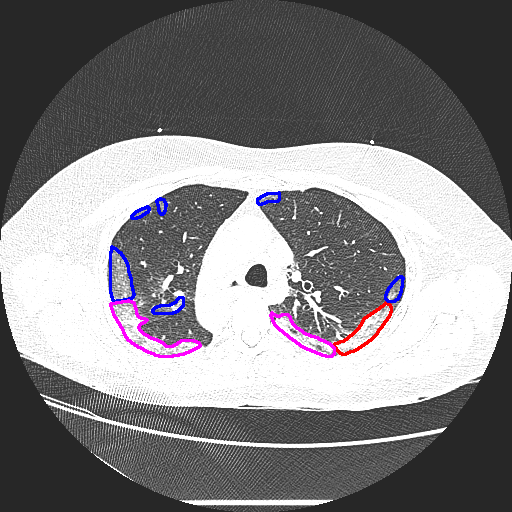} &  
        \includegraphics[width=\linewidth]{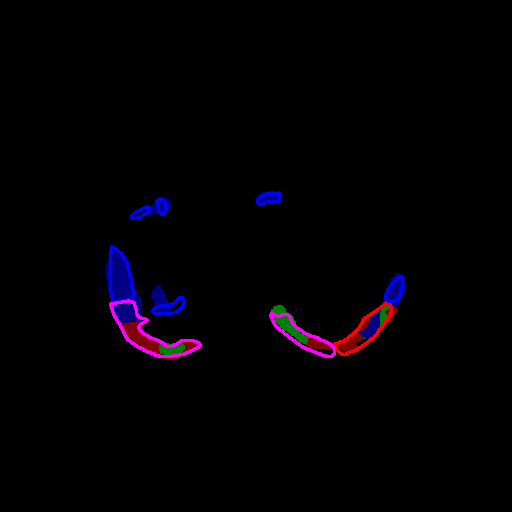} &  
        \includegraphics[width=\linewidth]{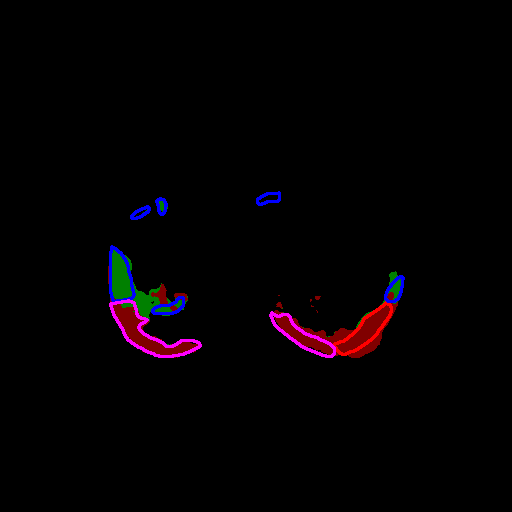} &  
        \includegraphics[width=\linewidth]{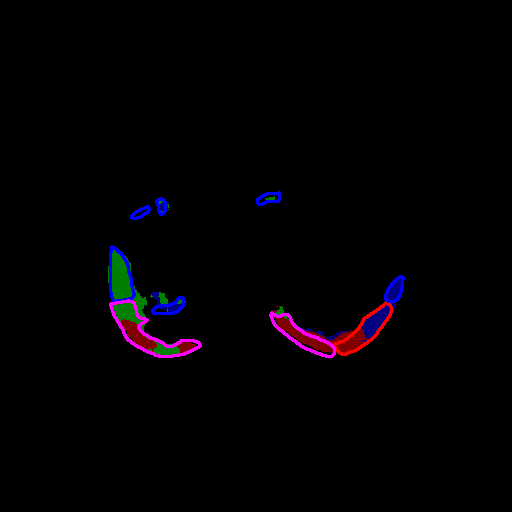} &  
        \includegraphics[width=\linewidth]{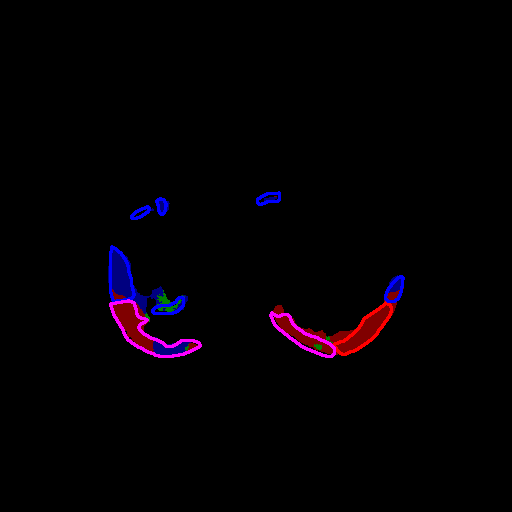} &  
        \includegraphics[width=\linewidth]{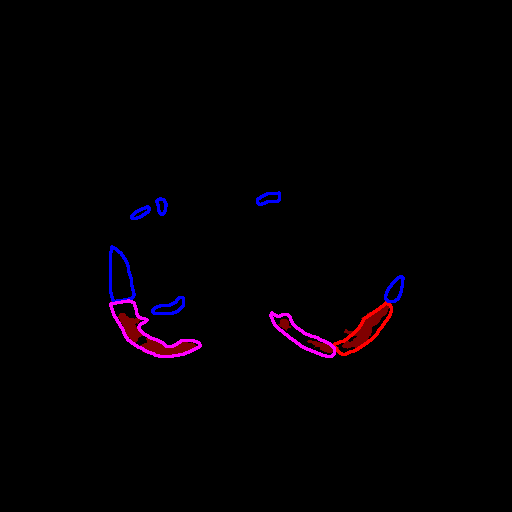} &  
        \includegraphics[width=\linewidth]{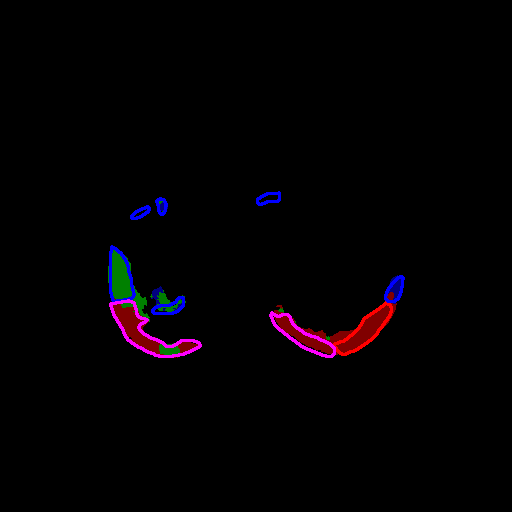}   \\ 
        
        \rotatebox{90}{\hspace{20pt} IQR75} &
        \includegraphics[width=\linewidth]{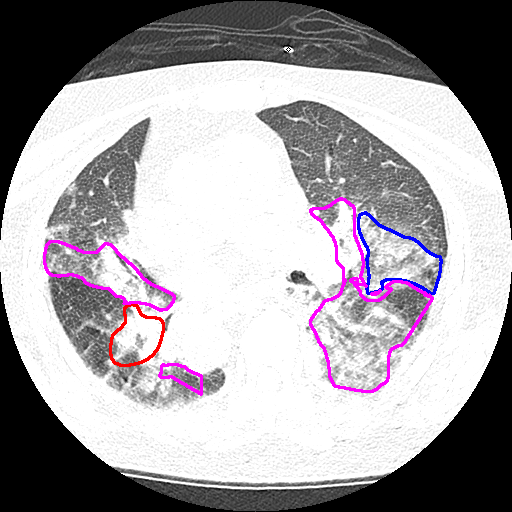} &  
        \includegraphics[width=\linewidth]{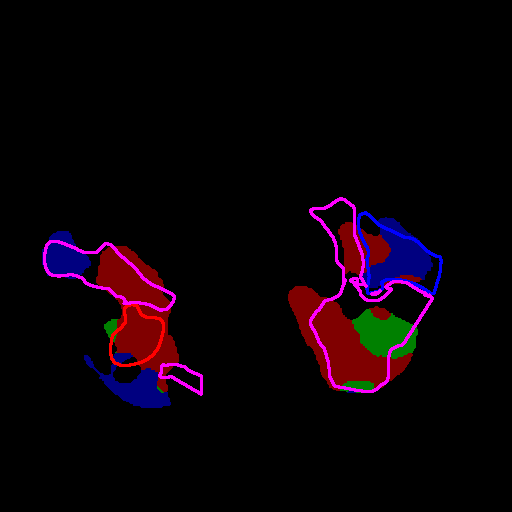} &  
        \includegraphics[width=\linewidth]{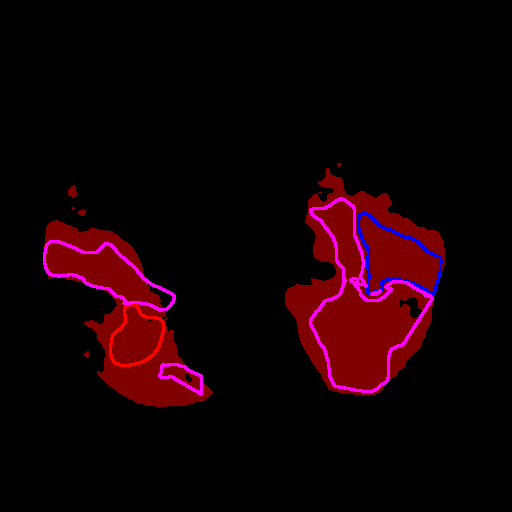} &  
        \includegraphics[width=\linewidth]{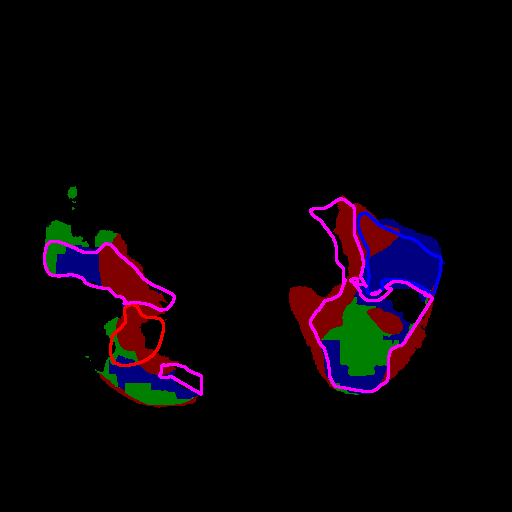} &  
        \includegraphics[width=\linewidth]{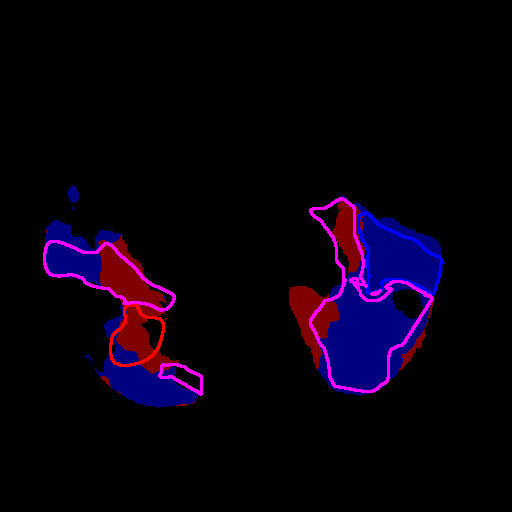} &  
        \includegraphics[width=\linewidth]{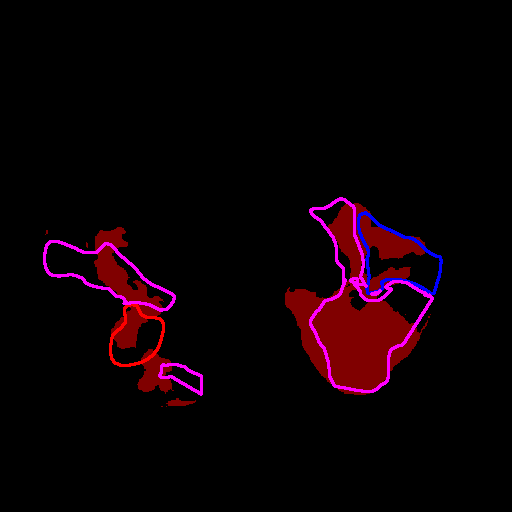} &  

        \includegraphics[width=\linewidth]{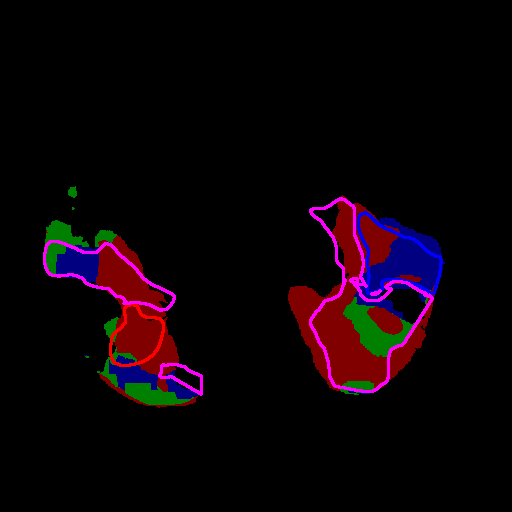}

    \end{tabularx}
    }
        \caption{A qualitative evaluation of the resulting predictions for multiclass lesion segmentation on the validation set from each method (columns, from left to right: original CT, and predictions from WASS, UNWM, DMmc, 2DRnx, InfNet and MAJ). The manual ground truth is depicted as red (CON), green (GGO), blue (CPP) and magenta (COM) contours on each image. The predictions are shown with the same color representation. The three rows are slices from different CT scans for which the mean AVD over the different classes is respectively the median, first and third quartile value.}
    \label{fig:master_qualitative_mc}
\end{figure*}

\section{Discussion}
The deep learning methods compared in this study depend on many hyperparameters. These include the choice of the architecture with different depths, number of layers and strategy to handle multi-scale information, the choice of the loss function, the choice of the optimizer and the optimization parameter such as the learning rate, the choice of addressing one task or several, and also the choice of the data preprocessing steps and the type of data augmentation operations used during training.
Different deep learning methods can give similarly good segmentations, but they are likely to have different biases and to make different mistakes.
In this case, the ensembling of diverse models can lead to averaging out the inconsistencies due to the hyperparameter and improve the segmentation performance and robustness~\citep{bishop2006pattern,kamnitsas2017ensembles}.
In several cases, this is confirmed by our results. Both on the validation and test set, and for each of the three main tasks, ranking the methods for each metric and averaging these ranks puts the majority voting method five times as number one and once as number three. There is no other method having a consistent top-three ranking.
%
In the lung segmentation cross-validation (Table~\ref{tab:master_lung}) the best DSC, HD95, ASD and AVD were reached by the ensemble method where the DSC of 0.978, HD95 of 2.43~mm and ASD of 0.559~mm were significantly superior. Majority voting also reaches the highest sensitivity for segmenting CPP. In the test set (Table~\ref{tab:master_lung_testset}) the ensemble method still achieves the best DSC (0.987) and AVD (30.3~ml), where the former is significantly better. However, JoHof achieves the best ASD (0.395~mm) and the significantly best HD95 (1.39~mm).
%
In the lesion segmentation the differences are less clear due to the more challenging nature of the problem. For the binary lesion segmentation in the validation set (Table~\ref{tab:master_bin}), majority voting obtains a significantly superior DSC of 0.554 and a lower HD95 of 42.6~mm than any of the individual models. On the test set (Table~\ref{tab:master_bin_testset}) the DSC of 0.724 and AVD of 91.3~ml are the best results where the former is significantly better. However, now the WASS method achieves the best HD95 (77.6~mm) and ASD (24.6~mm).
For the multiclass lesion segmentation, the best DSC in cross-validation (Table~\ref{tab:master_mc}) is achieved by the ensembling method for segmenting CPP, GGO, mean and GGO + CPP with values of 0.303, 0.339 (significantly better), 0.323 (significantly better) and 0.424 respectively. Significantly lower volume differences were achieved by WASS for segmenting CPP with 68.0~ml and mean with 70.5~ml. The ASD is significalty better for CON using 2DRnx (18.3~mm) and for GGO + CPP using InfNt (13.3~mm). There is no method that performs best for all metrics. In the test set, the superiority of majority voting is no longer observed. Ensembling only reaches the highest DSC for the segmentation of GGO (0.523) and mean (0.469). For CPP and mean WASS performs best regarding HD95 (125~mm and 163~mm respectively) and ASD (93.3 and 103 respectively). Though, the advantage is less obvious, ensembling is still the preferred method as it achieves the most consistent results for all metrics.

A qualitative assessment for the lung segmentation, binary lesion segmentation and multiclass lesion segmentation is given in Figures~\ref{fig:qualitative_master_lungs_testset}, \ref{fig:qualitative_master_bin_testset} and \ref{fig:master_qualitative_mc} respectively. These examples illustrate that, for certain regions, the methods indeed make different mistakes which are averaged out in the majority voting. However some areas are wrongly predicted to be a lesion or healthy tissue by most of the models or even all of them. Addition of more training data could reduce these errors.

Concerning clinical practice, the absolute volume difference is the most important metric as the percentage of lung involvement reflects the severity of the disease. However, from our results we notice that this AVD does not always correlate with other metrics commonly used to assess segmentations (DSC, HD, ASD). This is in line with the findings from~\cite{Bertels2019a, Bertels2019b}, respectively showing that soft Dice optimization successfully optimizes the Dice score at test time, but that it may lead to biased volume estimations due to the presence of inherent uncertainty in medical segmentation tasks. Furthermore, the fact that different methods were using different objective functions could explain the absence of a correlation between the AVD and other segmentation metrics across methods. For example, in the 5-fold cross-validation, the lowest AVD when segmenting CON (Table~\ref{tab:master_mc}), is 50.2~ml reached by the DMmc model (using cross-entropy optimization) but the DSC is lower than that of WASS, 2DRnx (both using dice optimization) and majority voting. 

Areas of consolidation in a peripheral subpleural location, which is a common finding in COVID-19, are difficult to delineate by automated lung segmentation methods because of the same high density as some chest wall structures.
Indeed, it was observed that all methods have some difficulty at the edge of the lungs where there is consolidation and especially at the back of the lungs. 
%
%
Since these lung masks are often used to constrain the lesion segmentation, this is potentially problematic as it can cause clinically important, missed consolidations in further steps of the pipeline. 
The sensitivity to include CON lies between 0.685 and 0.845 for the validation set, meaning that on approximately 20-30\% of consolidation volume is missed in further steps of the pipeline solely due to an incorrect lung mask. For CPP and GGO, all methods miss less then 0.5\% and 2.4\% of lesion, respectively.
Even though majority voting has the highest scores for all metrics for lung segmentation, it misses almost 5\% additional CON as compared to WASS and the difference in AVD is not statistically significant between the two methods. Since these two metrics are arguably more important from a clinical perspective, WASS could be the preferred choice for lung segmentation in COVID-19 patients.
On the test set, the values for the sensitivity for CON lie closer together, varying from 0.896 for CTA to 0.947 for UNWM and sensitivity for GGO is on average 99\%. Majority voting misses about 6~\% of consolidation lesions and only about 0.2~\% of ground glass opacities are omitted from the lung masks.

Concerning lung segmentation by CTA, a large difference in HD95 and ASD between the validation set and the test set is observed. HD95 increases from 16.0~mm to 168~mm and the ASD increases from 0.636~mm to 83.5~mm. For selected patients, air in the intestines was improperly segmented by CTA as part of the lungs, which led to outliers in the mean HD95 and ASD calculations. As the lung segmentation is the first step in the pipeline, a similar discrepancy can be observed between the validation and test set for the binary lesion segmentation.

For all validated methods, we noticed rather low dice scores for lesion segmentation and multiclass lesion segmentation. 
The delineation of the different tissue types is difficult and is subject to several sources of variability during the manual segmentation process:
the subjectivity of the radiologist,
the annotation tool and protocol used for the delineation,
the indistinct boundaries between different tissue types,
the variation in resolution of the CT scans across different centers,
and the presence of artefacts.
This underscores the need for automated methods for lesion segmentation, that can robustly produce more objective segmentation for CT scans from different centers.
For training, using CT scans that have been segmented by several raters can mitigate the overfitting of the network to a given rater.
Indeed, if the manual segmentations used for training differ depending on the raters, the neural network should retain and learn what is consistent across the different raters. 
To assure a certain amount of consensus, the manual ground truth in our dataset was reviewed by a different rater from the one who created the segmentation.

\begin{figure}[ht]
    \centering
    \includegraphics[width=\linewidth]{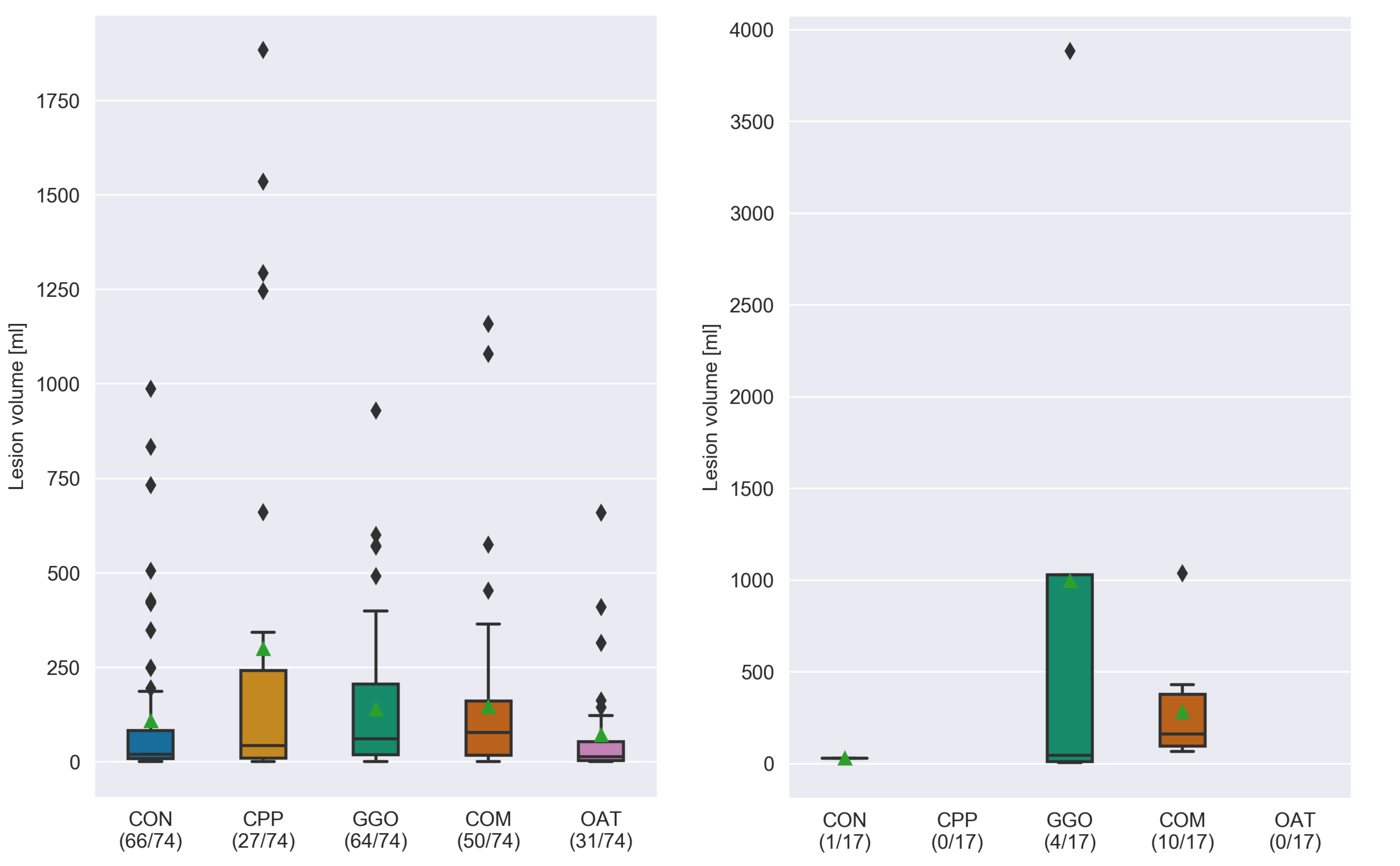}
    \caption{Boxplot of the lesion volumes per type for the validation set (left) and test set (right). Cases with a lesion size of 0~ml are discarded. Under each lesion type, the number of cases with such a lesion with respect to the total number of cases is indicated. The mean volumes are shown with a green triangle.}
    \label{fig:lesion_sizes}
\end{figure}

On top of the challenging nature of the ground truth delineation process, the lesion segmentation problem is highly imbalanced. Figure~\ref{fig:lesion_sizes} shows the distribution of lesion sizes per type, both for the validation set and the test set. In each boxplot, the cases which did not contain the considered lesion type were discarded. In this way, the boxplots represent the distributions of the actual lesion sizes without being influenced by the absence of a certain type. Per lesion type, the number of cases included are indicated. Within both sets, there is great variation in lesion size, moreover not all types are represented equally. Though this provides an additional difficulty for the deep learning methods, it provides a realistic representation of data collected in clinical practice. In addition, there remain numerous small lesions for which it is hard to obtain a high dice score due to the low number of potential true positives. This figure also shows that the test set in our study is fairly limited for the multiclass lesion segmentation. The large volume of COM can be explained by the public cases where only a binary label was present. 

The ground truth segmentations contain lesions with the label 'combined pattern'. For these lesions it was deemed either too difficult or time consuming to distinguish the substructures. Indeed, the different tissue types do not always have distinct boundaries since during the course of the disease, the lesions can evolve from one type to another and can therefore be in an intermediate state. The presented multiclass methods do not have this class and will predict for each voxel one of the lesion classes (CON, CPP or GGO). However, as mentioned in Section \ref{jeroen_method}, the probabilistic nature of CNNs also allows to form a combined tissue class if the probabilities of two or more classes are similar. Predictions of the voxels manually labeled as COM cannot be validated quantitatively since the ground truths for the separate lesion types in the combined pattern were not available. Although an initial qualitative evaluation by a radiologist was positive, further clinical investigation is needed to confirm this.

%
Any affected lung tissue that did not correspond to any of the lesion types (GGO, CON, CPP, LIN, RHS or COM), was labeled as 'other abnormal tissue'. This class was included in the binary lesion segmentation, but not in the multiclass segmentation. Hence, the binary methods segment all affected lung tissue while the multiclass models segment the lesions specific to COVID-19. The more challenging nature of the latter is reflected in the average DSC values: 0.724 for the binary segmentation and 0.523, 0.303 and 0.420 for GGO, CPP and CON respectively. Moreover, the 2-step approaches, DMmc and 2DRnx, use these binary segmentations and label the other abnormal tissue as CON, CPP or GGO. Although this affects the performance of the models, the effect is limited thanks to the relatively small size of this class (Figure~\ref{fig:lesion_sizes}). There is no clear distinction between the performance of the 1-step and 2-step methods.

We note a bias in our comparison on the validation set. For the internal methods (DMLu, WASS, 2DS, UNWM, 2DRnx and DMmc) the evaluation is performed through a 5-fold cross-validation, meaning that the training and validation set have the same distribution. However, the external methods (CTA, JoHof, InfNet, CovidENet) and DMLo were trained on different data. Thus, for the latter models, the validation set already acted as a test set, leading to a bias in the results. 
Independence of the test set towards internal methods was assured by using CT scans of different centers compared to the training set and by using a different annotation tool. Additionally, 10 public cases were included. 

As previously stated, for clinical use, the volume of the affected lung tissue is the most important parameter. However, the absolute differences between the segmented volumes and ground truths varies greatly between the various methods and between the binary and multiclass segmentation. With respect to the binary segmentation, the smallest AVD on the test is 91.3~ml, achieved by majority voting. With respect to multiclass segmentation, CON and CPP structures are segmented by the DMmc method with AVDs of 4.87~ml and 37.0~ml respectively. The lowest obtained AVD for GGO is 122~ml (UNWM method).
In the work of \cite{colombi2020well}, the estimation of the percentage of well aerated lung volume, i.e. the volume of lung without lesion divided by the total lung volume, is rounded to the nearest 5\%.
This level of accuracy allow them to predict ICU admission or death with an UAC of 0.83.
For an average total lung volume of approximately 6~L, our highest volume error of 122~ml corresponds to approximately 2\% of estimated percentage of volume error.
This suggests that our methods give a good estimation of the percentage of affected lung tissue, distinguish between the different types of lesions and aid in the assessment of the severity of the disease. The methods support radiologists in the labour-intensive lesion segmentation task and improve robustness. However, clinical supervision is still required. A version of the software that is permitted for clinical use in Europe and US, is available as icolung\footnote{\url{https://icometrix.com/services/icolung}}.

Future work should focus on expanding the test set in order to ameliorate errors associated with its limited size, particularly class imbalance. 

\section{Conclusion}
An increasing number of studies propose to use deep learning to provide fast and accurate quantification of COVID-19 using chest CT.
%
%
We have proposed the first systematic comparison of a large variety of deep learning methods for CT segmentation using a multi-center dataset.
Seven in-house methods and four public deep learning methods have been compared (Table~\ref{tab:overview}).

We found that all methods achieved a total lesion volume prediction with an average volume difference that is below the accuracy of human raters~\citep{colombi2020well} (see AVD in Table~\ref{tab:master_bin} and~\ref{tab:master_bin_testset}).

The differences between models on all different tasks remain small overall, and no method outperforms all the others for all metrics.
Ensembling different methods improves the performance for binary lung and binary lesion segmentation. 
%
%
Multiclass lesion segmentation is more difficult by nature but ensembling the different methods provides the most consistent results.
The code for all the methods developed in this paper will be publicly available. 

\section*{Acknowledgments}
We would like to thank MD.AI and Materialise for providing access to their platforms. We also thank the Materialise team for their effort in delineating images, in particular Sophie Deckx, Antoon Dierckx and the support engineers. Finally, we thank the many radiologists who volunteered to delineate datasets.

The computational resources and services used in this work were provided in part by the VSC (Flemish Supercomputer Center), funded by the Research Foundation Flanders (FWO) and the Flemish Government – department EWI. This research also received funding from the Flemish Government under the “Onderzoeksprogramma Artifici\"{e}le Intelligentie (AI) Vlaanderen” programme.

Lucas Fidon has received funding from the European Union's Horizon 2020 Research and Innovation Programme for this project under the Marie Sk{\l}odowska-Curie grant agreement TRABIT No 765148. Jeroen Bertels is part of NEXIS, a project that also has received funding from the European Union's Horizon 2020 Research and Innovations Programme under the grant agreement No 780026.
Tom Eelbode, Siri Willems and Sofie Tilborghs are supported by PhD fellowships of the Research Foundation - Flanders (FWO). David Robben is supported by an innovation mandate of Flanders Innovation \& Entrepreneurship (VLAIO).

The icovid project has received funding from the European Union’s Horizon 2020 Research and Innovation Programme under grant agreement No 101016131.



\bibliographystyle{model2-names.bst}\biboptions{authoryear}

\clearpage
\section*{Appendix}

\setcounter{section}{0} 
\setcounter{figure}{0}
\setcounter{equation}{0}

\def\rvy{{\mathbf{y}}}

\subsection*{Generalized Wasserstein Dice Loss}
\label{wasserstein_appendix}
We provide here more details about the Generalized Wasserstein Dice loss and the choice of the hyperparameter for the loss that were used in the method WASS. 

The Generalized Wasserstein Dice Loss used in the WASS method is a generalization of the Dice Loss for multi-class segmentation that can take advantage of prior knowledge about the set of classes~\citep{fidon2017generalised}.

In the WASS method, we used the Generalized Wasserstein Dice Loss to account for the presence of super class in the manual segementation: COM that corresponds to a mix of ground glass opacity, consolidation and crazy paving pattern, and OAT that corresponds to abnormal tissue that are not considered for prediction in this work.
When a voxel cannot be well classified, the Generalized Wasserstein Dice Loss favors mistakes that are semantically more plausible.

We used the following formulation for the generalized Wasserstein Dice loss between a predicted multi-class probability segmentation map 
$\hat{\rvy} = \left(\hat{y}_{i,l}\right)_{i,l}$ 
and the ground truth probability segmentation map 
$\rvy = \left(y_{i,l}\right)_{i,l}$
with $i \in \{1, \ldots, N\}$ the index for voxel, and $l \in \{1, \ldots, L\}$ the index for classes
\begin{equation}
\label{generalised_dice}
    \mathcal{L}_{WASS}(\hat{\rvy}, \rvy) = 
    \frac{ 
        2 \sum_{l=1}^L \alpha_l \sum_{i=1}^N y_{i,l}
        \left(1 - W^M(\hat{\rvy}_i, \rvy_{i})\right)
    }
    {
        \sum_{l=1}^L \alpha_l \sum_{i=1}^N y_{i,l}
        \left(2 - W^M(\hat{\rvy}_i, \rvy_{i})\right)
    }
\end{equation}
where $W^M\left(\hat{\rvy}_i, \rvy_i\right)$ is the Wasserstein distance between the predicted $\hat{\rvy_i}$ and the ground truth $\rvy_i$ one-hot probability distribution at voxel i, and where we used a weighting schema similar to the one used by \citep{sudre2017generalised} for the Generalized Dice Loss, with
\begin{equation}
    \forall l \in \{1, \ldots, L\}, \quad 
    \alpha_l = \frac{1}{1 + \sum_{i=1}^N y_{i,l}}
\end{equation}

The Wassertein distance $W^M$ depends on a distance matrix defined on the set of classes \{background, GGO, CON, CPP, COM, OAT, healthy lung\} as
\begin{equation}
    \label{eq:dist_mat}
    M =
    \left(m_{l,l'}\right)_{1 \leq l,\,l' \leq L} = 
    \left(
    \begin{array}{ccccccc}
        0  & 1   & 1   & 1   & 1  & 0  & 1\\
        1  & 0   & 0.8 & 0.8 & 0  & 0  & 1\\
        1  & 0.8 & 0   & 0.8 & 0  & 0  & 1\\
        1  & 0.8 & 0.8 & 0   & 0  & 0  & 1\\
        1  & 0   & 0   & 0   & 0  & 0  & 1\\
        0  & 0   & 0   & 0   & 0  & 0  & 1\\
        1  & 1   & 1   & 1   & 1  & 1  & 0\\
    \end{array}
    \right)
\end{equation}
The matrix M is symmetrical and has only positive values with zeros on the diagonal because it is a distance matrix.
In addition, to normalize M, we choose the maximum distance between classes to be equal to $1$.
We choose the distances between the lesion types GGO, CON, and CPP to be lesser than $1$, so that if the network cannot predict correctly a voxel with a given lesion type, at least it favors labelling this voxel with one of the lesion types.

For COM (column 5), the distances to the other lesions was put to zeros to reflect the fact that voxels labeled as COM can be either GGO, CON, or CPP.
By construction, the CNN of WASS cannot labelled the voxels as COM (it must make a choice).
However, thanks to the Generalized Wasserstein Dice loss and the definition of M \eqref{eq:dist_mat}, the CNN is not penalized for labelling the COM voxels as GGO, CON, or CPP.

Similarly, for OAT (column 6), that corresponds to other types of abnormal tissue that are not predicted by the CNN in WASS, we set the distance to all the other classes except \textit{healthy lung} to 0 to penalize less the CNN on those voxels.

It is worth noting, that since the ground truth segmentation map $\rvy$ is a one-hot segmentation map, for any voxel $i$, the Wasserstein distance has a simple form
\begin{equation}
    W^M\left(\hat{\rvy}_i, \rvy_i\right) = \sum_{l=1}^L\sum_{l'=1}^L m_{l,l'}y_{i,l}\hat{y}_{i,l'}
\end{equation}
\subsection*{Boxplots}
\label{boxplots_appendix}

\begin{figure*}[htbp]
    \resizebox*{!}{\dimexpr\textheight-4\baselineskip\relax}{
    \begin{tabularx}{\linewidth}{lYYYY}

        & DSC & HD95 & ASD & AVD \\
        
        \rotatebox{90}{\hspace{30pt} LUNG} &
        \includegraphics[width=\linewidth]{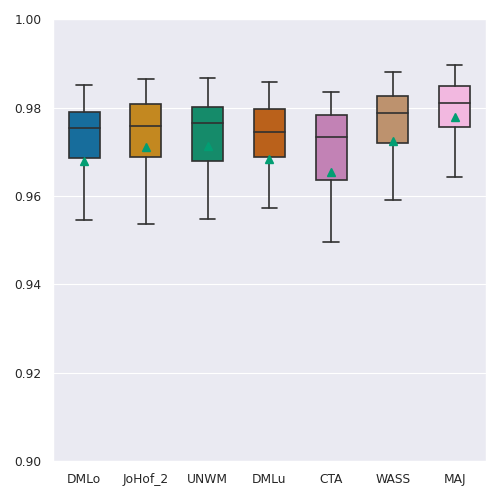} &  
        \includegraphics[width=\linewidth]{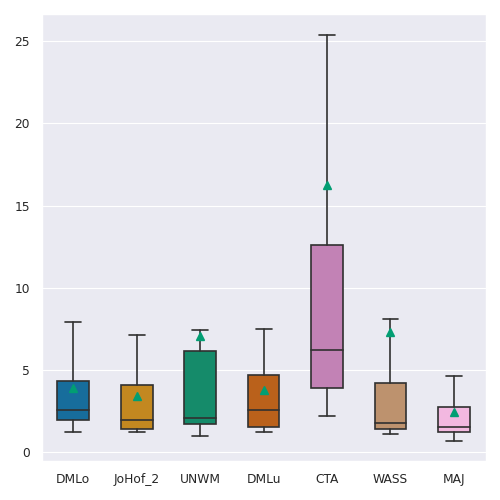}  &
        \includegraphics[width=\linewidth]{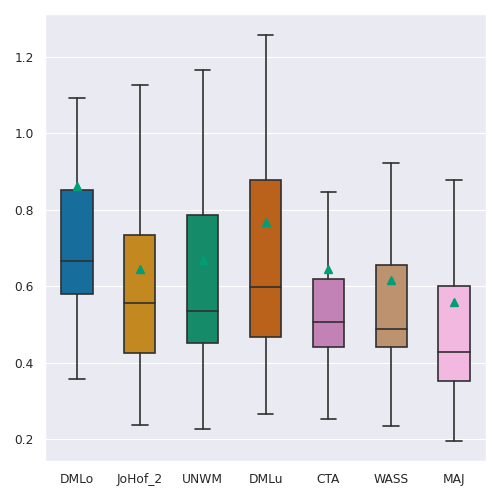} &
        \includegraphics[width=\linewidth]{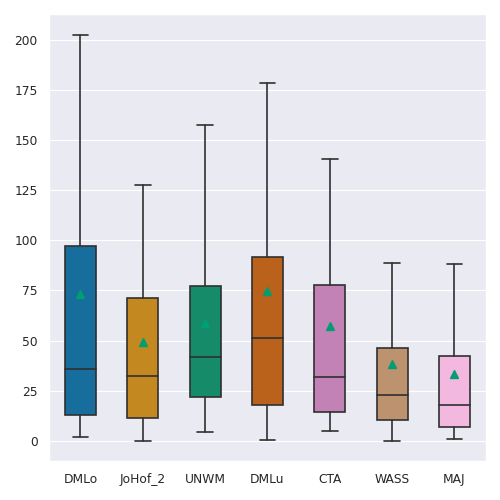}   \\ 
        
        \rotatebox{90}{\hspace{30pt} BIN} &
        \includegraphics[width=\linewidth]{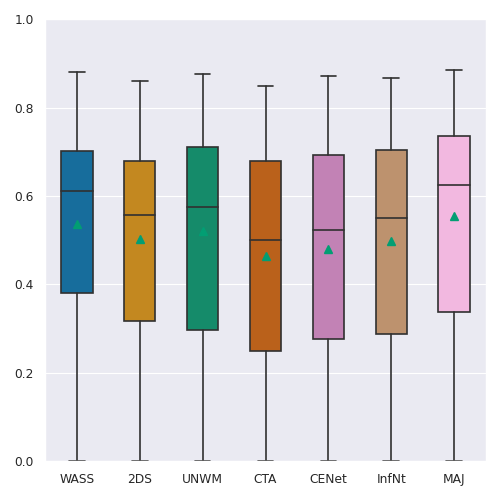} &  
        \includegraphics[width=\linewidth]{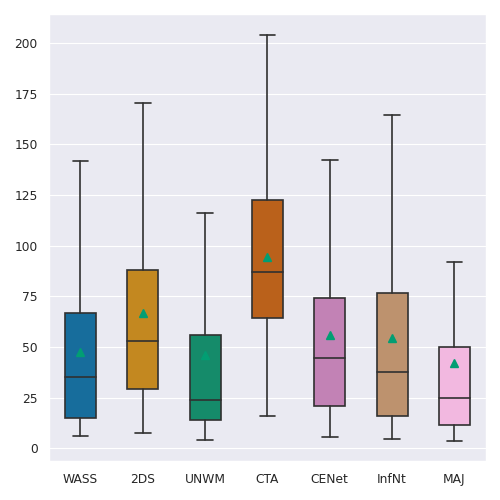} &
        \includegraphics[width=\linewidth]{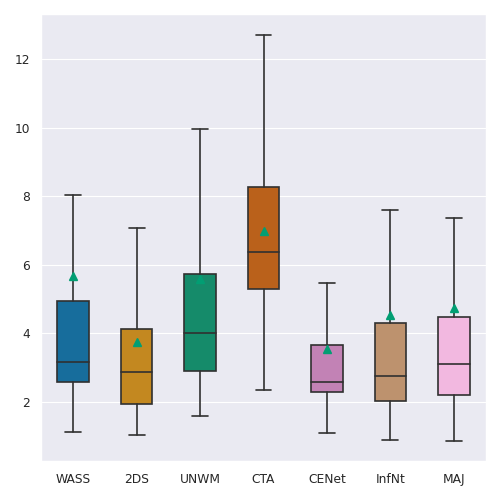} &
        \includegraphics[width=\linewidth]{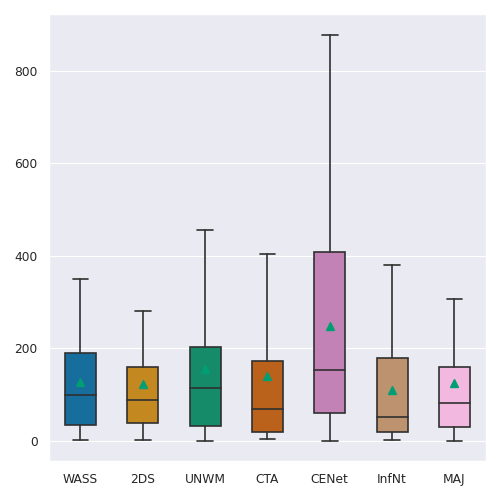}  \\
        
        \rotatebox{90}{\hspace{30pt} CON} &
        \includegraphics[width=\linewidth]{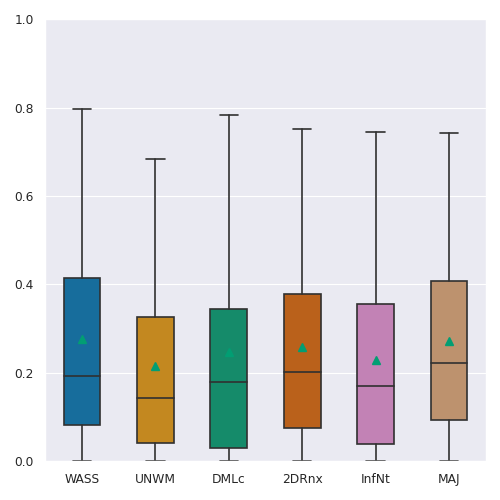} &  
        \includegraphics[width=\linewidth]{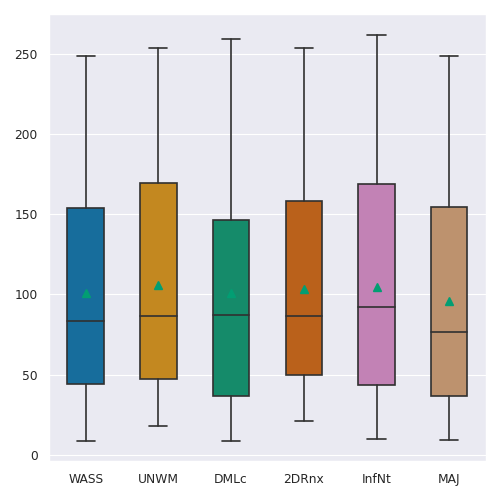} &
        \includegraphics[width=\linewidth]{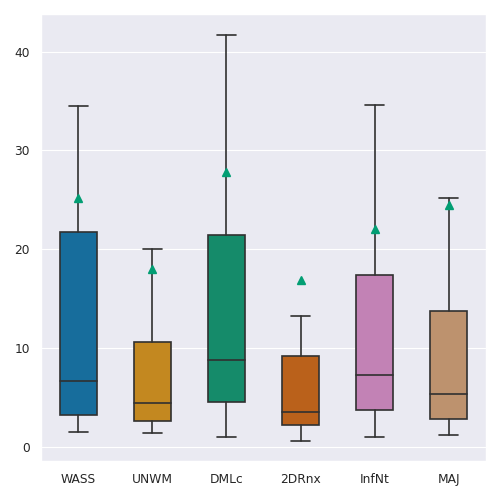} &
        \includegraphics[width=\linewidth]{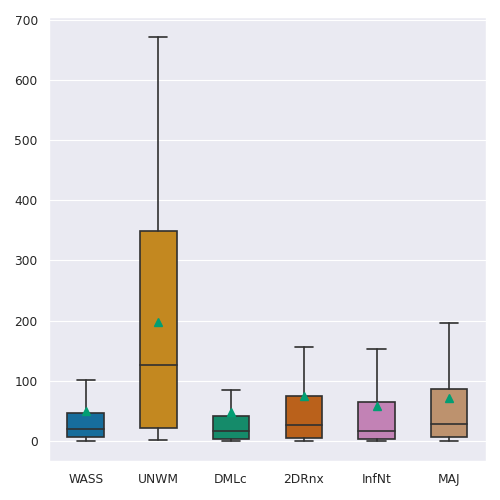} \\
        
        \rotatebox{90}{\hspace{30pt} CPP} &
        \includegraphics[width=\linewidth]{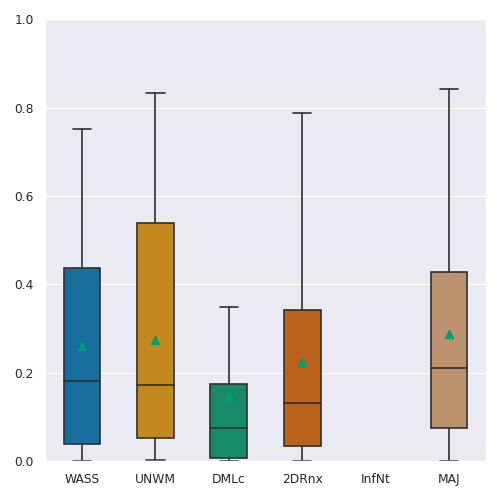} &  
        \includegraphics[width=\linewidth]{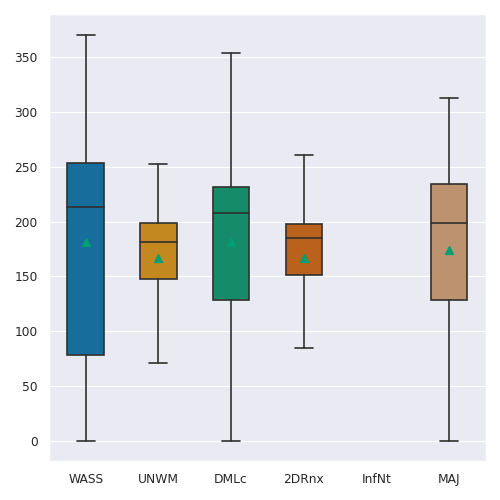} &
        \includegraphics[width=\linewidth]{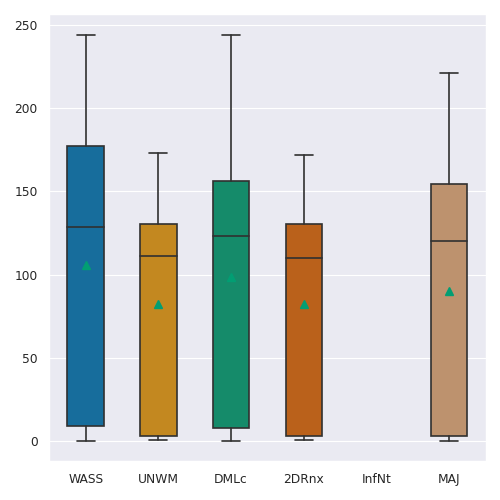} &
        \includegraphics[width=\linewidth]{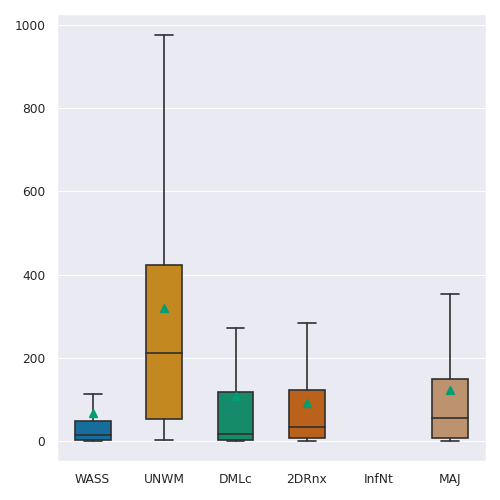} \\
        
        \rotatebox{90}{\hspace{30pt} GGO} &
        \includegraphics[width=\linewidth]{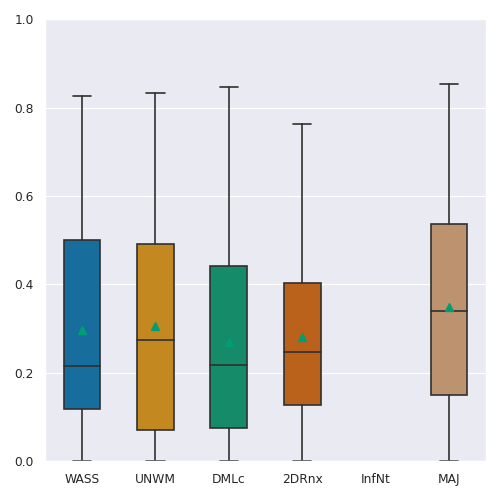} &  
        \includegraphics[width=\linewidth]{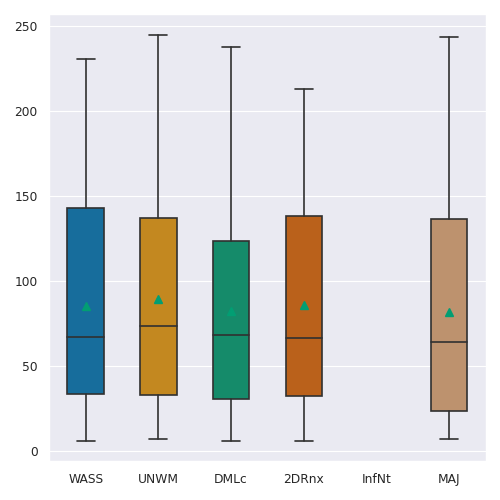} &
        \includegraphics[width=\linewidth]{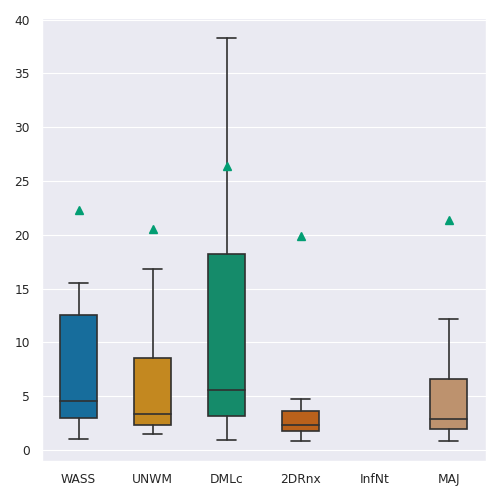} &
        \includegraphics[width=\linewidth]{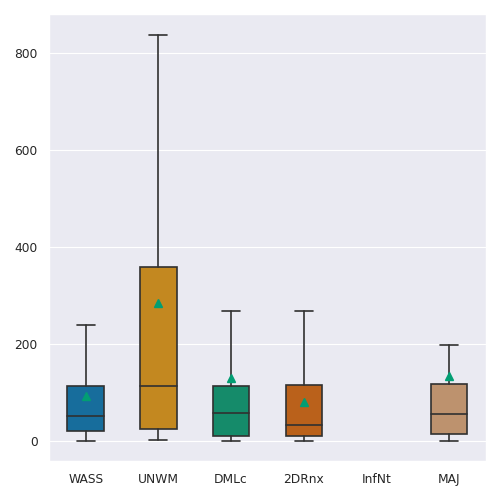} \\ 
        
        \rotatebox{90}{\hspace{30pt} MEAN} &
        \includegraphics[width=\linewidth]{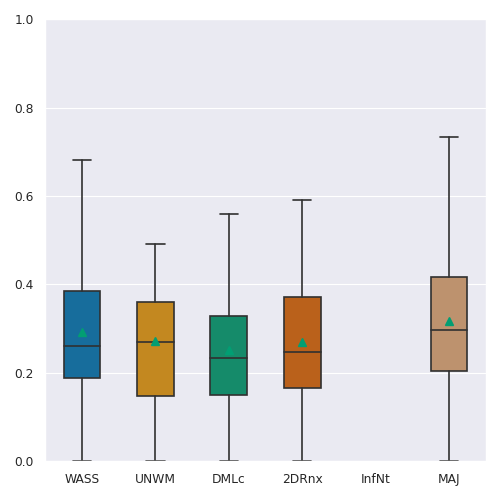} &  
        \includegraphics[width=\linewidth]{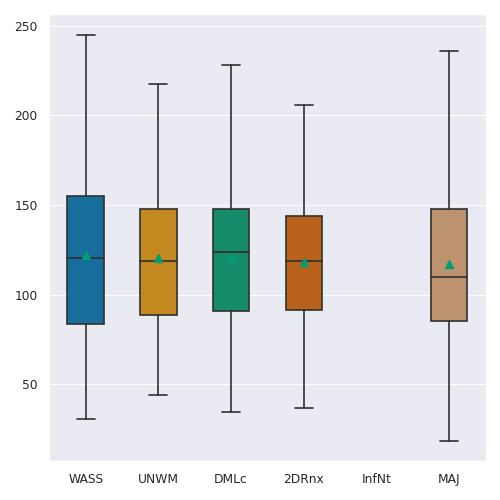} &
        \includegraphics[width=\linewidth]{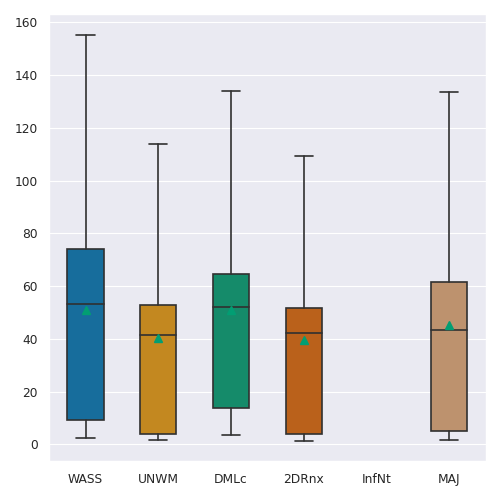} &
        \includegraphics[width=\linewidth]{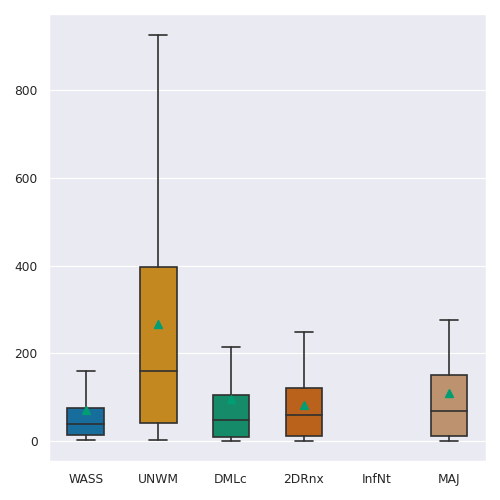} \\
        
        \rotatebox{90}{\hspace{30pt} GGO + CPP} &
        \includegraphics[width=\linewidth]{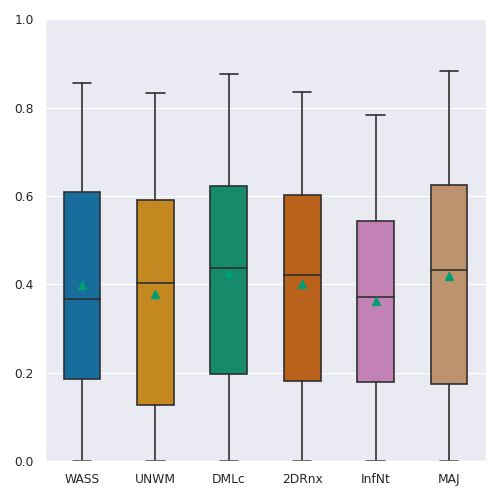} &  
        \includegraphics[width=\linewidth]{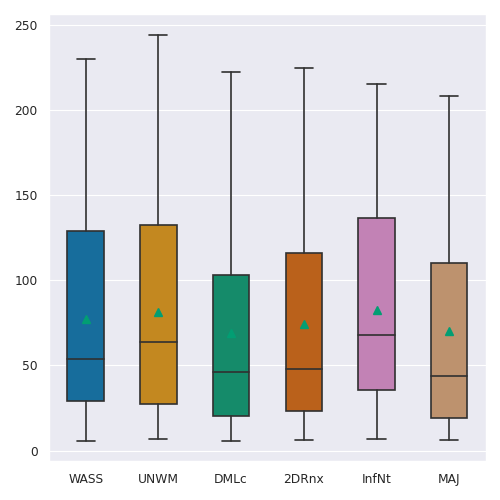} &
        \includegraphics[width=\linewidth]{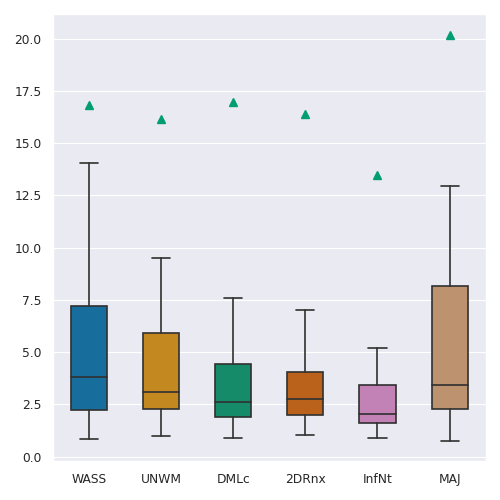} &
        \includegraphics[width=\linewidth]{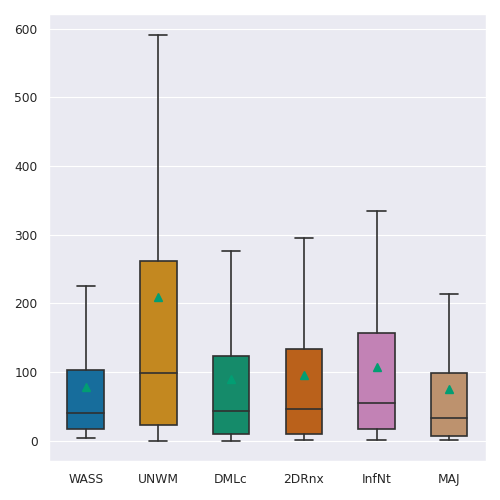} 
        
    \end{tabularx}
    }
    \caption{Boxplots for each of the reported metrics (colums, from left to right: DSC, HD95, ASD and AVD) and challenge tasks (rows, from top to bottom: LUNGS, BIN, CON, CPP, GGO, MEAN of classes and GGO+CPP). The mean of the metric for each method is additionally shown on each boxplot by a green triangle and outliers are not shown. }
    \label{fig:boxplots_master}
\end{figure*}

\begin{figure*}[htbp]
    \resizebox*{!}{\dimexpr\textheight-4\baselineskip\relax}{
    \begin{tabularx}{\linewidth}{lYYYY}

        & DSC & HD95 & ASD & AVD \\
        
        \rotatebox{90}{\hspace{30pt} LUNG} &
        \includegraphics[width=\linewidth]{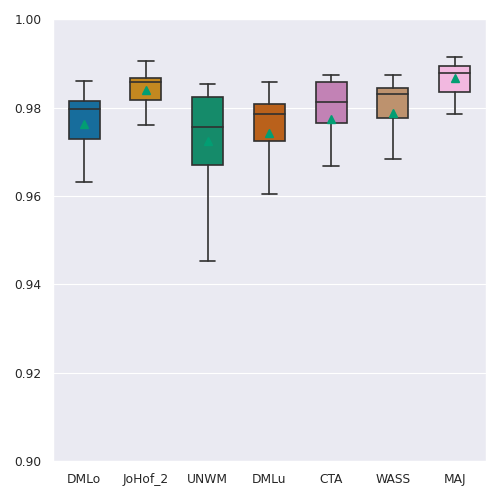} &  
        \includegraphics[width=\linewidth]{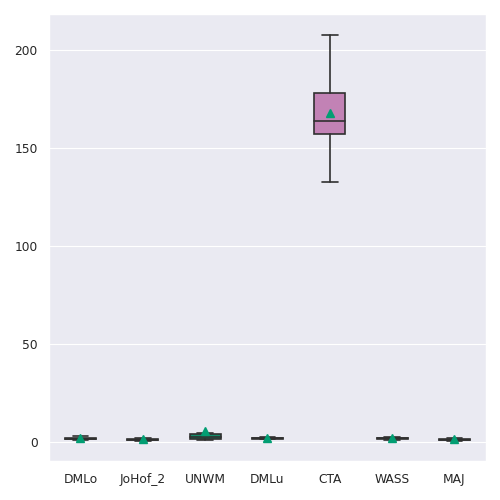}  &
        \includegraphics[width=\linewidth]{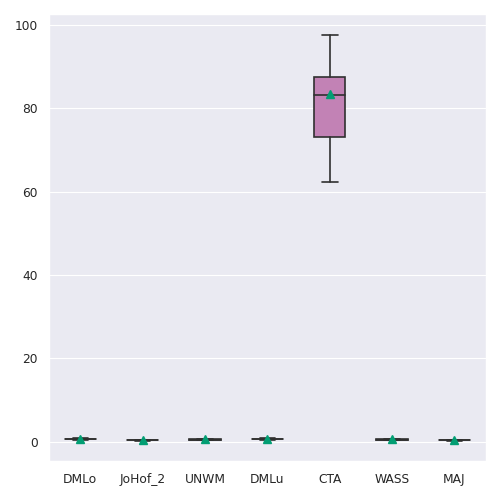} &
        \includegraphics[width=\linewidth]{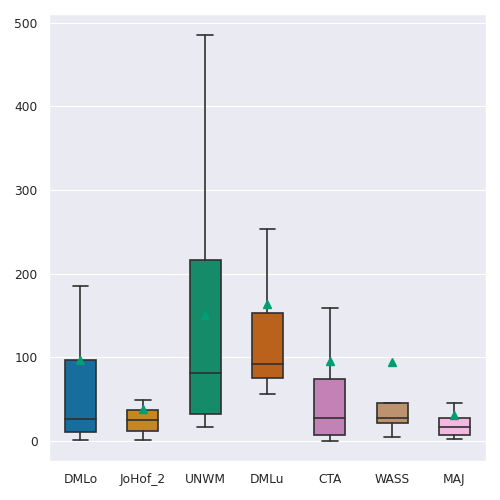}   \\ 
        
        \rotatebox{90}{\hspace{30pt} BIN} &
        \includegraphics[width=\linewidth]{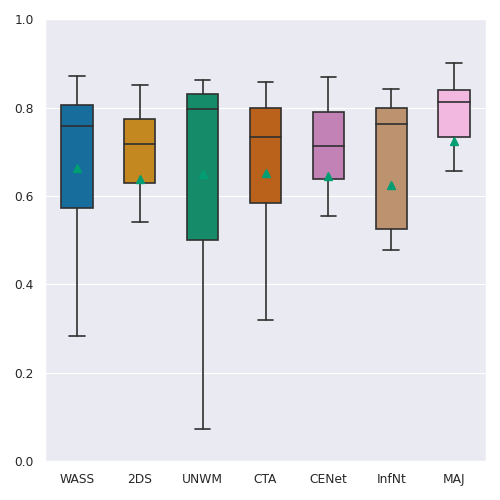} &  
        \includegraphics[width=\linewidth]{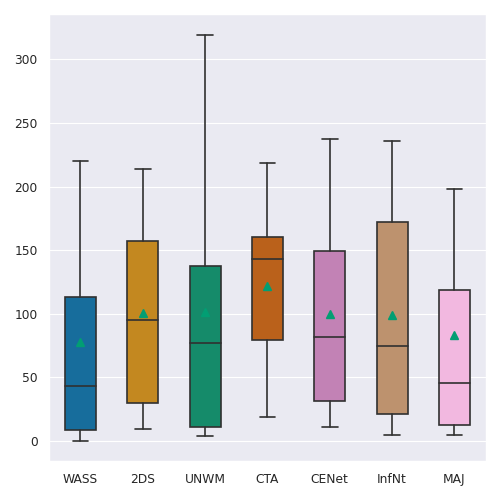} &
        \includegraphics[width=\linewidth]{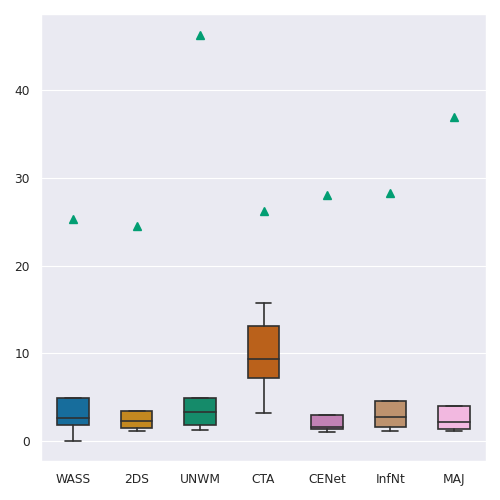} &
        \includegraphics[width=\linewidth]{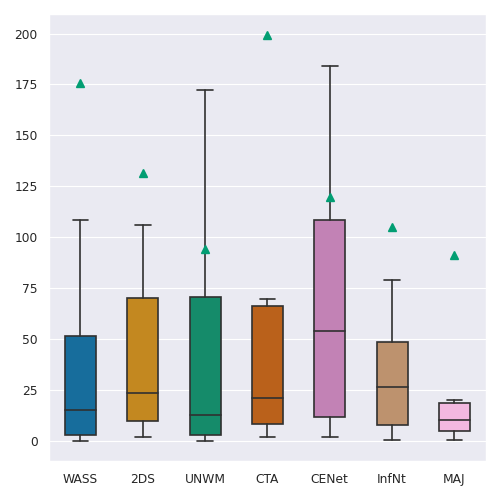}  \\
        
        \rotatebox{90}{\hspace{30pt} CON} &
        \includegraphics[width=\linewidth]{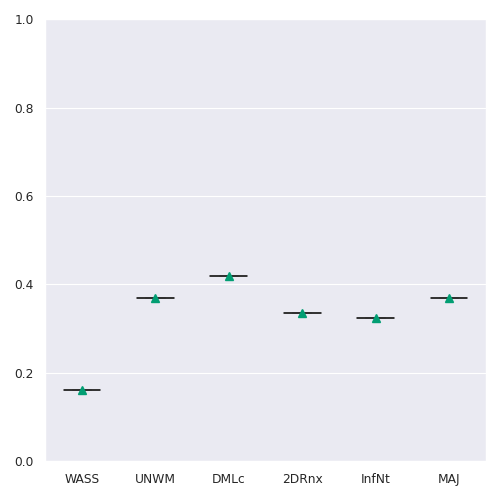} &  
        \includegraphics[width=\linewidth]{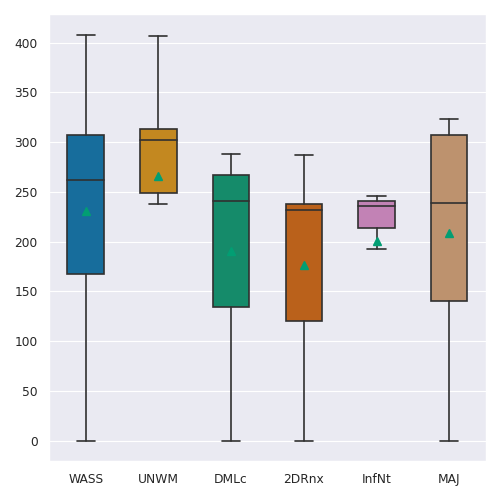} &
        \includegraphics[width=\linewidth]{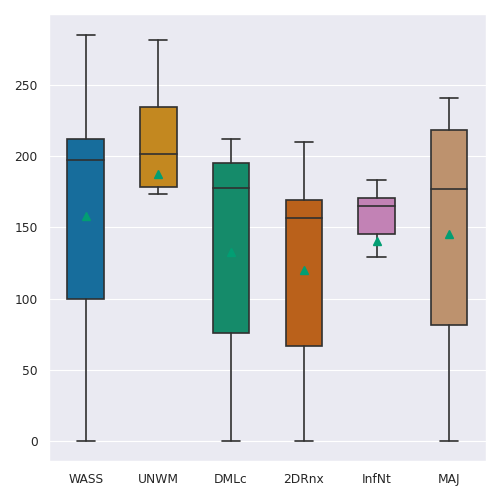} &
        \includegraphics[width=\linewidth]{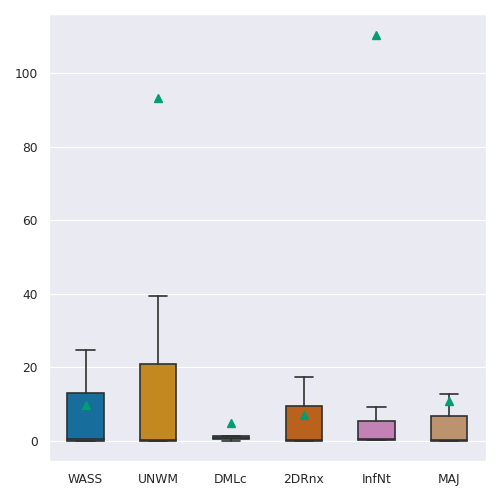} \\
        
        \rotatebox{90}{\hspace{30pt} CPP} &
        \includegraphics[width=\linewidth]{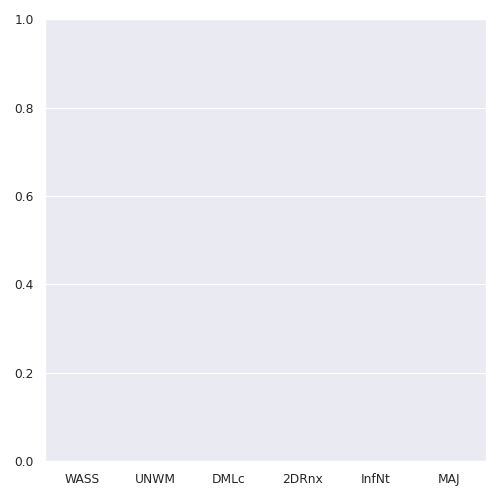} &  
        \includegraphics[width=\linewidth]{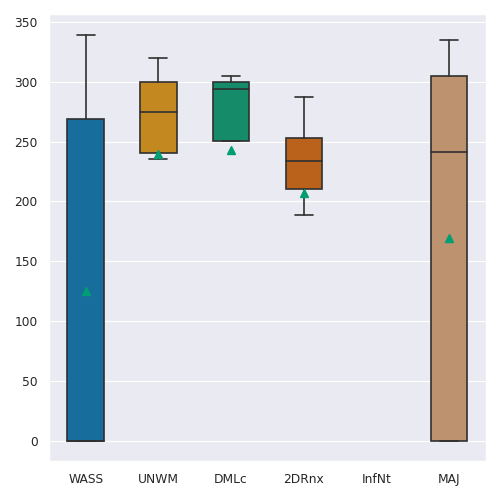} &
        \includegraphics[width=\linewidth]{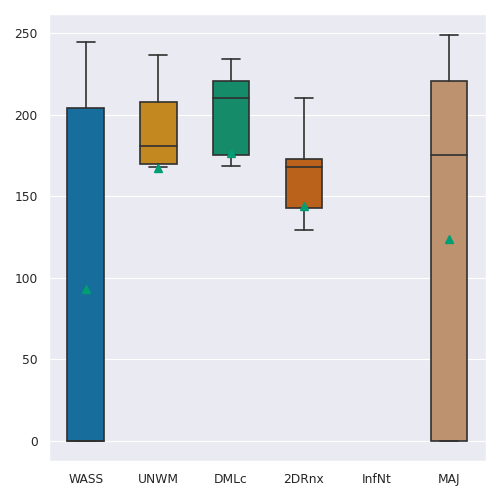} &
        \includegraphics[width=\linewidth]{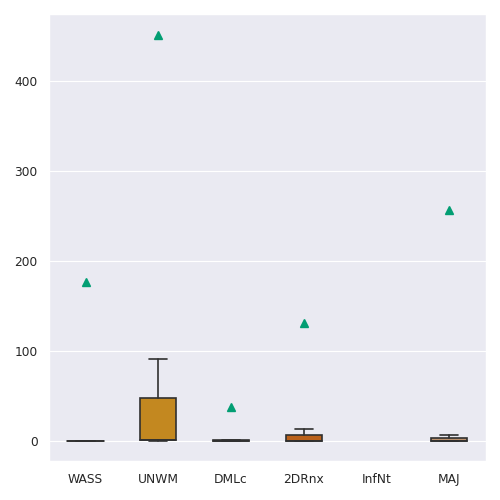} \\
        
        \rotatebox{90}{\hspace{30pt} GGO} &
        \includegraphics[width=\linewidth]{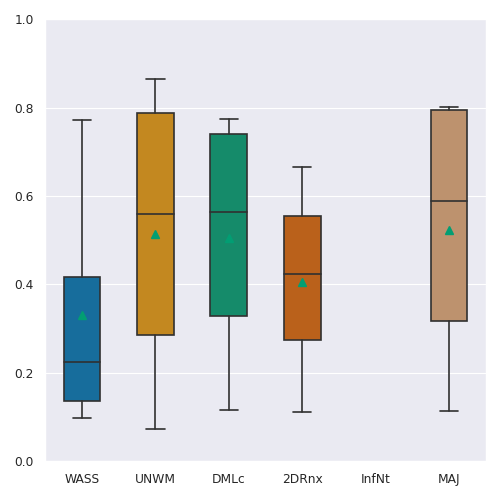} &  
        \includegraphics[width=\linewidth]{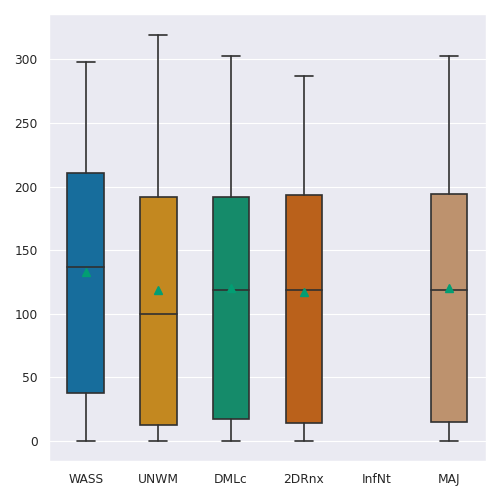} &
        \includegraphics[width=\linewidth]{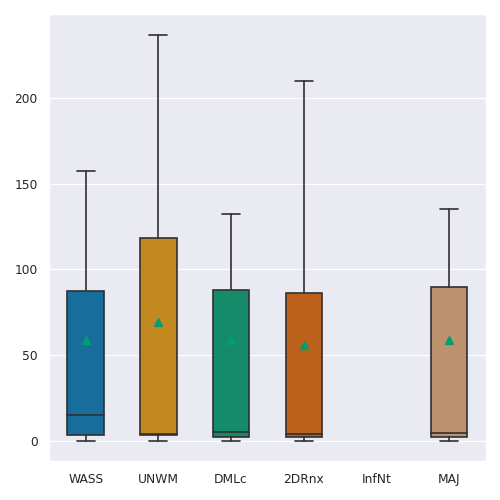} &
        \includegraphics[width=\linewidth]{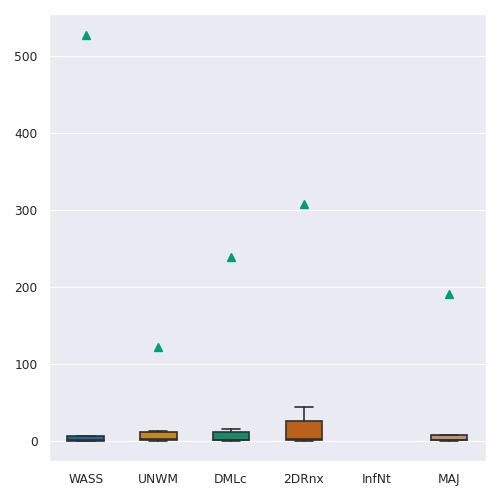} \\ 
        
        \rotatebox{90}{\hspace{30pt} MEAN} &
        \includegraphics[width=\linewidth]{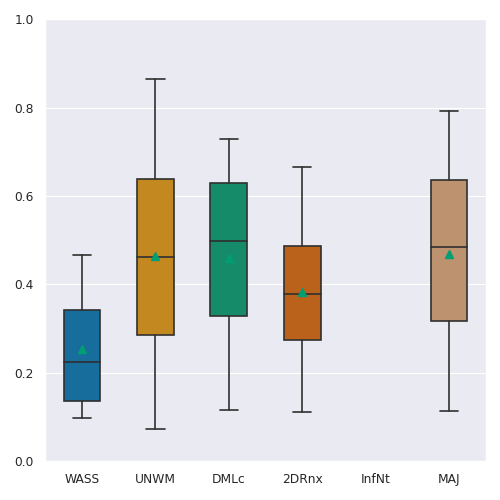} &  
        \includegraphics[width=\linewidth]{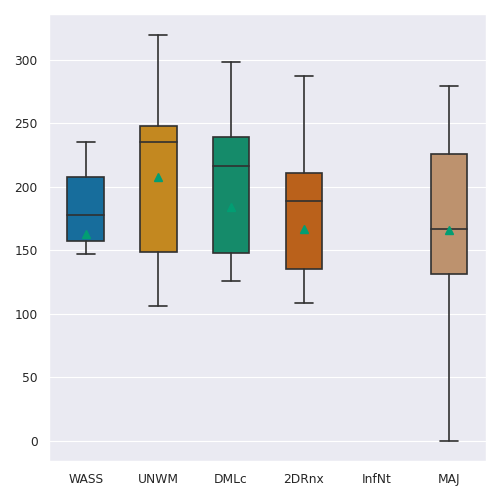} &
        \includegraphics[width=\linewidth]{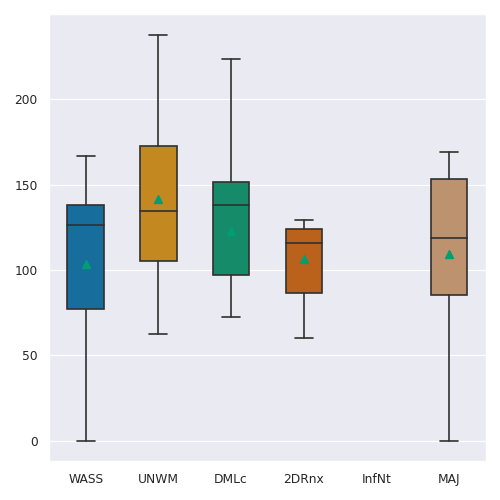} &
        \includegraphics[width=\linewidth]{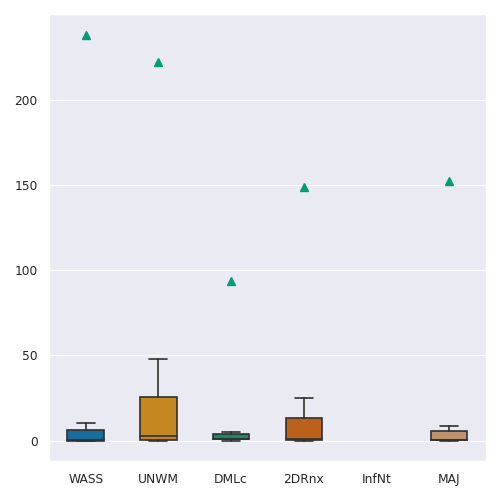} \\
        
        \rotatebox{90}{\hspace{30pt} GGO + CPP} &
        \includegraphics[width=\linewidth]{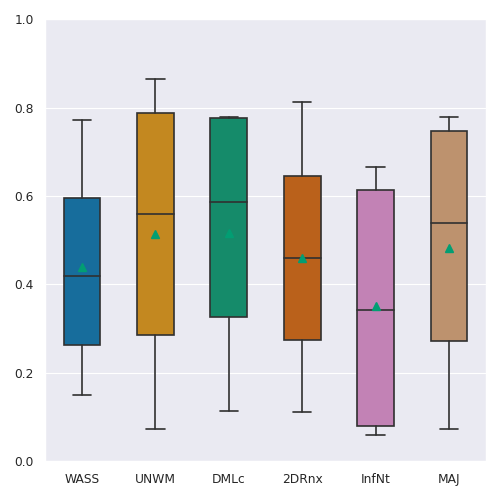} &  
        \includegraphics[width=\linewidth]{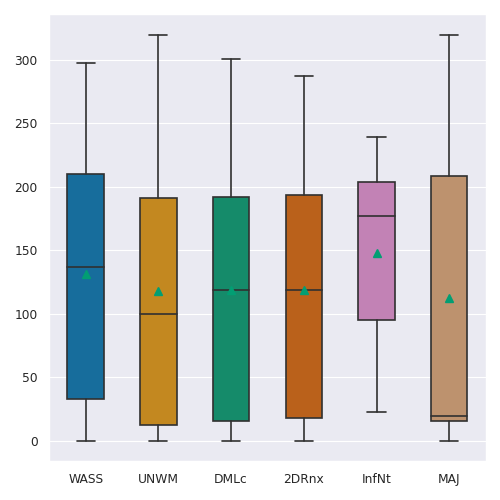} &
        \includegraphics[width=\linewidth]{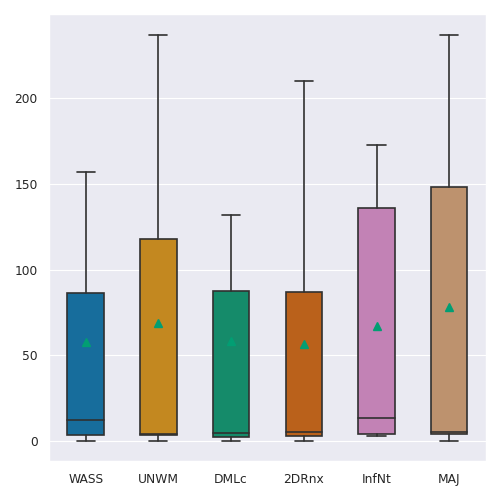} &
        \includegraphics[width=\linewidth]{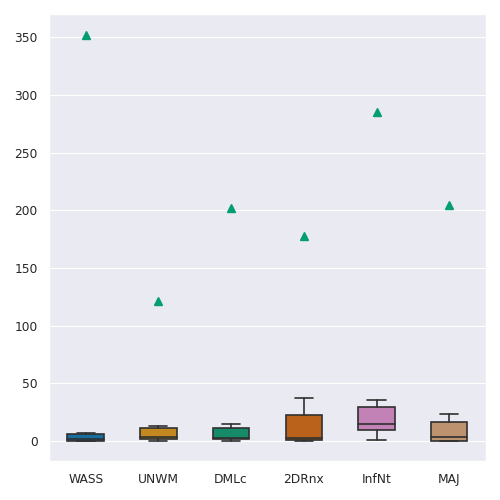} 
        
    \end{tabularx}
    }
    \caption{Boxplots for each of the reported metrics (colums, from left to right: DSC, HD95, ASD and AVD) and challenge tasks (rows, from top to bottom: LUNGS, BIN, CON, CPP, GGO, MEAN of classes and GGO+CPP) on the test set. The mean of the metric for each method is additionally shown on each boxplot by a green triangle and outliers are not shown. }
    \label{fig:boxplots_master_testset}
\end{figure*}

\begin{figure}[H]

    \def\arraystretch{0}
    \setlength{\tabcolsep}{0pt}
    \begin{tabularx}{\linewidth}{YYY}
    
    \includegraphics[width=\linewidth]{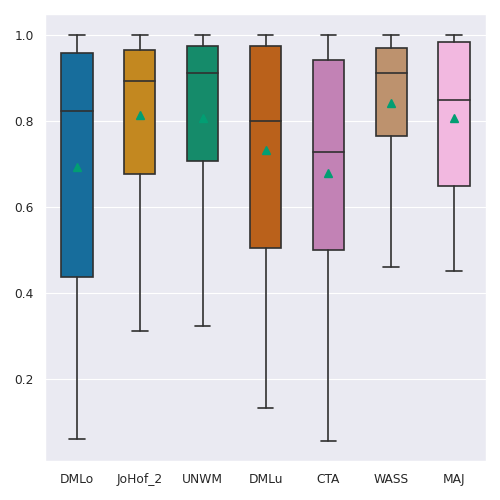} &
    \includegraphics[width=\linewidth]{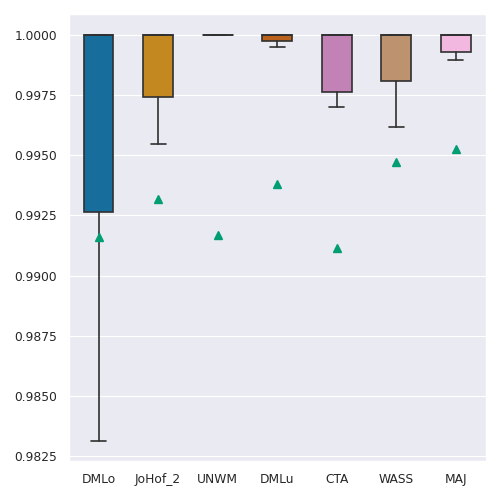} &
    \includegraphics[width=\linewidth]{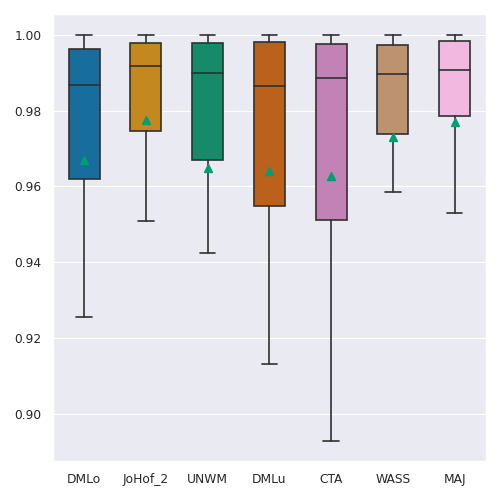} 

    \end{tabularx}
    \caption{Boxplots of the sensitivity of the automated lung masks for each of the lesion classes (CON, CPP and GGO).}
    \label{fig:consolidation_lung}
\end{figure}

\begin{figure}[H]

    \def\arraystretch{0}
    \setlength{\tabcolsep}{0pt}
    \begin{tabularx}{\linewidth}{YYY}
    
    \includegraphics[width=\linewidth]{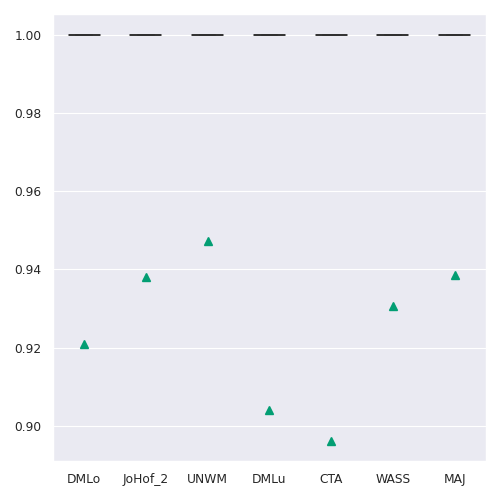} &
    \includegraphics[width=\linewidth]{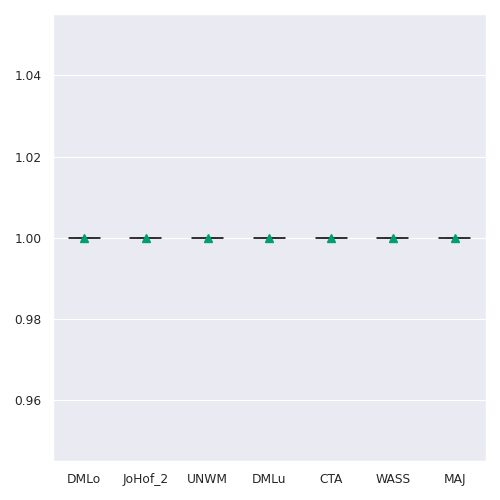} &
    \includegraphics[width=\linewidth]{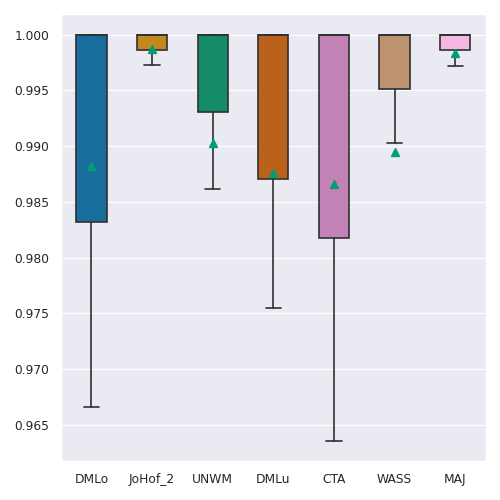} 

    \end{tabularx}
    \caption{Boxplots of the sensitivity of the automated lung masks for each of the lesion classes (CON, CPP and GGO) on the test set.}
    \label{fig:consolidation_lung_testset}
\end{figure}

In this section, boxplots are provided by Figures~\ref{fig:boxplots_master},~\ref{fig:boxplots_master_testset},~\ref{fig:consolidation_lung} and~\ref{fig:consolidation_lung_testset} for each table in the main text. Complementary information such as data distribution, spread and rankings can be assessed more visually from these graphs. 

\subsection*{Confidence intervals}
\label{confidence_intervals_appendix}

In this section, two complementary tables are provided for each table in the main text. These two tables contain respectively the lower and upper 95\% confidence interval boundaries for each mean metric value provided. For lung segmentation Tables~\ref{tab:ci95_lo_lungs}, ~\ref{tab:ci95_hi_lungs}, ~\ref{tab:ci95_lo_lungs_testset} and ~\ref{tab:ci95_hi_lungs_testset}, for binary lesion segmentation Tables~\ref{tab:ci95_lo_bin}, ~\ref{tab:ci95_hi_bin}, ~\ref{tab:ci95_lo_bin_testset} and ~\ref{tab:ci95_hi_bin_testset} and for multiclass lesion segmentation Tables~\ref{tab:ci95_lo_mc}, ~\ref{tab:ci95_hi_mc}, ~\ref{tab:ci95_lo_mc_testset} and ~\ref{tab:ci95_hi_mc_testset}.

\begin{table}[]
    \caption{Lower 95\% confidence interval boundary of metrics for lung segmentation task for cross-validation set. Methods indicated with \# did not use the dataset in Section~\ref{subsec:training_data} for training.}
    \begin{tabularx}{\linewidth}{l|YYYYYYY}
    \toprule
    \emph{method} $\rightarrow$ & DMLo\textsuperscript{\#}    &      & UNWM   &   & CTA\textsuperscript{\#}   &   & MAJ   \\
    metric                      &       &  JoHof\textsuperscript{\#} &       &  DMLu  &       & WASS      &       \\
    \midrule
    DSC &   0.959 &   0.966 &   0.966 &   0.961 &   0.958 &   0.960 &   0.972 \\
    HD95 &   3.01 &   2.07 &   3.26 &   2.79 &   8.35 &   3.39 &   2.01 \\
    ASD &   0.688 &   0.545 &   0.571 &   0.600 &   0.504 &   0.526 &   0.456 \\
    AVD &  42.2 &  33.129 &  45.2 &  46.4 &  33.8 &  17.5 &  25.1 \\
    \midrule
    SEN\textsuperscript{CON} &   0.609 &   0.750 &   0.741 &   0.657 &   0.598 &   0.785 &   0.716 \\
    SEN\textsuperscript{CPP} &   0.986 &   0.989 &   0.984 &   0.988 &   0.983 &   0.991 &   0.991 \\
    SEN\textsuperscript{GGO} &   0.947 &   0.964 &   0.942 &   0.942 &   0.938 &   0.954 &   0.957 \\
    \bottomrule
\end{tabularx}

    \label{tab:ci95_lo_lungs}
\end{table}

\begin{table}[]
    \caption{Upper 95\% confidence interval boundary of metrics for lung segmentation task for cross-validation set. Methods indicated with \# did not use the dataset in Section~\ref{subsec:training_data} for training.}
    \begin{tabularx}{\linewidth}{l|YYYYYYY}
    \toprule
    \emph{method} $\rightarrow$ & DMLo\textsuperscript{\#}    &      & UNWM   &   & CTA\textsuperscript{\#}   &   & MAJ   \\
    metric                      &       &  JoHof\textsuperscript{\#} &       &  DMLu  &       & WASS      &       \\
    \midrule
    DSC &   0.976 &   0.977 &   0.975 &   0.977 &   0.974 &   0.978 &   0.982 \\
    HD95 &   4.98 &   4.796 &  11.3 &   4.79 &  23.7 &  14.0 &   3.53 \\
    ASD &   1.032 &   0.743 &   0.783 &   0.916 &   0.768 &   0.727 &   0.706 \\
    AVD & 104 &  97.6 &  65.5 & 102 &  79.5 &  95.7 &  55.7 \\
    \midrule
    SEN\textsuperscript{CON} &   0.794 &   0.875 &   0.880 &   0.821 &   0.774 &   0.904 &   0.854 \\
    SEN\textsuperscript{CPP} &   0.997 &   0.998 &   0.999 &   0.997 &   0.998 &   0.998 &   0.998 \\
    SEN\textsuperscript{GGO} &   0.988 &   0.991 &   0.989 &   0.988 &   0.989 &   0.993 &   0.994 \\
    \bottomrule
\end{tabularx}

    \label{tab:ci95_hi_lungs}
\end{table}

\begin{table}[]
    \caption{Lower 95\% confidence interval boundary of metrics for lung segmentation task for test set. Sensitivity for CPP is not applicable.}
    \begin{tabularx}{\linewidth}{l|YYYYYYY}
    \toprule
    \emph{method} $\rightarrow$ & DMLo    &      & UNWM   &   & CTA   &   & MAJ   \\
    metric                      &       &  JoHof &       &  DMLu  &       & WASS      &       \\
    \midrule
    DSC &   0.972 &   0.982 &   0.966 &   0.969 &   0.972 &   0.973 &   0.977 \\
    HD95 &   1.75 &   0.987 &   0.635 &   1.81 & 157 &   1.66 &   1.33 \\
    ASD &   0.501 &   0.297 &   0.435 &   0.504 &  76.386 &   0.403 &   0.305 \\
    AVD &  30.3 &  16.382 &  76.3 &  95.8 &  23.2 &  23.0 &  21.9 \\
    \midrule
    SEN\textsuperscript{CON} &   0.766 &   0.817 &   0.844 &   0.716 &   0.693 &   0.795 &   0.757 \\
    SEN\textsuperscript{CPP} & - &  - & - & - & - & - & - \\
    SEN\textsuperscript{GGO} & 0.973 &   0.997 &   0.975 &   0.970 &   0.969 &   0.972 &   0.977 \\
    \bottomrule
\end{tabularx}
    \label{tab:ci95_lo_lungs_testset}
\end{table}

\begin{table}[]
    \caption{Upper 95\% confidence interval boundary of metrics for lung segmentation task for test set. Sensitivity for CPP is not applicable.}
    \begin{tabularx}{\linewidth}{l|YYYYYYY}
    \toprule
    \emph{method} $\rightarrow$ & DMLo    &      & UNWM   &   & CTA   &   & MAJ   \\
    metric                      &       &  JoHof &       &  DMLu  &       & WASS      &       \\
    \midrule
    DSC &   0.981 &   0.986 &   0.978 &   0.980 &   0.983 &   0.984 &   0.987 \\
    HD95 &   2.20 &   1.78 &  10.9 &   2.29 & 180 &   2.20 &   1.83 \\
    ASD &   0.786 &   0.492 &   0.824 &   0.863 &  90.527 &   0.745 &   0.651 \\
    AVD & 164 & 183 & 59.2 & 232 & 168 & 165 & 156 \\
    \midrule
    SEN\textsuperscript{CON} &   1.000 &   1.000 &   1.000 &   1.000 &   1.000 &   1.000 &   1.000 \\
    SEN\textsuperscript{CPP} & - &  - & - & - & - & - & - \\
    SEN\textsuperscript{GGO} &   1.000 &   1.000 &   1.000 &   1.000 &   1.000 &   1.000 &   1.000 \\
    \bottomrule
\end{tabularx}
    \label{tab:ci95_hi_lungs_testset}
\end{table}

\begin{table}[]
    \caption{Lower 95\% confidence interval boundary of metrics for lesion segmentation task for cross-validation set. Methods indicated with \# did not use the dataset in Section~\ref{subsec:training_data} for training.}
    \begin{tabularx}{\linewidth}{l|YYYYYYY}
    \toprule
    \emph{method} $\rightarrow$ &    WASS &      &    UNWM &      &   CENet\textsuperscript{\#} &    &     MAJ \\
    metric &         &    2DS     &         &    CTA\textsuperscript{\#}     &         &    InfNt\textsuperscript{\#}     &         \\
    \midrule
    DSC &   0.486 &   0.444 &   0.466 &   0.411 &   0.426 &   0.442 &   0.502 \\
    HD95 &  38.2 &  56.2 &  35.6 &  83.1 &  46.0 &  43.3 &  32.4 \\
    ASD &   3.82 &   3.04 &   4.36 &   6.41 &   2.84 &   3.37 &   3.37 \\
    AVD &  99.7 &  94.3 & 122 &  98.7 & 196 &  82.7 &  95.3 \\
    \bottomrule
\end{tabularx}
    \label{tab:ci95_lo_bin}
\end{table}

\begin{table}[]
    \caption{Upper 95\% confidence interval boundary of metrics for lesion segmentation task for cross-validation set. Methods indicated with \# did not use the dataset in Section~\ref{subsec:training_data} for training.}
    \begin{tabularx}{\linewidth}{l|YYYYYYY}
    \toprule
    \emph{method} $\rightarrow$ &    WASS &      &    UNWM &      &   CENet\textsuperscript{\#} &    &     MAJ \\
    metric &         &    2DS     &         &    CTA\textsuperscript{\#}     &         &    InfNt\textsuperscript{\#}     &         \\
    \midrule
    DSC &   0.586 &   0.551 &   0.574 &   0.522 &   0.537 &   0.553 &   0.607 \\
    HD95 &  57.9 &  78.4 &  57.8 & 105 &  66.0 &  65.8 &  52.9 \\
    ASD &   7.42 &   4.45 &   6.75 &   7.59 &   4.24 &   5.63 &   6.04 \\
    AVD & 158 & 188 & 192 & 177 & 303 & 141 & 157 \\
    \bottomrule
\end{tabularx}

    \label{tab:ci95_hi_bin}
\end{table}

\begin{table}[]
    \caption{Lower 95\% confidence interval boundary of metrics for lesion segmentation task for test set.}
    \begin{tabularx}{\linewidth}{l|YYYYYYY}
    \toprule
    \emph{method} $\rightarrow$ &    WASS &      &    UNWM &      &   CENet &    &     MAJ \\
    metric &         &    2DS     &         &    CTA     &         &    InfNt     &         \\
    \midrule
    DSC &   0.549 &   0.514 &   0.520 &    0.541 &   0.519 &   0.488 &   0.605 \\
    HD95 &  35.055 &  66.643 &  44.077 &   94.534 &  65.640 &  63.122 &  37.159 \\
    ASD &  0 &   0.843 &   3.211 &    7.170 &   1.045 &   1.235 &   1.714 \\
    AVD & 0 & 0 &  0 & 0 &  15.002 & 0 & 0 \\
    \bottomrule
\end{tabularx}

    \label{tab:ci95_lo_bin_testset}
\end{table}

\begin{table}[]
    \caption{Upper 95\% confidence interval boundary of metrics for lesion segmentation task for test set.}
    \begin{tabularx}{\linewidth}{l|YYYYYYY}
    \toprule
    \emph{method} $\rightarrow$ &    WASS &      &    UNWM &      &   CENet &    &     MAJ \\
    metric &         &    2DS     &         &    CTA     &         &    InfNt     &         \\
    \midrule
    DSC &   0.778 &   0.764 &   0.778 &   0.764 &   0.774 &   0.763 &   0.843 \\
    HD95 & 120.1 & 134 & 159 & 149 & 134 & 135 & 130 \\
    ASD &  54.3 &  48.3 &  89.5 &  45.4 &  55.1 &  55.4 &  72.2 \\
    AVD & 451 & 305 & 189 & 503 & 224 & 235 & 225 \\
    \bottomrule
\end{tabularx}
    \label{tab:ci95_hi_bin_testset}
\end{table}

\begin{table}[]
    \caption{Lower 95\% confidence interval boundary of metrics for multiclass lesion segmentation task for cross-validation set. Methods indicated with \# did not use the dataset in Section~\ref{subsec:training_data} for training.}
    \begin{tabularx}{\linewidth}{ll|YYYYYY}
    \toprule
     & \emph{method} $\rightarrow$   &    WASS &      &  DMmc &   &     InfNt\textsuperscript{\#} & \\
    class & metric &         &  UNWM       &         &   2DRnx      &        & MAJ \\
    \midrule
    CON & DSC &   0.219 &   0.165 &   0.188 &   0.205 &   0.174 &   0.217 \\
         & HD95 &  86.5 &  91.4 &  84.8 &  90.3 &  89.8 &  81.3 \\
         & ASD &  16.9 &  10.6 &  18.3 &   9.51 &  14.7 &  15.1 \\
         & AVD &  23.9 & 151 &  27.1 &  46.8 &  39.5 &  48.3 \\
    CPP & DSC &   0.176 &   0.176 &   0.077 &   0.144 &     - &   0.204 \\
         & HD95 & 155 & 152 & 160 & 153 &     - & 153 \\
         & ASD &  84.4 &  66.8 &  80.4 &  66.9 &     - &  71.9 \\
         & AVD &  35.5 & 242.4 &  55.4 &  61.1 &     - &  82.8 \\
    GGO & DSC &   0.244 &   0.244 &   0.216 &   0.233 &     - &   0.289 \\
         & HD95 &  72.0 &  75.8 &  69.1 &  72.9 &     - &  67.9 \\
         & ASD &  14.0 &  12.3 &  17.5 &  11.6 &     - &  12.6 \\
         & AVD &  66.1 & 192 &  65.1 &  57.9 &     - &  71.0 \\
    MEAN & DSC &   0.256 &   0.237 &   0.219 &   0.240 &     - &   0.283 \\
         & HD95 & 111 & 111 & 112 & 110 &     - & 107 \\
         & ASD &  42.3 &  33.4 &  42.8 &  32.9 &     - &  37.0 \\
         & AVD &  47.7 & 206 &  52.5 &  60.1 &     - &  72.6 \\
    \midrule
    GGO & DSC &   0.342 &   0.317 &   0.366 &   0.347 &   0.309 &   0.361 \\
    +     & HD95 &  62.9 &  66.8 &  54.6 &  59.7 &  69.1 &  54.7 \\
    CPP     & ASD &   8.41 &   7.66 &   7.92 &   7.68 &   5.86 &  10.2 \\
         & AVD &  54.8 & 148 &  62.7 &  66.6 &  79.9 &  48.5 \\

    \bottomrule
\end{tabularx}

    \label{tab:ci95_lo_mc}
\end{table}

\begin{table}[]
    \caption{Upper 95\% confidence interval boundary of metrics for multiclass lesion segmentation task for cross-validation set. Methods indicated with \# did not use the dataset in Section~\ref{subsec:training_data} for training.}
    \begin{tabularx}{\linewidth}{ll|YYYYYY}
    \toprule
     & \emph{method} $\rightarrow$   &    WASS &      &  DMmc &   &     InfNt\textsuperscript{\#} & \\
    class & metric &         &  UNWM       &         &   2DRnx      &        & MAJ \\
    \midrule
    CON & DSC &   0.335 &   0.264 &   0.307 &   0.314 &   0.283 &   0.327 \\
         & HD95 & 118 & 122 & 119 & 119 & 122 & 113 \\
         & ASD &  36.1 &  28.2 &  39.9 &  27.0 &  32.2 &  36.5 \\
         & AVD &  79.1 & 250 &  73.3 & 103 &  78.6 &  98.7 \\
    CPP & DSC &   0.364 &   0.387 &   0.242 &   0.327 &     - &   0.402 \\
         & HD95 & 203 & 178 & 199 & 178. &     - & 192 \\
         & ASD & 124 &  95.2 & 114 &  95.5 &     - & 106 \\
         & AVD & 100 & 397 & 159 & 122 &     - & 160 \\
    GGO & DSC &   0.352 &   0.368 &   0.324 &   0.331 &     - &   0.406 \\
         & HD95 & 101 & 105 &  98.1 & 101 &     - &  98.0 \\
         & ASD &  33.2 &  31.5 &  38.1 &  30.8 &     - &  32.8 \\
         & AVD & 119 & 383 & 194 & 104 &     - & 200 \\
    MEAN & DSC &   0.333 &   0.313 &   0.291 &   0.309 &     - &   0.363 \\
         & HD95 & 134 & 130 & 131 & 128 &     - & 128 \\
         & ASD &  60.6 &  48.0 &  59.7 &  47.4 &     - &  54.5 \\
         & AVD &  93.4 & 332 & 139 & 105 &     - & 147 \\
    \midrule
    GGO & DSC &   0.464 &   0.444 &   0.490 &   0.461 &   0.419 &   0.487 \\
    +     & HD95 &  91.2 &  96.1 &  82.7 &  88.8 &  95.7 &  85.1 \\
    CPP     & ASD &  24.8 &  24.3 &  25.6 &  24.7 &  20.8 &  29.7 \\
         & AVD & 103 & 275 & 118 & 126 & 135 & 103 \\

    \bottomrule
\end{tabularx}

    \label{tab:ci95_hi_mc}
\end{table}

\begin{table}[]
    \caption{Lower 95\% confidence interval boundary of metrics for multiclass lesion segmentation task for test set.}
    \begin{tabularx}{\linewidth}{ll|YYYYYY}
    \toprule
     & \emph{method} $\rightarrow$   &    WASS &      &  DMmc &   &     InfNt & \\
    class & metric &         &  UNWM       &         &   2DRnx      &        & MAJ \\
    \midrule
    CON  & DSC &      - &      - &      - &      - &      - &      - \\
         & HD95 &  126 &  180 &  106 &   96.5 &  139 &  110 \\
         & ASD &   76.9 &  122 &   64.9 &   57.6 &   94.3 &   69.7 \\
         & AVD &   0 &  0 &   0 &   0 & 0 &   0 \\
    CPP  & DSC &      - &      - &      - &      - &      - &      - \\
         & HD95 &    7.89 &  158 &  162 &  136 &      - &   50.3 \\
         & ASD &    5.88 &  110 &  116 &   93.8 &      - &   36.4 \\
         & AVD & 0 & 0 &  0 & 0 &      - & 0 \\
    GGO  & DSC &    0.028 &    0.154 &    0.202 &    0.171 &      - &    0.195 \\
         & HD95 &   49.8 &   25.6 &   34.6 &   34.3 &      - &   34.0 \\
         & ASD &   0 &   0 &   0 &   0 &      - &   0 \\
         & AVD & 0 & 0 & 0 & 0 &      - & 0 \\
    MEAN & DSC &    0.090 &    0.137 &    0.199 &    0.158 &      - &    0.185 \\
         & HD95 &  105 &  152 &  111 &   98.9 &      - &   97.8 \\
         & ASD &   61.3 &   98.3 &   70.9 &   58.8 &      - &   65.5 \\
         & AVD & 0 & 0 &  0 & 0 &      - & 0 \\
    \midrule
    GGO  & DSC &    0.172 &    0.154 &    0.201 &    0.161 &    0.035 &    0.151 \\
    +    & HD95 &   47.6 &   25.6 &   33.9 &   36.1 &   89.3 &    5.46 \\
    CPP  & ASD &   0 &   0 &   0 &   0 &    9.428 &   0 \\
         & AVD & 0 & 0 & 0 & 0 & 0 & 0 \\
    \bottomrule
\end{tabularx}

    \label{tab:ci95_lo_mc_testset}
\end{table}

\begin{table}[]
    \caption{Upper 95\% confidence interval boundary of metrics for multiclass lesion segmentation task for test set.}
    \begin{tabularx}{\linewidth}{ll|YYYYYY}
    \toprule
     & \emph{method} $\rightarrow$   &    WASS &      &  DMmc &   &     InfNt & \\
    class & metric &         &  UNWM       &         &   2DRnx      &        & MAJ \\
    \midrule
    CON  & DSC &      - &      - &     - &     - &     - &     - \\
         & HD95 &  336 &  353 & 275 & 256 & 263 & 306 \\
         & ASD &  239 &  254 & 201 & 182 & 187 & 221 \\
         & AVD &   22.3 &  263 &  13.1 &  15.9 & 323 &  27.7 \\
    CPP  & DSC &      - &      - &     - &     - &     - &     - \\
         & HD95 &  243 &  321 & 324 & 278 &     - & 289 \\
         & ASD &  181 &  225 & 237 & 195 &     - & 211 \\
         & AVD &  521 & 1304 & 109 & 384 &     - & 755 \\
    GGO  & DSC &    0.631 &    0.873 &   0.807 &   0.641 &     - &   0.852 \\
         & HD95 &  216 &  211 & 205 & 201 &     - & 206 \\
         & ASD &  124 &  141 & 123 & 117 &     - & 124 \\
         & AVD & 1559 &  351 & 699 & 893 &     - & 558 \\
    MEAN & DSC &    0.416 &    0.792 &   0.721 &   0.608 &     - &   0.753 \\
         & HD95 &  221 &  264 & 257 & 235 &     - & 234 \\
         & ASD &  145 &  185 & 175 & 154 &     - & 153 \\
         & AVD &  700 &  639 & 274 & 430 &     - & 447 \\
    \midrule
    GGO  & DSC &    0.708 &    0.873 &   0.833 &   0.761 &   0.670 &   0.813 \\
    +    & HD95 &  216 &  211 & 204 & 201 & 207 & 220 \\
    CPP  & ASD &  124 &  140 & 123 & 118 & 124 & 158 \\
         & AVD & 1039 &  350 & 590 & 511 & 814 & 593 \\
    \bottomrule
\end{tabularx}

    \label{tab:ci95_hi_mc_testset}
\end{table}

\end{document}